\documentclass[twocolumn,floatfix]{aastex63}
\def\lowzr{$1.30<z<1.50$}
\def\highzr{$1.92<z<2.28$}
\def\mr{$M_{\rm dyn}/M_{\rm *,\rm c}$}
\def\mdyn{$M_{\rm dyn}$}
\def\niiha{[N\,{\sc ii}]/H$\alpha$}

\shorttitle{The Heavy Metal Survey}
\shortauthors{Kriek et al.}
\usepackage{natbib}
\usepackage{threeparttablex}
\usepackage{tablefootnote}
\usepackage{textgreek}

\usepackage{lineno}

\begin{document}
  
\title{The Heavy Metal Survey: Star Formation Constraints and Dynamical Masses of 21 Massive Quiescent Galaxies at $z=1.3-2.3$}
 
\author[0000-0002-7613-9872]{Mariska Kriek}
\affiliation{Leiden Observatory, Leiden University, P.O. Box 9513, 2300 RA Leiden, The Netherlands}

\author[0000-0002-9861-4515]{Aliza G. Beverage}
\affiliation{Astronomy Department, University of California, Berkeley, CA 94720, USA}

\author[0000-0002-0108-4176]{Sedona H. Price}
\affiliation{Department of Physics \& Astronomy and PITT PACC, University of Pittsburgh, Pittsburgh, PA 15260, USA}

\author[0000-0002-1714-1905]{Katherine A. Suess}
\affiliation{Kavli Institute for Particle Astrophysics and Cosmology and Department of Physics, Stanford University, Stanford, CA 94305, USA}

\author[0000-0001-6813-875X]{Guillermo Barro}
\affiliation{University of the Pacific, Stockton, CA 90340 USA}

\author[0000-0001-5063-8254]{Rachel S. Bezanson}
\affiliation{Department of Physics \& Astronomy and PITT PACC, University of Pittsburgh, Pittsburgh, PA 15260, USA}

\author[0000-0002-1590-8551]{Charlie Conroy}
\affiliation{Center for Astrophysics \textbar\ Harvard \& Smithsonian, Cambridge, MA, 02138, USA}

\author[0000-0000-0000-0000]{Sam E. Cutler}
\affiliation{Department of Astronomy, University of Massachusetts, Amherst, MA 01003, USA}

\author[0000-0002-8871-3026]{Marijn Franx}
\affiliation{Leiden Observatory, Leiden University, P.O. Box 9513, 2300 RA Leiden, The Netherlands}

\author{Jamie Lin}
\affiliation{Department of Physics and Astronomy, Tufts University, 574 Boston Avenue, Medford, MA 02155, USA}

\author[0000-0002-5337-5856]{Brian Lorenz}
\affiliation{Astronomy Department, University of California, Berkeley, CA 94720, USA}

\author[0000-0002-0463-9528]{Yilun Ma}
\affiliation{Department of Astrophysical Sciences, Princeton University, Princeton, NJ 08544, USA}

\author[0000-0003-1665-2073]{Ivelina G. Momcheva}
\affiliation{Max-Planck-Institut f{\"u}r Astronomie, K{\"o}nigstuhl 17, D-69117, Heidelberg, Germany}

\author[0000-0002-8530-9765]{Lamiya A. Mowla}
\affiliation{Astronomy Department, Whitin Observatory, Wellesley College, 106 Central Street, Wellesley, MA 02481, USA}

\author[0000-0002-7075-9931]{Imad Pasha}
\affiliation{Department of Astronomy, Yale University, New Haven, CT 06511, USA}

\author[0000-0002-8282-9888]{Pieter van Dokkum}
\affiliation{Department of Astronomy, Yale University, New Haven, CT 06511, USA}

\author[0000-0001-7160-3632]{Katherine E. Whitaker}
\affiliation{Department of Astronomy, University of Massachusetts, Amherst, MA 01003, USA}
\affiliation{Cosmic Dawn Center (DAWN), Niels Bohr Institute, University of Copenhagen, Jagtvej 128, K{\o}benhavn N, DK-2200, Denmark}

\begin{abstract}
In this paper, we present the Heavy Metal Survey, which obtained ultradeep medium-resolution spectra of 21 massive quiescent galaxies at $1.3<z<2.3$ with \textit{Keck}/LRIS and MOSFIRE. With integration times of up to 16\,hr per band per galaxy, we observe numerous Balmer and metal absorption lines in atmospheric windows. We successfully derive spectroscopic redshifts for all 21 galaxies and for 19 we also measure stellar velocity dispersions ($\sigma_v$), ages, and elemental abundances, as detailed in an accompanying paper. Except for one emission-line active galactic nucleus, all galaxies are confirmed as quiescent through their faint or absent H$\alpha$ emission and evolved stellar spectra. For most galaxies exhibiting faint H$\alpha$, elevated \niiha\ suggests a non-star-forming origin. We calculate dynamical masses (\mdyn) by combining $\sigma_v$ with structural parameters obtained from \textit{HST}/COSMOS(-DASH), and compare them with stellar masses ($M_*$) derived using spectrophotometric modeling, considering various assumptions. For a fixed initial mass function (IMF), we observe a strong correlation between \mdyn/$M_*$ and $\sigma_v$. This correlation may suggest that a varying IMF, with high-$\sigma_v$ galaxies being more bottom heavy, was already in place at $z\sim2$. When implementing the $\sigma_v$-dependent IMF found in the cores of nearby early-type galaxies \textit{and} correcting for biases in our stellar mass and size measurements, we find a low scatter in \mdyn/$M_*$ of 0.14\,dex. However, these assumptions result in unphysical stellar masses, which exceed the dynamical masses by 34\%. This tension suggests that distant quiescent galaxies do not simply grow inside-out into today's massive early-type galaxies and the evolution is more complicated.
\end{abstract}

\keywords{Galaxy evolution --- Galaxy formation}

\section{INTRODUCTION}\label{sec:int}

The majority of stars in today’s universe live in early-type galaxies with quiescent stellar populations \citep[e.g.,][]{AMuzzin2013b}. These galaxies are massive, large, exhibit little-to-no rotation, and are thought to have formed the majority of their stars at high redshifts \citep[e.g.,][]{DThomas2005, RMcDermid2015}. Nonetheless, despite the wealth of information from low-redshift studies, the formation histories of massive early-type galaxies are still poorly understood.

To quantify the growth of massive galaxies and understand the physical processes driving this evolution, it is imperative to directly observe them during their early stages.
Such studies find that massive galaxies with quiescent stellar populations already exist when the Universe was only a fraction of its current age. These distant quiescent galaxies were first identified almost two decades ago
\citep[e.g.,][]{MFranx2003,ACimatti2004,KGlazebrook2004}  and have been found to dominate the massive end of the galaxy distribution out to $z\sim2.5$ \citep[e.g.,][]{AMuzzin2013b,ATomczak2014,DMcLeod2021}. Galaxy formation models originally failed to predict this quiescent galaxy population and -- almost two decades later -- are still struggling to explain their presence. 

Our poor understanding of this galaxy population is primarily due to the difficulty of obtaining high-quality spectra. Quiescent galaxies typically do not have bright emission lines and thus, we rely on faint stellar absorption features to measure redshifts and learn about their stellar, chemical, and kinematic properties. Obtaining such spectra is even more challenging at $z\gtrsim 1$, as the bulk of the stellar spectrum is shifted to near-IR wavelengths.

A few years after their initial discovery, the first spectra for $z\sim 2$ quiescent galaxies were obtained, showing broad continuum features \citep[e.g.,][]{EDaddi2005,MKriek2006}. The first direct detection of Balmer and metal absorption lines took nearly 30\,hr with GNIRS (Gemini-South) on one galaxy and the lines were still only marginally detected 
\citep[e.g.,][]{MKriek2009b,PvanDokkum2009}. 
With the advent of more efficient near-IR spectrographs such as X-Shooter \citep[VLT; ][]{JVernet2011}, KMOS \citep[VLT; ][]{RSharples2013} and MOSFIRE \citep[Keck Observatory;][]{IMcLean2012}, spectroscopic redshifts, robust velocity dispersion measurements, and dynamical mass estimates became available for substantial samples or stacks of distant quiescent galaxies \citep[e.g.,][]{JvandeSande2011,JvandeSande2013,RBezanson2013,SBelli2014,SBelli2017,ILonoce2015,MOnodera2015, JMendel2015,PSaracco2019,ACarnall2022,ZZhuang2023,MPark2023}.

In recent years, spectroscopic studies have pushed to even higher redshifts \citep[$z>3$; e.g.,][]{KGlazebrook2017, CSchreiber2018, MTanaka2019, BForrest2020, JEsdaile2021,ACarnall2023,JAntwi-Danso2023}. At the same time, deeper observations have enabled the first measurements of chemical abundances and resolved stellar kinematics at $z>2$. These initial studies show intriguing results and demonstrate the power of using such measurement to gain insights into the formation mechanisms of distant quiescent galaxies. First, they have old ages and extreme chemical abundance patterns \citep{MKriek2016, MJafariyazani2020}, indicating that they formed their stars in early vigorous bursts, followed by an efficient quenching process. Second, they appear to be rotationally supported \citep{ANewman2015,ANewman2018b,SToft2017}. 

However, these studies are based on few very massive and/or lensed galaxies, and many questions remain. We do not know how these galaxies became so massive at such early epochs, when and how fast they formed and assembled their mass, whether they are supported by rotation or random motions, which physical processes are responsible for halting their star formation, and how they evolved into the massive early-type galaxies in the today’s universe. Addressing these questions requires statistical samples of distant quiescent galaxies with ultra-deep spectra covering several Balmer and metal absorption lines, which enable stellar, chemical, and kinematic measurements.

In order to obtain such a spectroscopic galaxy sample, we have conducted the Heavy Metal Survey with MOSFIRE \citep{IMcLean2012} and LRIS \citep{JOke1995} on the Keck I telescope.  The Heavy Metal survey observes 21  ``bright'' quiescent galaxies selected to be in two redshift intervals, \lowzr\ and \highzr, as well as many more star-forming and fainter quiescent galaxies at similar redshifts. With integration times of up to 16\,hr per filter per mask, we observe numerous Balmer and metal absorption lines. While this distant quiescent galaxy sample is not as large as the sample by \citet[][$\sim$30 galaxies at $z>1.35$]{SBelli2017,SBelli2019}, it is unique for its wavelength coverage and the only survey so far that obtains ultradeep spectra at rest-frame $\sim4800-5400$\,\AA\ for a sample of distant quiescent galaxies. This wavelength range targets the strongest $\alpha$-element absorption line (i.e., Mgb) in the rest-frame optical, as well as several prominent Fe lines.  

In this paper we present our survey design and observational strategy, data reduction and overview (Section~\ref{sec:obs}), methods to derive spectral properties (Section~\ref{sec:measurements}) and characteristic of the galaxy sample (Section~\ref{sec:results}), and discuss the implications of our finding for galaxy evolution studies (Section~\ref{sec:discussion}). The primary science applications of this data set, the chemical abundance measurements, will be presented in an accompanying paper \citep{ABeverage2024}. The spectra and chemical abundances for the primary galaxies in first Heavy Metal mask were also presented in \cite{MKriek2019}. Several other science applications including molecular gas properties and active galactic nuclei (AGN) outflows will be presented in future papers (K. Suess et al. in prep; Y. Ma et al in prep). 

Throughout this work we assume a $\Lambda$CDM cosmology with $\Omega_{\rm m}= 0.3$, $\Omega_\Lambda=0.7$, and $H_0 =70\rm \, km s^{-1} \,Mpc^{-1}$. All magnitudes are given in the AB-magnitude system \citep{JOke1983}. The wavelengths of all emission and absorption lines are given in vacuum.

\begin{figure}
  \begin{center}  
    \includegraphics[width=0.48\textwidth]{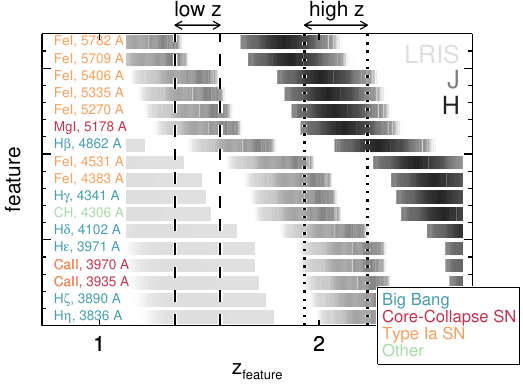}
    \caption{LRIS and MOSFIRE visibility of various rest-frame optical absorption features as a function of redshift. Each row represents a different spectral feature, as indicated on the left. The color of the feature reflects the primary origin of the chemical element, as indicated in the bottom-right box. The gray bars indicate whether a feature is visible in a certain filter, with the different shades of gray corresponding to the different filters (as indicated in the top right) and the gradations for each filter indicating the throughput. The Heavy Metal low (\lowzr) and high (\highzr) redshift intervals are indicated by the dashed and dotted vertical lines, respectively.\label{fig:features}}
    \end{center}  
\end{figure}

\section{Observations and Data}\label{sec:obs}
\subsection{Survey Design}
The Heavy Metal survey aims to study the formation histories of massive quiescent galaxies using stellar, chemical, and kinematic measurements. Achieving this goal requires (i) a statistically significant sample of $\sim20$ distant quiescent galaxies with (ii) ultradeep rest-frame optical spectroscopy covering several hydrogen, iron and $\alpha$-element absorption features, and (iii) ancillary datasets including ultradeep multiwavelength photometry and high-resolution imaging.

\begin{table*}  
\begin{center}
\caption{Overview Observations and Data \label{tab:obs}}
\begin{tabular}{l l c c c c l c c c}
\hline\hline
Mask & Instrument & R.A. & Decl. & P.A. & $N_{\rm gal}$ & Filter & Semesters & $t$ & FWHM \\
&  & (hh:mm:ss) & (dd:mm:ss) & (deg) & ($N_{\rm prim}$) & & & (hr) & seeing\\
\hline
{\bf \noindent Heavy Metal 1} & LRIS & 10:00:42.61 & 02:34:43.97 & 93.0 & 19(5) & red & 2016B & 4.4 & 0\farcs9 \\
(HM1, $z\sim1.4$)  & MOSFIRE & 10:00:42.01 & 02:34:54.30 & 86.7 & 29(5) & J & 2017A & 11.8 & 0\farcs63 \\
   & & & & & &  H & 2017A & 0.86 & 0\farcs75 \\
\hline
{\bf Heavy Metal 2} & LRIS & 10:00:44.88 & 01:45:14.24 & -75.0 & 18(6) & red & 2021AB & 6.3/6.5$^{\rm a}$ & 0\farcs9\\
(HM2, $z\sim1.4$) & MOSFIRE & 10:00:42.59 & 01:45:17.24 & 127.3 & 22(6) & J & 2018B & 11.5 & 0\farcs66 \\
   & & & & & &  H & 2018B & 0.92 & 0\farcs76 \\
\hline
{\bf Heavy Metal 3} & MOSFIRE & 10:00:44.70 & 02:01:35.48 & -23.1 & 20(5) & J & 2018B, 2019A, 2020B & 14.4 & 0\farcs83\\
(HM3, $z\sim2.1$) & & & & & &  H & 2018B, 2019A &16.7 & $0\farcs66$ \\
 & & & & & &  K &  2018B, 2019A &  2.2 & 0\farcs77 \\
\hline
{\bf Heavy Metal 4} & MOSFIRE & 09:59:13.80 & 01:48:58.90 & 77.6 & 22(5) & J & 2019AB, 2021A & 12.4 & 0\farcs73\\
(HM4, $z\sim2.1$) & & & & & & H & 2019AB, 2020B & 15.5 & 0\farcs59\\
& & & & & & K & 2019B, 2021A & 2.6 & 0\farcs89\\
\hline\hline
\end{tabular}
\end{center}
\footnotesize
$^{\rm a}$In 2021A one of the red detectors was not working. To observe all galaxies, we used two masks, which differed by 180 degrees, and only one of the detectors. The masks were observed for 3.5 and 3.75 hr. In 2021B a new detector had been installed, and we reobserved the original mask for 2.75 hr. Hence, the integration times varied per galaxy, depending on the location in the mask.
\normalsize
\end{table*}

To that end, we executed the Heavy Metal survey in the overlapping area of the UltraVISTA \citep{HMcCracken2012}, COSMOS \citep{NScoville2007}, and COSMOS-DASH \citep{IMomcheva2017,LMowla2018} surveys, using the LRIS and MOSFIRE spectrometers on the Keck I telescope. The UltraVISTA survey provides deep multiwavelength photometry, while the F814W and F160W imaging from COSMOS and COSMOS-DASH reveals the rest-frame optical structures of distant galaxies. For our selection we used the COSMOS UltraVISTA v4.1 catalog by \citet{AMuzzin2013a}. Quiescent galaxies were identified by their rest-frame $U-V$ and $V-J$ colors \citep[e.g.,][]{SWuyts2007,RWilliams2009}. In this work, we use the $UVJ$ criteria by \cite{AMuzzin2013b}. 

We select the targets to be at  \lowzr\ or \highzr.  These specific redshift intervals are chosen such that we observe MgI at 5178\,\AA\ and several FeI and Balmer absorption lines in atmospheric windows, as illustrated in Figure~\ref{fig:features}. Furthermore, by using two redshift intervals, combined with deep spectroscopic surveys at lower redshifts such as LEGA-C at $0.5<z<1.0$ \citep{AvanderWel2016,AvanderWel2021}, we can study evolutionary trends. 
For the \lowzr\ galaxies, we use LRIS-RED and MOSFIRE J-band to observe the 4000\,\AA\ break region and the region around MgI at 5178\,\AA, respectively. For the \highzr\ galaxies, we target these same regions with MOSFIRE in the J and H bands. We also obtained shallower spectra in the H and K bands for the low- and high-redshift masks, respectively, to obtain additional constraints on several emission lines (i.e., H$\alpha$, [NII]). 

Both LRIS and MOSFIRE are among the most efficient spectrographs at their respective wavelengths. Nonetheless, even with unprecedented integration times, only the brightest galaxies are within reach.  The \lowzr\ galaxies are selected to be brighter than $J=21.6$ and the galaxies at \highzr\ are selected to be brighter than $H=21.8$.  These magnitudes limits, combined with the long integration times (see next section), ensure sufficient signal-to-noise ratios (S/N) to facilitate the anticipated science. There are $\sim100$ and $\sim50$ quiescent galaxy candidates that meet our criteria in the low- and high-redshift intervals, respectively.

\begin{figure*}
  \begin{center}  
  \includegraphics[width=1.\textwidth]{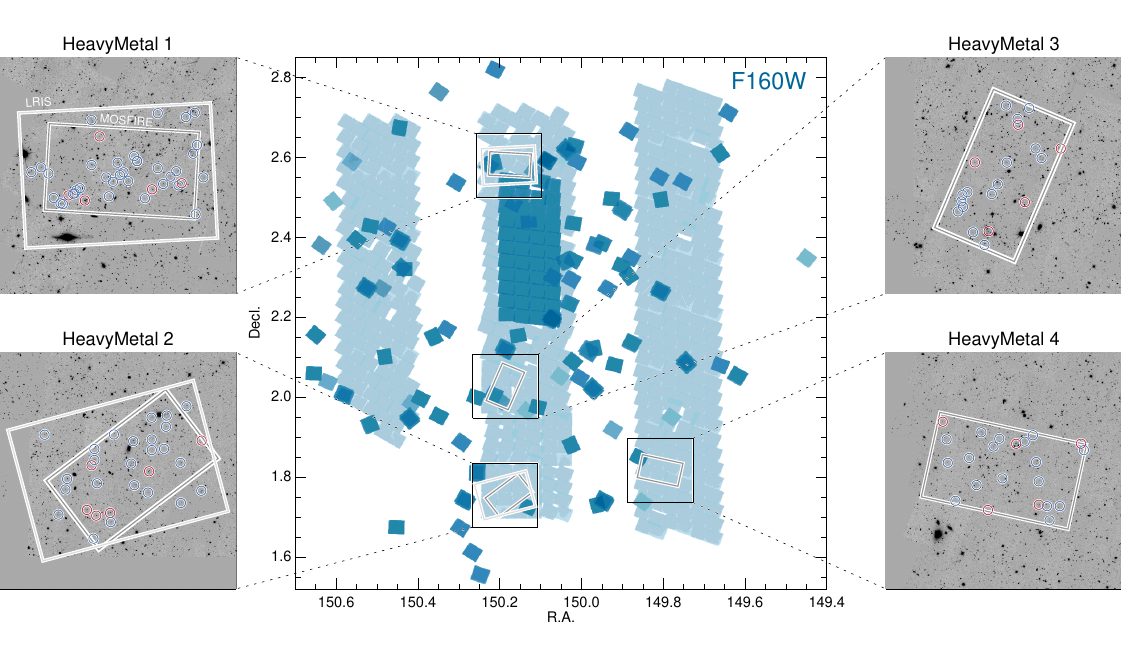}  
  \caption{Footprints of the Heavy Metal observations in the larger COSMOS field. The middle panel shows the weight map of all publicly available HST/F160W imaging, constructed by the COSMOS-DASH collaboration \citep{IMomcheva2017}. The dark-blue contiguous area represents the  CANDELS survey \citep{AKoekemoer2011,NGrogin2011}. The three larger and lighter stripes represent the shallower COSMOS-DASH survey, which overlaps with the deep UltraVISTA stripes. The smaller dark-gray rectangles represents the MOSFIRE field of view for all four Heavy Metal pointings. For Heavy Metal 1 and 2, both targeting lower redshifts ($z\sim1.4$) we also show the LRIS field of view by the larger, light-gray rectangles. For each pointing we show the zoom-in panels to the left or right of the primary panel. In the zoom panels, we show the COSMOS-DASH/F160W images and indicate the primary (red circles) and filler (blue circles) targets. }\label{fig:pointing}
    \end{center}  
\end{figure*}

The large survey area enabled us to identify four pointings for which we observe at least bright five distant quiescent galaxies, simultaneously. Two pointings target galaxies at \lowzr\ and the other two target galaxies at \highzr. In total, we have 21 primary targets. No other pointing allowed the observation of at least 5 primary targets within a single MOSFIRE field of view. The four pointings are shown in Figure~\ref{fig:pointing}, in comparison to the photometric coverage in HST/WFC3-IR F160W. In Table~\ref{tab:obs} we list the mask parameters of all LRIS and MOSFIRE masks. This sample size is an improvement of an order of magnitude compared to the 2 galaxies at $z\sim2$ for which spectra of comparable depth and wavelength coverage were previously available \citep{MKriek2006,MJafariyazani2020}.

The remaining slits were placed on fainter quiescent and star-forming galaxies. We prioritized galaxies at similar redshift. For the Heavy Metal 3 and 4 masks, we also added quiescent galaxies at $z\sim1.4$, though for these galaxies we lack LRIS observations, which target the most prominent absorption lines for these redshifts.  

\subsection{Observing Strategy}
The Heavy Metal survey was executed over eight semesters, ranging from 2016B to 2021B.  In total, 26 nights were allocated, though half of the nights were lost due to bad weather or technical problems. The primary goal of the Heavy Metal survey is to measure faint absorption lines, in particular around 5000 \AA. This regions is targeted by MOSFIRE J-band and H-band for the $z\sim1.4$ and $z\sim2.1$ pointings, respectively. We require integration times of $\sim$12 and $\sim$16\,hr, respectively, for $z\sim1.4$ (J-band) and $z\sim2.1$ (H-band), and used our best imaging conditions for these observations. Second priority is the Balmer/4000~\AA\ break region, which has more prominent features and thus requires slightly shorter integration times. This region was observed for $\sim4$ and $\sim12$ hr, respectively, with LRIS and MOSFIRE J-band for the $z\sim1.4$ and $z\sim2.1$ pointings. Finally, for all four pointings we took shorter integrations ($\sim$1-2 hr) of the wavelength regions around H$\alpha$, to assess whether the galaxies have any nebular line emission. This wavelength region is observed with MOSFIRE H-band and K-band for $z\sim1.4$ and $z\sim2.1$, respectively. These observations were planned to be taken under our least-favorable seeing conditions. In Table~\ref{tab:obs}, we summarize the observing settings and integration times for all masks and filters.

The MOSFIRE slits were configured with a width of $0\farcs7$, and have a minimum length of 7$\arcsec$. The LRIS slits were 1$\arcsec$ wide, with a minimum length of 10$\arcsec$. For all masks we used a minimum of five stars for the alignment. With MOSFIRE, the galaxies were observed using an ABA$'$B$'$ dither pattern, and with the longer LRIS slits we used an ABC dither pattern. Both dither patterns are preferred over an ABBA dither pattern as they result in better background subtraction and higher S/N \citep[see Appendix A in][]{MKriek2016}.

In all masks we observed at least one star in a slit. These ``slit star'' observations have three advantages. First, they enable us to monitor the seeing and possible drifts while observing. Second, the profiles and positions of the slit stars aided the data reduction, such that we could accurately register and weigh the individual science frames. Third, the slit star was used in the flux calibration, as explained in \cite{MKriek2016} and in the next section. 

\subsection{Data Reduction}\label{sec:reduction}
The MOSFIRE data are reduced using a custom software package that was originally developed for the MOSDEF survey \citep{MKriek2015}. This package is all automated, working with a single parameter file input, indicating the mask and target name, directories to raw frames and mask files, filter to be reduced, and path and name of photometric catalog to be used for the flux calibration. The first step is to read in all headers and identify the science and calibration frames. Next, a master dome flat frame is made, which is used to correct all science frames for pixel-to-pixel sensitivity variations and to trace the edges of all spectra. Next, we do an initial background subtraction of all science frames by subtracting the average of the previous and following frame. For the first and last science exposure, we only use one adjacent science frame as sky frame. 

The next step is to derive the wavelength solution using bright isolated sky lines. For this step we use the edge solutions from the master flat frame. This procedure is all automatic, as the position of the slit gives us a rough position of where to expect the sky lines. For the $K$ band we also use the arc lamp frames, allowing for an offset (i.e., flexure) between the sky lines and the arc lines. The final ingredient for the rectification is the position of the galaxies in the spectra. The exact position is a combination of the assigned dither position and the observed drift \citep[about 1 pixel hr$^{-1}$, see][]{MKriek2015}. We use the wavelength and edge solution to derive this position in all science frames for the slit star. Thus, for each frame we collapse the slit star spectrum along the wavelength direction and measure the position, FWHM of the seeing, and throughput. This position, combined with the wavelength and edge solutions, now gives us a transformation from the raw to the reduced frame for each science exposure.

Using the transformations derived in the previous step, we now perform an additional background subtraction on the (unrectified) science frames. We do this step before resampling, so we can better model the remaining sky. We run L.A. Cosmic \citep{PvanDokkum2001} on the cleaned frames and combine the cosmic-ray map with the available MOSFIRE bad pixel map.
The ``cleaned'' frames are now resampled to the final frame in a single transformation. We apply this same transformation to the sky and mask frames for each science exposure. Finally, we combine all science frames for each galaxy and filter, while weighing the frames using the throughput and seeing, and excluding all masked pixels. We also make a final weight map for each object and filter, as well as two noise frames, one based on the frame-to-frame variations and one on the sky and read-out noise. For more details on these steps, see \cite{MKriek2015}.

All spectra are calibrated for the relative response using telluric standards. Instead of observing new telluric standards for each science exposure, we make use of the library collected by the MOSDEF survey. For each mask and filter we construct a response spectrum from multiple telluric standards observed at similar airmass, combined with the stellar spectrum of a star of the same spectral type. The telluric spectra are reduced using a similar procedure as the science spectra. See \cite{MKriek2015} for more information on the construction of the response spectra and the motivation for using this procedure.  

Lastly, we generate one-dimensional (1D) science and error spectra for both primary and filler targets through an optimal weighing technique, as outlined by \citet{KHorne1986}, followed by absolute flux calibration.
Our employed MOSDEF software initially conducts absolute flux calibration for each galaxy by applying a scaling factor that is derived by comparing the 1D spectrum of a slit star to its integrated photometry. This step effectively performs a slit-loss correction for point sources. However, for all primary galaxies we detect the stellar continuum, and thus we directly scale the spectra to their respective broadband photometry (see Sect.~\ref{sec:fit}).

For the LRIS reduction we follow a similar procedure as for the MOSFIRE spectra. The only major difference is the calibration, as we do not have a library of telluric standards. To correct for atmospheric transmission features, we use the slit star spectrum combined with a theoretical sky spectrum. Furthermore, we calibrate each 1D science spectrum individually using the photometric data in the overlapping wavelength regime. See \cite{MKriek2019} for more information. 

\begin{figure*}
  \begin{center}  
  \includegraphics[width=0.95\textwidth]{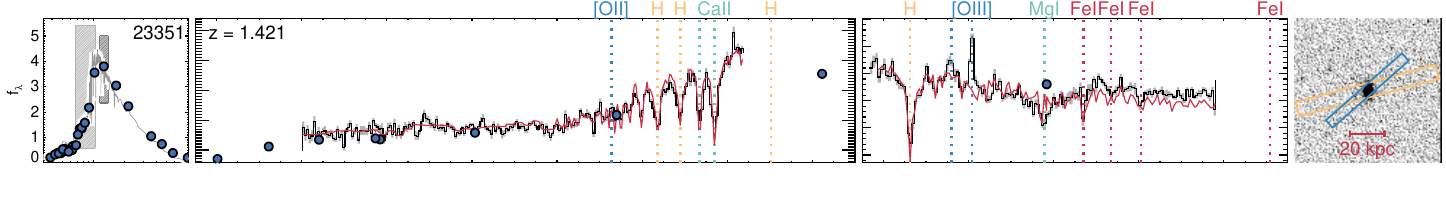}\vspace{-0.28in}
  \includegraphics[width=0.95\textwidth]{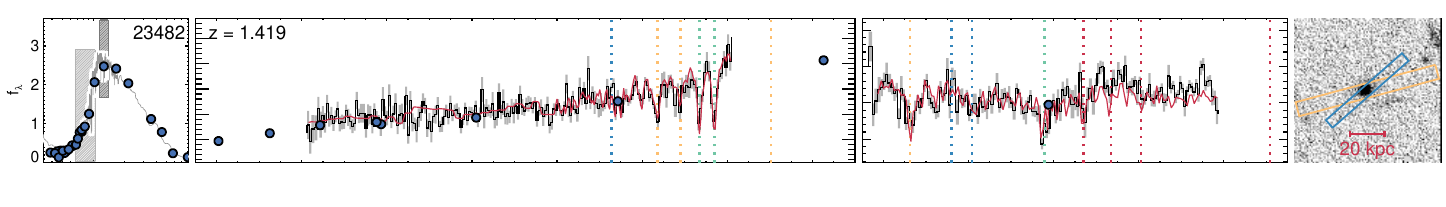}\vspace{-0.28in}
  \includegraphics[width=0.95\textwidth]{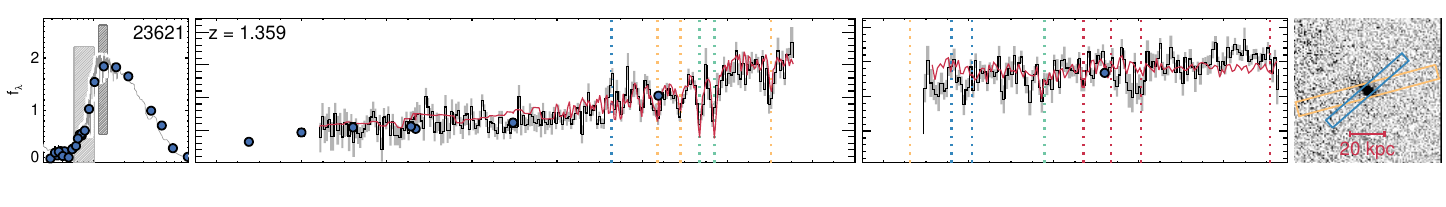}\vspace{-0.28in}
  \includegraphics[width=0.95\textwidth]{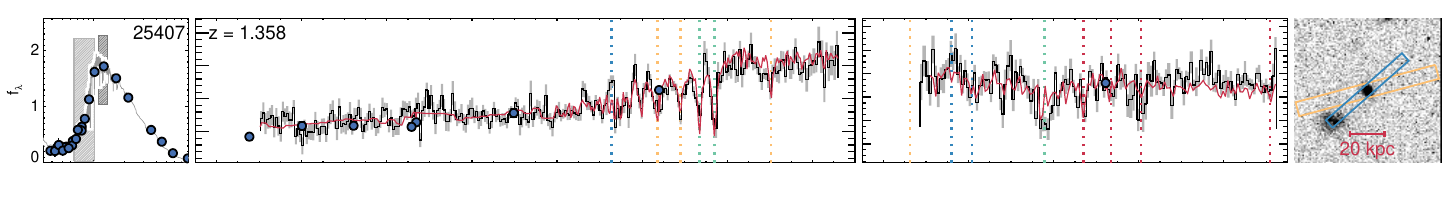}\vspace{-0.28in}
  \includegraphics[width=0.95\textwidth]{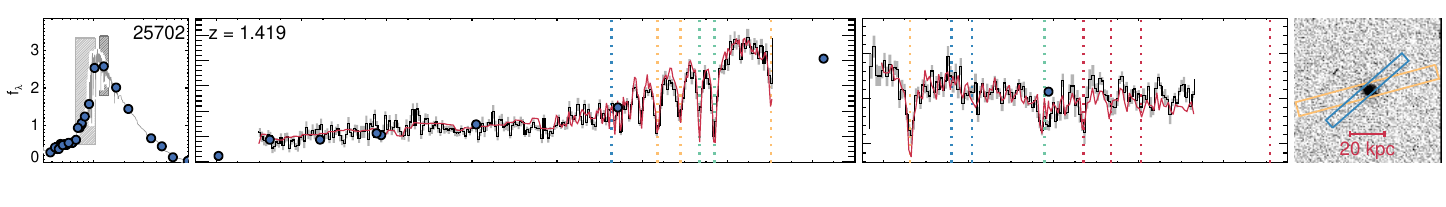}\vspace{-0.28in}
  \includegraphics[width=0.95\textwidth]{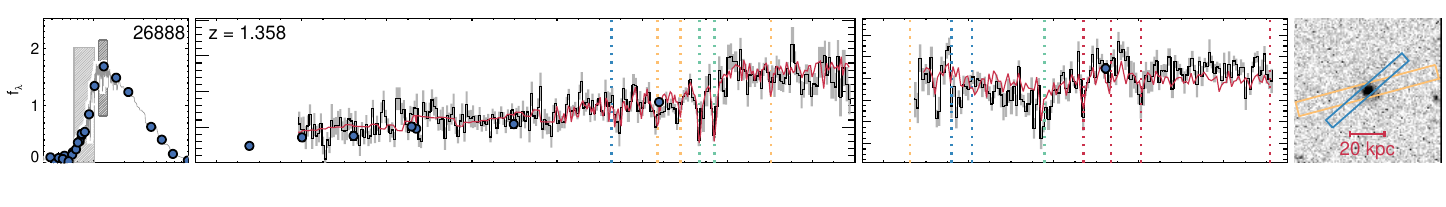}\vspace{-0.28in}
  \includegraphics[width=0.95\textwidth]{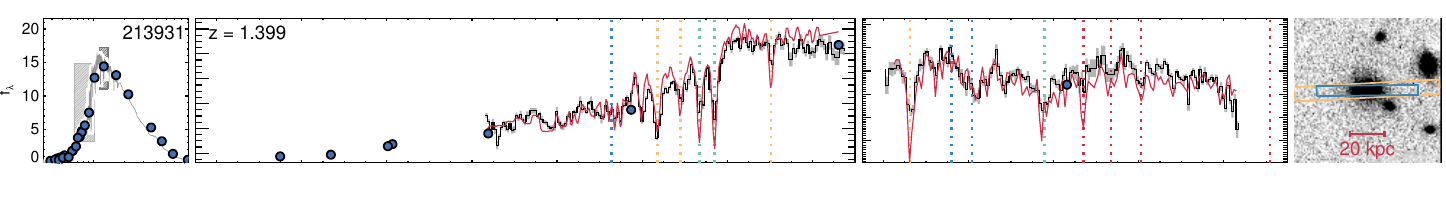}\vspace{-0.28in}
  \includegraphics[width=0.95\textwidth]{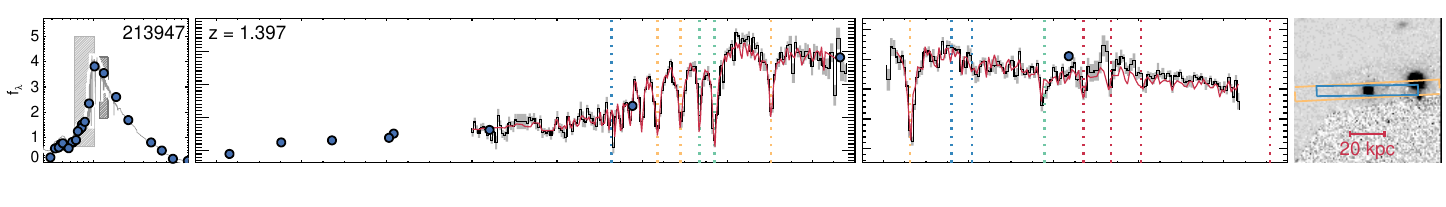}\vspace{-0.28in}
  \includegraphics[width=0.95\textwidth]{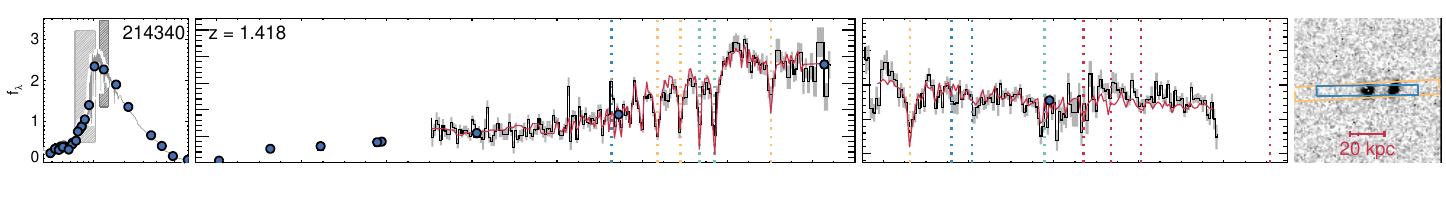}\vspace{-0.28in}
  \includegraphics[width=0.95\textwidth]{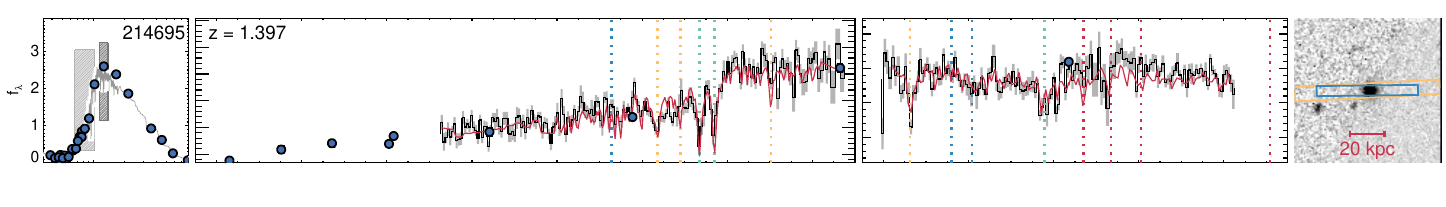}\vspace{-0.28in}
  \includegraphics[width=0.95\textwidth]{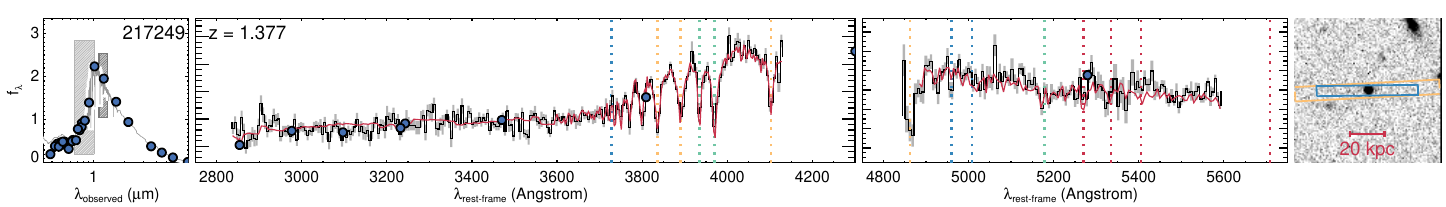}\vspace{-0.05in}
  \caption{Overview of UltraVISTA photometric SEDs (left), spectra (middle), and HST-F160W images (right) of distant quiescent galaxies at $z\sim1.4$. The LRIS (4-6 hr) and MOSFIRE J band spectra (12 hr) are shown in the middle-left and middle-right column, respectively. The spectra are binned by 15 and 10 pixels, respectively, such that each bin corresponds to $\sim$5\,\AA\ in rest frame. Flux densities ($f_\lambda$) are in $10^{-18}\rm erg\, s^{-1}\, cm^{-2}\, \AA^{-1}$. The extent of the LRIS and MOSFIRE-J panels are indicated by the light- and middle-gray rectangles in the left panels (arrows indicate that full range exceeds panel). The best-fit \texttt{FSPS} models to the combined photometry and spectra are shown in gray (left panel) and red (middle panel). Prominent spectral lines are indicated by the dotted vertical lines. The orientation of the MOSFIRE (blue) and LRIS (orange) slits are indicated in the right panel.\label{fig:speclz}}
  \end{center}  
\end{figure*}

\begin{figure*}
  \begin{center}  
    \includegraphics[width=0.95\textwidth]{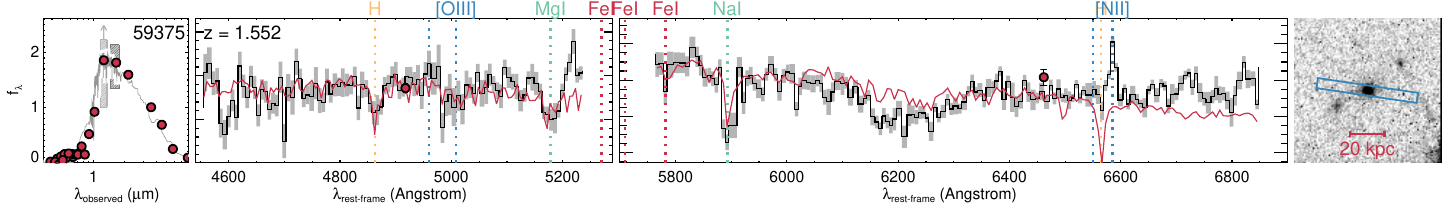}\vspace{0.1in}
    \includegraphics[width=0.95\textwidth]{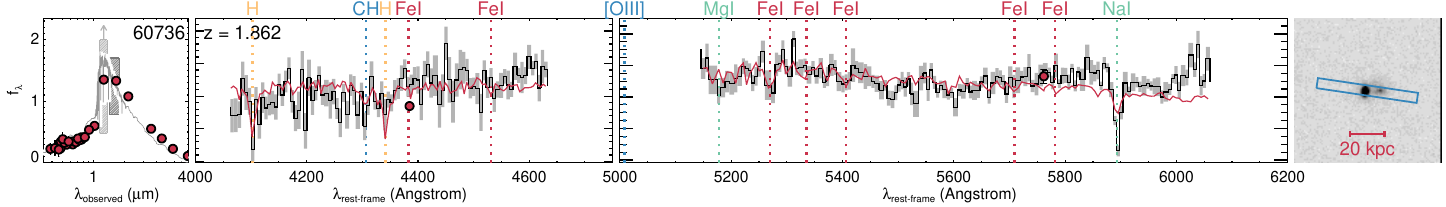}\vspace{0.1in}
    \includegraphics[width=0.95\textwidth]{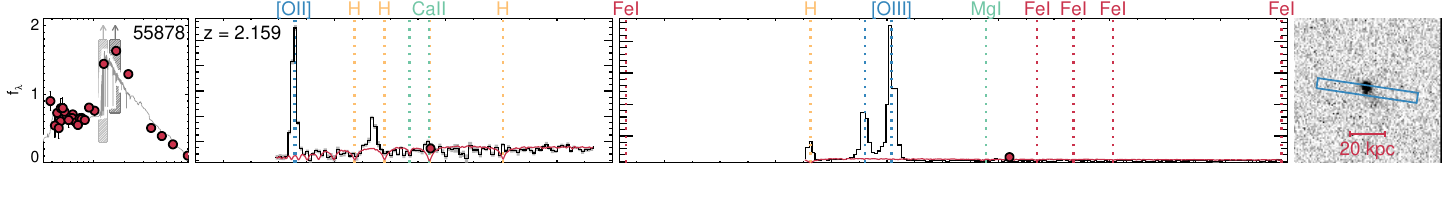}\vspace{-0.25in}
    \includegraphics[width=0.95\textwidth]{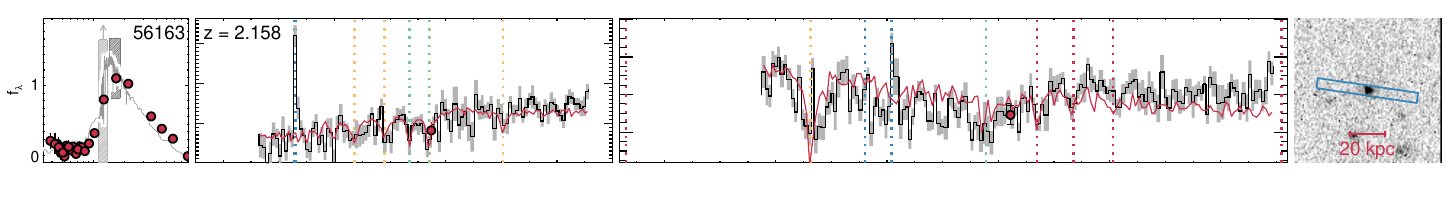}\vspace{-0.25in}    \includegraphics[width=0.95\textwidth]{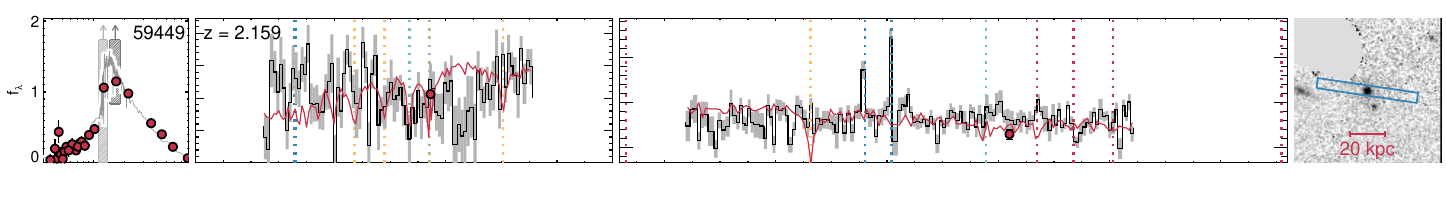}\vspace{-0.25in}
    \includegraphics[width=0.95\textwidth]{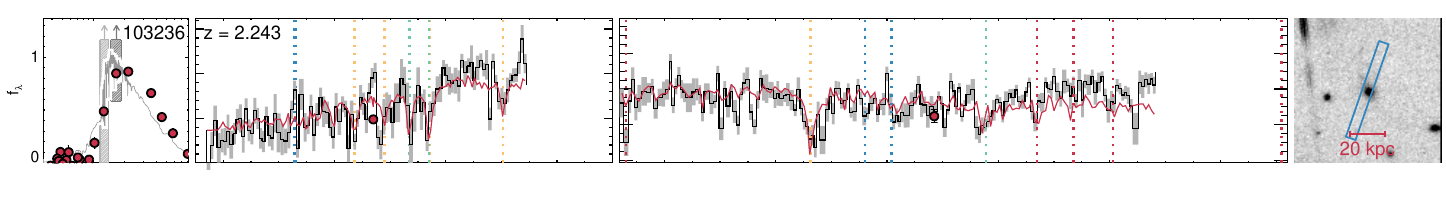}\vspace{-0.25in}
    \includegraphics[width=0.95\textwidth]{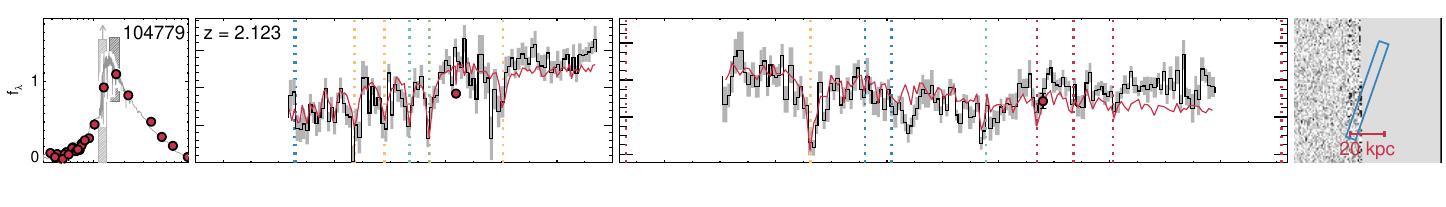}\vspace{-0.25in}
    \includegraphics[width=0.95\textwidth]{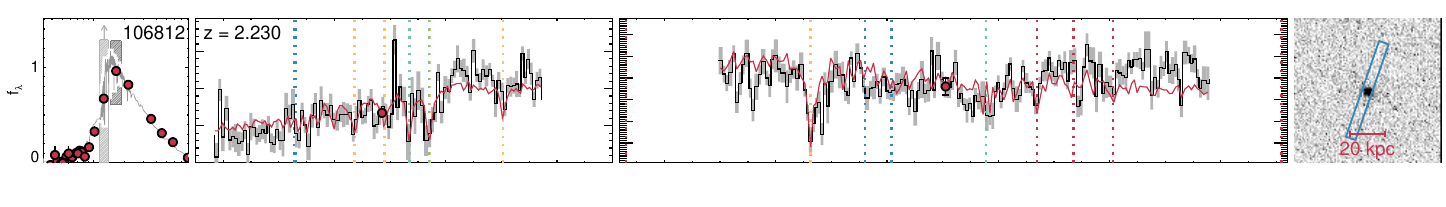}\vspace{-0.25in}
    \includegraphics[width=0.95\textwidth]{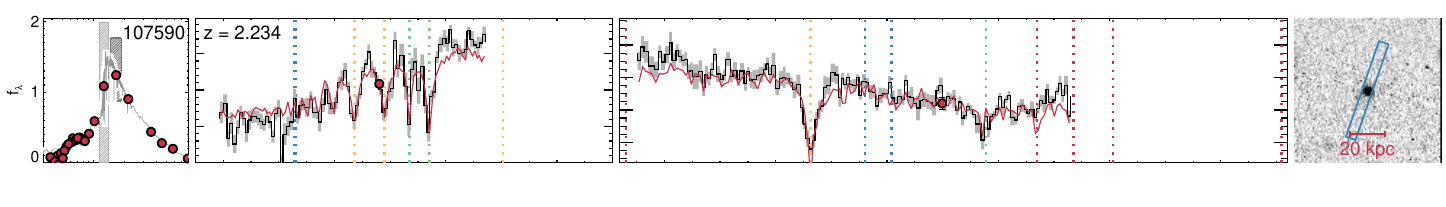}\vspace{-0.25in}
    \includegraphics[width=0.95\textwidth]{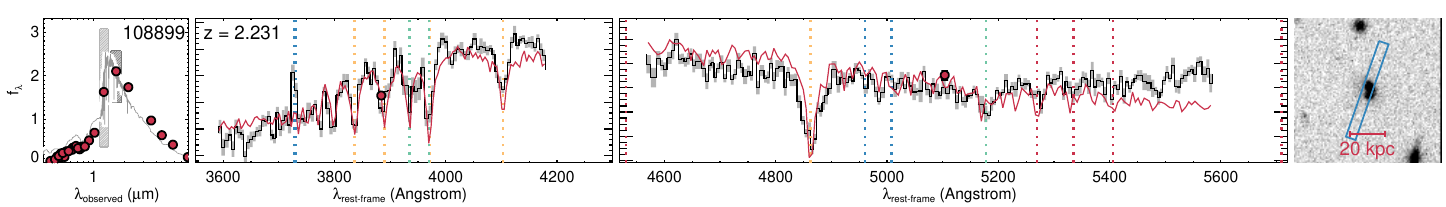}\vspace{-0.05in}
    \caption{Overview of UltraVISTA photometric SEDs (left), spectra (middle), and HST-F160W images (right) of distant quiescent galaxies at $z\sim2.1$. The MOSFIRE J-band (12-14 hr) and H-band spectra (16-17 hr) are shown in the middle-left and middle-right column, respectively. The spectra are binned by 13 and 10 pixels, respectively, such that each bin corresponds to $\sim$5\,\AA\ in rest-frame. Flux densities ($f_\lambda$) are in $10^{-18}\rm erg\, s^{-1}\, cm^{-2}\, \AA^{-1}$ and the extent of the MOSFIRE J-band and H-band panels are indicated by the light- and middle-gray rectangles in the left panels. The best-fit {\sc fsps} models to the combined photometry and spectra are shown in gray (left panels) and red (middle panels). Prominent spectral lines are indicated by the dotted vertical lines.\label{fig:spechz}}
    \end{center}    
\end{figure*}

\begin{figure*}
  \begin{flushleft}  
    \includegraphics[width=0.21\textwidth]{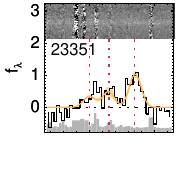}\hspace{-0.18in}
    \includegraphics[width=0.21\textwidth]{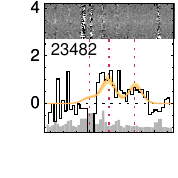}\hspace{-0.18in}
    \includegraphics[width=0.21\textwidth]{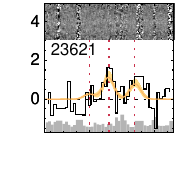}\hspace{-0.18in}
    \includegraphics[width=0.21\textwidth]{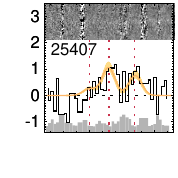}\hspace{-0.18in}
    \includegraphics[width=0.21\textwidth]{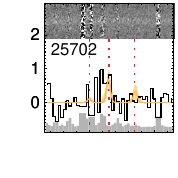}\vspace{-0.2in}\\
    \includegraphics[width=0.21\textwidth]{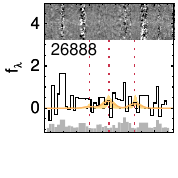}\hspace{-0.18in}
    \includegraphics[width=0.21\textwidth]{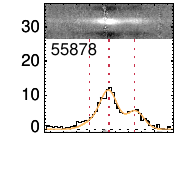}\hspace{-0.18in}
    \includegraphics[width=0.21\textwidth]{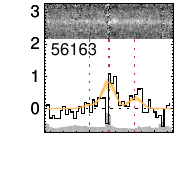}\hspace{-0.18in}  \includegraphics[width=0.21\textwidth]{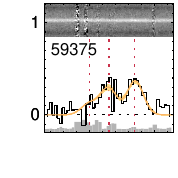}\hspace{-0.18in}
    \includegraphics[width=0.21\textwidth]{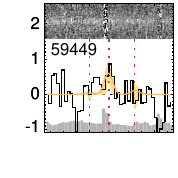}\vspace{-0.2in}\\
      \includegraphics[width=0.21\textwidth]{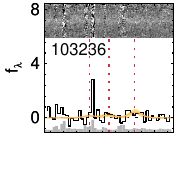}\hspace{-0.18in}
    \includegraphics[width=0.21\textwidth]{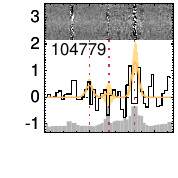}\hspace{-0.18in}
    \includegraphics[width=0.21\textwidth]{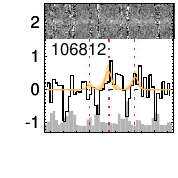}\hspace{-0.18in} 
    \includegraphics[width=0.21\textwidth]{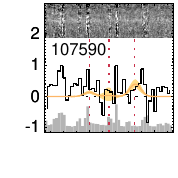}\hspace{-0.18in}
    \includegraphics[width=0.21\textwidth]{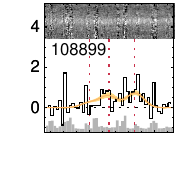}\\
    \includegraphics[width=0.21\textwidth]  
    {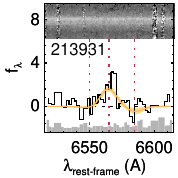}\hspace{-0.18in}
    \includegraphics[width=0.21\textwidth]
    {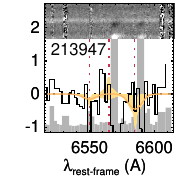}\hspace{-0.18in}
    \includegraphics[width=0.21\textwidth]{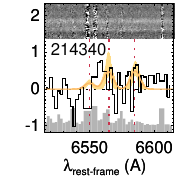}\hspace{-0.18in}
    \includegraphics[width=0.21\textwidth]{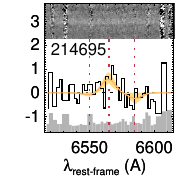}\hspace{-0.18in}
    \includegraphics[width=0.21\textwidth]{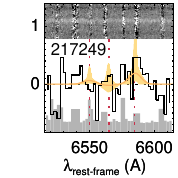}\hspace{-0.18in}
  \caption{The 2D and 1D spectra in wavelength regions around the H$\alpha$ spectral feature for all primary Heavy Metal galaxies. For each 1D spectrum the continuum has been removed. The 1D spectra are shown in black, binned to 3 (unmasked) pixels, and in gray we show the corresponding error spectrum. The yellow fit presents the best-fit emission-line model to H$\alpha$ and the two [N\,{\sc ii}] lines, with the 68\% uncertainty shown by the shaded yellow region. The vertical red dotted lines indicate the location of the three emission lines. Galaxy 55878 is the only one with strong H$\alpha$ in emission and will be discussed in Y. Ma et al (in preparation). We have 8 additional galaxies with marginal ($>3\,\sigma$) H$\alpha $ detections (see Table~\ref{tab:sample}) and 11 galaxies for which we derive upper limits on the H$\alpha$ flux. One galaxy (59375) has a significantly deeper spectrum; due to a incorrect photometric redshift, the line was observed in a different filter ($H$) than expected ($K$).     \label{fig:halpha}}
    \end{flushleft}  
\end{figure*}

\subsection{Data Overview}

In Figures~\ref{fig:speclz} and \ref{fig:spechz} we present an overview of the UltraVISTA photometric spectral energy distributions \citep[SEDs][left column]{AMuzzin2013a}, 1D spectra (middle two columns), and the F160W image from COSMOS-DASH \citep{LMowla2018,IMomcheva2017} for all primary targets. 
The position of the LRIS (yellow) and MOSFIRE (blue) slits are shown in the images, as well. 

Figure~\ref{fig:speclz} shows the galaxies in Heavy Metal 1 and 2, targeting $z\sim1.4$. For these two masks, we show the LRIS and MOSFIRE J-band spectra, all shifted (in wavelength only) to rest frame. We observe multiple Balmer absorption lines (yellow dotted lines) and the two CaII lines (green) around 4000\,\AA\ for all 11 galaxies. Most galaxies also show clear MgI and several FeI lines (red) in their MOSFIRE spectra. None of the targets show Balmer emission lines in their LRIS and MOSFIRE-J spectra. Nonetheless, three galaxies have either [O\,{\sc ii}] or [O\,{\sc iii}] in emission. We will further discuss these emission lines in Section~\ref{sec:results}. 

Figure~\ref{fig:spechz} shows the primary quiescent galaxies at $z\sim2.1$  targeted by the Heavy Metal 3 and 4 masks. Instead of LRIS and MOSFIRE J-band, we now show MOSFIRE J-band and H-band in the middle two columns. Two of the primary targets (59375 and 60736) scatter out the intended redshift regime (\highzr), and thus their spectra do not cover all targeted absorption lines. Nonetheless, we  detect additional absorption lines, such as Na for these galaxies.

Considering the remaining eight targets, six of them show several Balmer absorption and metal lines in their spectra. Two galaxies, 55878 and 59449, do not show any clear absorption lines, but their emission lines do reveal their redshifts. Galaxy 55878 has strong asymmetric emission lines, most likely originating from an AGN, and no absorption lines are detected. This galaxy will be discussed in detail in Ma et al. (in preparation). Galaxy 59449 shows two [O\,{\sc iii}] emission lines in its spectrum, but no absorption lines are detected. We would have expected to detect some continuum features, and thus the line and continuum emission may not originate from the same galaxy. However, we could not identify a redshift solution just from the continuum emission.

In Figure~\ref{fig:halpha}, we present spectra in the H$\alpha$ region for all primary targets. To illustrate whether H$\alpha$ is detected, we zoom in on a small wavelength region and show the continuum-subtracted 1D spectra (see Sect.~\ref{sec:halpha}). This spectral range does not encompass critical absorption features; these observations were taken to assess whether the galaxies have any H$\alpha$ emission. Hence, the spectra acquired in this band are shallower compared to the deeper spectra shown in Figures~\ref{fig:speclz} and \ref{fig:spechz} (refer to Table~\ref{tab:obs} for details). It is worth noting that the spectrum of 59375 is significantly deeper, as H$\alpha$ falls within the H band, where ultradeep observations were conducted, rather than the K band. Galaxy 60736's spectrum is not included, as there is no coverage of H$\alpha$.

Except for 55878 (an AGN), none of the galaxies exhibits strong H$\alpha$ emission in their 2D or 1D spectra. Nonetheless, several galaxies show very faint H$\alpha$ and [N\,{\sc ii}] emission lines, in particular after the continuum removal, as this step  corrects for the underlying Balmer absorption feature. In Section~\ref{sec:halpha} we describe our methodology to measure all H$\alpha$ lines in order to derive constraints on the star formation rates (SFRs).

\begin{table*}
\centering
\caption{Overview of primary quiescent galaxy sample\label{tab:sample}}
\begin{tabular}{l l c c c  c c c c c c}
\hline \hline

\multicolumn{1}{c}{ID} & \multicolumn{2}{c}{\underline{Coordinates}} & \multicolumn{2}{c}{Observed}  & \multicolumn{1}{c}{$z_{\rm spec}$} &  \multicolumn{2}{c}{Rest-frame} & \multicolumn{3}{c}{\underline{SPS fitting parameters}} \\
& \multicolumn{1}{c}{R.A.} & 
\multicolumn{1}{c}{Decl.} &  \multicolumn{2}{c}{\underline{magnitude}} &  & \multicolumn{2}{c}{\underline{colors}} & \multicolumn{1}{c}{log $M_*$\footnote{Typical uncertainty on stellar mass is 0.1 dex.}} & \multicolumn{1}{c}{log SFR$_{\rm cont}$\footnote{Typical uncertainty on SFR is 0.2 dex.}} & \multicolumn{1}{c}{$A_V$\footnote{Typical uncertainty on $A_V$ is 0.1 mag.}} \\
& (hh:mm:ss) & (dd:mm:ss) & \multicolumn{1}{c}{$J$} & \multicolumn{1}{c}{$H$} & & $U-V$ & $V-J$ & $M_\odot$ &  M$_\odot$/yr\\
\hline
HM1-213931 & 10:00:48.02 & 02:33:42.95 & 19.2 & & 1.399 & 1.78 & 0.97 &  11.63 & -1.17 & 0.0 \\
HM1-213947 & 10:00:50.68 & 02:33:55.37 & 20.8 & & 1.397 & 1.42 & 0.55 & 10.87 & -0.31 & 0.0 \\
HM1-214340 & 10:00:37.09 & 02:34:10.28 & 21.2 & &  1.418 & 1.48 & 0.78 & 10.80 & -0.47 & 0.0 \\
HM1-214695 & 10:00:32.28 & 02:34:25.12 & 21.1 & & 1.397 & 1.87 & 0.95 & 11.06 & -0.48 & 0.0 \\
HM1-217249 & 10:00:45.62 & 02:36:19.30 & 21.4 & & 1.377 & 1.40 & 0.47 &  10.61 & -0.24 & 0.0 \\
HM2-23351 & 10:00:48.44 & 01:43:24.74 & 20.7 & & 1.421 & 1.59 & 0.72 & 10.99 &  -0.66 & 0.0 \\
HM2-23482 & 10:00:46.19 & 01:43:31.31 & 21.1 & & 1.419 & 1.73 & 1.16 &  11.15 & -0.07 & 0.0 \\
HM2-23621 & 10:00:49.95 & 01:43:38.67 & 21.5 & & 1.359 & 1.87 & 1.34 & 11.05 & -0.17 & 0.2\\
HM2-25407 & 10:00:39.85 & 01:45:12.98 & 21.5 & & 1.358 & 1.62 & 0.92 & 10.70 & -0.34 & 0.0 \\
HM2-25702 & 10:00:49.20 & 01:45:27.47 & 21.1 & & 1.419 & 1.48 & 0.65 & 10.85 & -0.42 & 0.0 \\
HM2-26888 & 10:00:31.30 & 01:46:27.80 & 21.6 & & 1.358 & 1.97 & 1.03 & 10.93 & -1.09 & 0.0 \\
HM3-103236 & 10:00:47.18 & 01:59:19.56 & & 21.7 & 2.243 & 2.05 & 1.25 & 11.13 & -1.06 & 0.2\\
HM3-104779 & 10:00:41.29 & 02:00:30.36 & & 21.4 & 2.123 & 1.48 & 0.81 &  11.01 & -0.64 & 0.0\\
HM3-106812 & 10:00:49.42 & 02:02:07.23 & & 21.6 & 2.230 & 1.65 & 0.79 & 11.07 & -1.13 & 0.0\\
HM3-107590 & 10:00:35.45 & 02:02:41.24 & & 21.3 & 2.234 & 1.28 & 0.54 & 11.07 & -0.11 & 0.0 \\
HM3-108899 & 10:00:42.38 & 02:03:39.19 & & 20.7 & 2.231 & 1.51 & 0.77 & 11.28 & 0.10 & 0.0 \\
HM4-55878 & 09:59:16.40 & 01:47:23.18 & & 21.0 & 2.159 & 1.29 & 0.45 & 11.16 & 1.75 & 1.0 \\
HM4-56163 & 09:59:08.14 & 01:47:35.96 & & 21.4 & 2.158 & 1.63 & 1.00 &  11.08 & -0.19 & 0.0 \\
HM4-59375 & 09:59:11.82 & 01:50:04.61 & & 20.9 & 1.552 & 2.28 & 1.27 & 11.18 & -6.12 & 0.0 \\
HM4-59449 & 09:59:01.28 & 01:50:04.84 & & 21.4 & 2.159 & 1.39 & 0.89 & 11.10 & -0.56 & 0.0\\
HM4-60736 & 09:59:23.74 & 01:59:58.44 & & 21.2 & 1.862 & 1.53 & 0.80 & 10.88 & -0.30 & 0.0 \\
\hline\hline
\end{tabular}
\end{table*}

\section{Methodology}\label{sec:measurements}

In this section, we outline the methods we employed to determine the spectral, photometric, and structural properties of the Heavy Metal galaxies. We begin by deriving the spectroscopic redshifts, emission-line fluxes, stellar population characteristics, and rest-frame $UVJ$ colors for both the primary and filler galaxies (Sect.~\ref{sec:fit}). In Section ~\ref{sec:halpha}, we detail our approach to measuring the H$\alpha$ emission-line fluxes and subsequently calculating the SFRs for our primary, quiescent targets. Lastly, in Section~\ref{sec:dynamical_masses}, we present the methodology used to derive the galaxy structures and estimate dynamical masses for the primary Heavy Metal galaxies.

\subsection{Redshifts and stellar population properties}\label{sec:fit}
For all primary quiescent galaxies, we derive a spectroscopic redshift and stellar population properties by simultaneously fitting the spectra and the UltraVISTA photometry with the Flexible Stellar Population Synthesis models \citep[\texttt{FSPS};][]{CConroy2009,CConroy2010}. We assume an exponentially delayed star formation history, the average \cite{MKriek2013} dust attenuation law, and the \cite{GChabrier2003} initial mass function (IMF). We use a custom version of the {\sc fast} fitting code \citep{MKriek2009b}, in which the automatic scaling of the spectra to the photometry has been improved\footnote{In the original {\sc fast} release, the spectra were scaled to the photometry by convolving the spectra (in $f_\nu$) by the transmission curve of the overlapping filters. However, the MOSFIRE spectra only partially overlap with the filter curves, and thus this method does not work for most galaxies. Instead, the spectra, similar to the photometry, were scaled to the models using least-square scaling.}. To facilitate comparison with the full galaxy distribution from which the galaxies are selected, we assume solar metallicity. {\sc fast} does not fit for the absorption-line broadening, and thus we fit binned spectra.

For galaxy HM4-55878 the emission lines are very strong and affect the broadband spectral shape. Thus, for this galaxy we first correct the photometry for the emission-line fluxes (see Sect.~\ref{fig:halpha}). Furthermore, we mask the wavelength regions affected by emission lines while fitting. The strong lines also affect the absolute calibration of our spectra, and our default method does not work (see Sect.~\ref{sec:reduction}). Instead, for this galaxy we use the filter curves and integrated broadband magnitudes, corrected for the partial overlap between the spectra and filter curve.

The resulting best-fit redshifts, stellar masses, SFRs, and magnitudes of dust attenuation ($A_V$) are listed in Table~\ref{tab:sample}. The typical uncertainties on the stellar mass, SFR, and $A_V$ are 0.1\,dex, 0.2\,dex, and 0.1\,mag, respectively. These uncertainties include the flux uncertainties as well as variations in the various assumptions (except for the IMF) and the stellar population synthesis model \citep[{\sc fsps};][]{GBruzual2003,CMaraston2005}. The best-fit models are shown in Figures~\ref{fig:speclz} and ~\ref{fig:spechz}. For displaying purposes, we show the original models convolved to the velocity dispersion of the spectra, as derived by \citet{ABeverage2024}.

While most stellar continuum fits look reasonable, there are a few exceptions. First, for HM1-213931 the fit is quite poor, probably because it is a blended spectrum of multiple galaxies, which have a velocity offset. Though we cannot deblend the spectra of the sources, we find a different spectrum when assuming different weighing profiles for the extraction \citep[see][for more information on the implications]{ABeverage2024}. For HM4-59449 we do not see any clear absorption lines, and the redshift is based on the faint [O\,{\sc iii}] emission lines. 

For the filler galaxies, we derive spectroscopic redshifts by fitting the emission lines. The majority of the filler targets show multiple emission lines, resulting in robust spectroscopic redshifts. 
For the $z\sim1.4$ masks, we observed different  filler galaxies for the LRIS and MOSFIRE masks. This strategy results in a larger number of filler galaxies, but a lower success rate of confirming the spectroscopic redshift. For the $z\sim2.1$ masks, we use the same filler galaxies in the different settings and thus had more wavelength coverage to detect possible spectral features. 

Finally, for all galaxies in the observed Heavy Metal masks we determine rest-frame $U$, $V$, and $J$ colors using the \textsc{EA$z$Y} \citep{GBrammer2008} code. When available, we assume the spectroscopic redshift, otherwise we adopt the photometric redshifts provided by \citep{AMuzzin2013a}. Each rest-frame magnitude is determined individually, using a fit to just the surrounding photometric datapoints. Hence, the colors are not based on the best-fit stellar population model to the full spectrum.

\begin{table}
\caption{Emission line properties$^{\rm a}$\label{tab:emissionlines}}
\centering
\begin{tabular}{l | c c c}
\hline \hline
ID  & \multicolumn{1}{c}{$F_{\rm H\alpha}$} & \multicolumn{1}{c}{SFR$_{{\rm H}\alpha}$} & \multicolumn{1}{c}{[N\,{\sc ii}]/H$\alpha$} \\ & $10^{-17}$ erg s$^{-1}$ cm$^{-2}$ & M$_\odot$/yr & \\
\hline
HM1-213931 & $6.0^{+1.4}_{-1.5}$ & $4.0^{+0.9}_{-1.0}$ & $<0.24$ \\
HM1-213947 & $<1.8$ & $<1.2$ & | \\
HM1-214340 & $<2.2$ & $<1.5$ & | \\
HM1-214695 & $<3.2$ & $<2.1$ & | \\
HM1-217249 & $<1.1$ & $<0.7$ & | \\
HM2-23351 & $1.4^{+0.4}_{-0.3}$ & $0.9^{+0.3}_{-0.2}$ & $2.1^{+0.7}_{-0.4}$\\
HM2-23482 & $3.3^{+0.5}_{-0.6}$ & $2.2^{+0.4}_{-0.4}$ & $0.7^{+0.2}_{-0.1}$\\
HM2-23621 & $2.8^{+0.6}_{-0.6}$ & $2.0^{+0.4}_{-0.4}$ & $0.7^{+0.2}_{-0.2}$\\
HM2-25407 & $3.1^{+0.4}_{-0.6}$ & $1.9^{+0.2}_{-0.4}$ & $0.7^{+0.2}_{-0.1}$\\
HM2-25702 & $<1.7$ & $<1.2$ & | \\
HM2-26888 & $<1.9$ & $<1.1$ & | \\
HM3-103236 & $<2.0$ & $<4.8$ & | \\
HM3-104779 & $<2.9$ & $<5.3$ & | \\
HM3-106812 & $<2.7$ & $<5.6$ & | \\
HM3-107590 & $<2.5$ & $<5.3$ & | \\
HM3-108899 & $3.2^{+0.6}_{-0.6}$ & $6.6^{+1.2}_{-1.3}$ & $1.3^{+0.3}_{-0.3}$\\
HM4-55878 & $60.0^{+0.6}_{-0.9}$ & $465^{+5}_{-7}$ & $0.48^{+0.01}_{-0.02}$\\
HM4-56163 & $3.2^{+0.5}_{-0.5}$ & $6.1^{+0.9}_{-1.0}$ & $0.4^{+0.1}_{-0.1}$\\
HM4-59375 & $1.2^{+0.1}_{-0.1}$ & $1.0^{+0.1}_{-0.1}$ & $1.3^{+0.2}_{-0.1}$ \\
HM4-59449 & $<3.2$ & $<6.2$ & | \\
HM4-60736 & | & | & | \\
\hline\hline
\multicolumn{3}{l}{$^{\rm a}$All limits are $3\,\sigma$}
\end{tabular}
\end{table}

\begin{table*}
\caption{Structural and kinematic properties\label{tab:struct}}
\centering
\begin{tabular}{l c c c c c c c c c c}
\hline \hline
\multicolumn{1}{c}{ID} & \multicolumn{3}{c}{\underline{F814W structural properties}} &
\multicolumn{3}{c}{\underline{F160W structural properties$^{\rm a}$}} & \multicolumn{1}{c}{$R_{\rm e, major}$} & \multicolumn{1}{c}{log\,$M_{\rm *,c}^{\rm b}$} & \multicolumn{1}{c}{$\sigma_{v}^{\rm c}$} &
\multicolumn{1}{c}{log\,$M_{\rm dyn}$} \\
 & \multicolumn{1}{c}{$R_{\rm e, major}$} & \multicolumn{1}{c}{$n$} & \multicolumn{1}{c}{$q$} & \multicolumn{1}{c}{$R_{\rm e, major}$} & \multicolumn{1}{c}{$n$} & \multicolumn{1}{c}{$q$} & @ 5000\,\AA & $M_\odot$ & 
\multicolumn{1}{c}{km s$^{-1}$} &
\multicolumn{1}{c}{$M_\odot$} \\ & kpc & & & kpc & & & kpc &  \\
\hline
HM1-213931 & 3.72$\pm$0.17 & 5.7$\pm$0.2 & 0.75$\pm$0.01 & 3.79$\pm$0.17 & 4.3$\pm$0.2 & 0.88$\pm$0.01 & 3.76$\pm$0.22 & 11.35 & $371^{+13}_{-13}$ & $11.81^{+ 0.04}_{-0.04}$ \\
HM1-213947 & 0.81$\pm$0.01 & 2.1$\pm$0.1 & 0.75$\pm$0.01 & 0.57$^{\rm d}$ & 6.0 & 0.95 & 0.69$\pm$0.04 & 10.73 & $216^{+16}_{-13}$ & $10.71^{+0.07}_{-0.07}$ \\
HM1-214340 & 3.16$\pm$0.17 & 3.7$\pm$0.2 & 0.76$\pm$0.02 & 2.13$\pm$0.11 & 2.4$\pm$0.2 & 0.78$\pm$0.03 & 2.46$\pm$0.16 & 10.81 & $149^{+19}_{-23}$ & $10.91^{+0.10}_{-0.16}$ \\
HM1-214695 & 1.99$\pm$0.11 & 2.3$\pm$0.1 & 0.83$\pm$0.03 & 1.94$\pm$0.14 & 2.5$\pm$0.3 & 0.98$\pm$0.04 & 1.96$\pm$0.16 & 11.02 & $210^{+24}_{-25}$ & $11.16^{+0.10}_{-0.11}$ \\
HM1-217249 & 0.59$\pm$0.01 & 2.4$\pm$0.1 & 0.57$\pm$0.01 & 0.60$\pm$0.02 & 1.3$\pm$0.1 & 0.56$\pm$0.03 & 0.59$\pm$0.03 & 10.60 & $185^{+21}_{-24}$ & $10.46^{+0.09}_{-0.12}$ \\
HM2-23351 & 1.82$\pm$0.03 & 2.2$\pm$0.1 & 0.31$\pm0.01$ & 1.58$\pm$0.05 & 2.3$\pm$0.1 & 0.34$\pm$0.02 & 1.66$\pm$0.09 & 10.96 & $257^{+14}_{-13}$ & $11.05^{+0.05}_{-0.06}$ \\
HM2-23482 & 2.16$\pm$0.10 & 2.2$\pm$0.1 & 0.69$\pm$0.02 & 3.25$\pm$ 0.33 & 5.5$\pm$0.7 & 0.71$\pm$0.04 & 2.80$\pm$0.23 &     11.11 & $289^{+38}_{-35}$ & $11.45^{+0.11}_{-0.10}$ \\
HM2-23621 & 5.62$\pm$0.31 & 1.5$\pm$0.1 & 0.36$\pm$0.01 & 3.75$\pm$0.24 & 1.7$\pm$0.2 & 0.40$\pm$0.04 & 4.42$\pm$0.29 &  10.98 & $199^{+38}_{-41}$ & $11.31^{+0.14}_{-0.23}$ \\
HM2-25407 & 1.36$\pm$0.14 & 5.3$\pm$0.5 & 0.95$\pm$0.03 & 1.42$\pm$0.08 & 1.7$\pm$0.2 & 0.73$\pm$0.04 & 1.40$\pm$0.10 & 10.66 & $201^{+27}_{-28}$ & $10.92^{+0.13}_{-0.13}$ \\
HM2-25702 & 1.46$\pm$0.12 & 5.8$\pm$0.4 & 0.74$\pm$0.02 & 1.41$\pm$0.08 & 3.9$\pm$0.4 & 0.79$\pm$0.03 & 1.43$\pm$0.10 &     10.84 & $243^{+21}_{-20}$ & $11.01^{+0.08}_{-0.08}$ \\
HM2-26888 & 2.34$\pm$0.27 & 2.9$\pm$0.3 & 0.75$\pm$0.04 & 2.46$\pm$0.19 & 4.1$\pm$0.5 & 0.89$\pm$0.04 & 2.41$\pm$0.20 & 10.94 & $166^{+33}_{-27}$ & $10.97^{+0.16}_{-0.16}$ \\
HM3-103236 & | & | & | & 2.41$^{\rm d}$ & 2.6 & 0.58 & 2.36$\pm$0.59 & 11.08 & $197^{+37}_{-35}$ & $11.08^{+0.17}_{-0.25}$ \\
HM3-104779 & | & | & | & | & | & | & | & | & $186^{+30}_{-31}$ & | \\
HM3-106812 & | & | & | &  3.64$\pm$0.63 & 5.8$\pm$1.3 & 0.70$\pm$0.07 & 3.58$\pm$0.65 & 11.14 & $187^{+41}_{-45}$ & $11.11^{+0.22}_{-0.24}$ \\
HM3-107590 & | & | & | & 1.00$\pm$0.07 & 4.3$\pm$0.5 & 0.83$\pm$0.04 & 0.98$\pm$0.09 & 11.05 & $173^{+25}_{-23}$ & $10.59^{+0.12}_{-0.13}$ \\
HM3-108899 & | & | & | & 3.46$\pm$0.17 & 3.6$\pm$0.3 & 0.51$\pm$0.02 & 3.40$\pm$0.24 & 11.25 & $306^{+24}_{-23}$ & $11.55^{+0.07}_{-0.08}$ \\
HM4-55878 & | & | & | &  2.36$\pm$0.21 & 3.4$\pm$0.5 & 0.67$\pm$0.04 & 2.34$\pm$0.24 & 11.20 & | & |\\
HM4-56163 & | & | & | &  3.55$^{\rm d}$ & 6.0 & 0.93 & 3.51$\pm$0.88 & 11.05 & $213^{+54}_{-48}$ & $11.26^{+0.24}_{-0.27}$ \\
HM4-59375 & | & | & | & | & | & | & | & | & $333^{+42}_{-43}$ & | \\
HM4-59449 & | & | & | & | & | & | & | & | & | & | \\
HM4-60736 & | & | & | &  1.58$\pm$0.05 & 2.0$\pm$0.2 & 0.44$\pm$0.02 & 1.62$\pm$0.10 & 10.70 & $270^{+30}_{-27}$ & $11.16^{+0.08}_{-0.10}$ \\
\hline\hline
\multicolumn{11}{l}{$^{\rm a}$Adopted from \cite{SCutler2022}.}\\
\multicolumn{11}{l}{$^{\rm b}$Stellar masses corrected for the total magnitude difference between the photometric catalog and {\sc Galfit}. The typical}\\
\multicolumn{11}{l}{~ uncertainties are 0.1 dex, excluding variations in the IMF.} \\
\multicolumn{11}{l}{$^{\rm c}$Adopted from \cite{ABeverage2024}.}\\
\multicolumn{11}{l}{$^{\rm d}$Bad {\sc Galfit} fit, no uncertainties available. We 
adopted uncertainties of 25\%, $\pm$1 and $\pm$0.1 for $R_e$, $n$ and 
and $q$, respectively.}\\
\end{tabular}
\end{table*}

\subsection{H$\alpha$ SFR measurements}\label{sec:halpha}

For the primary quiescent targets, we measure the H$\alpha$ emission-line flux. We use the best-fit stellar population model convolved to the best-fit velocity dispersion as the continuum model. This approach ensures that we incorporate the underlying Balmer absorption. To derive the fluxes and correct for emission-line blending, we fit H$\alpha$ and the two [N\,{\sc ii}] at 6548\,\AA\ and 6584\,\AA, simultaneously. For our model spectrum, we use three Gaussians with the same velocity dispersion and a fixed ratio between the two [N\,{\sc ii}] lines of a factor of 3. The redshift is fixed to the best-fit absorption-line redshift. If none of the lines are clearly visible, the velocity dispersion of the emission lines cannot exceed the stellar velocity dispersion by more than 1\,$\sigma$. The minimum allowed velocity dispersion is set by the spectral resolution. 

We derive the uncertainties on the emission-line flux measurements using Monte Carlo simulations. We make 500 realizations of the spectrum around H$\alpha$, by perturbing the fluxes following the error spectrum. For each realization we also allow for variations in the subtracted continuum, assuming an uncertainty on the H$\alpha$ absorption line strength of 5\%. For each realization we fit all three lines using the same method as for the actual spectrum. We derive the 16\% and 84\% confidence intervals on the emission-line fluxes from the resulting distribution. In Figure~\ref{fig:halpha} we show these fits and confidence intervals for all 20 galaxies with coverage of H$\alpha$. For galaxies that do not have a 3$\sigma$ detection for H$\alpha$, we derive the 3$\sigma$ upper limit. All values are given in Table~\ref{tab:emissionlines}.

For galaxy 213947 we have to do additional masking to derive the H$\alpha$ flux, as the 2D spectrum partially overlaps with the (negative) spectrum of a close galaxy. This nearby galaxy only has emission lines and no continuum emission, and thus only a small wavelength range is affected. Thus, for this galaxy, we mask the wavelengths that are contaminated by the emission lines of the nearby galaxy. 

We convert the integrated H$\alpha$ flux to the integrated luminosity using the spectroscopic redshift. In order to correct this line for dust attenuation, we ideally would use the Balmer decrement (H$\alpha$/H$\beta$). However, with the exception of 55878, H$\beta$ is too faint to yield a useful Balmer decrement measurement, and thus we use the stellar attenuation for the dust correction, instead. We do note, however, that nearly all galaxies have a best-fit $A_V=0$. For 55878, we do use the Balmer decrement (H$\alpha$/H$\beta$=4.43) for the dust correction. 
Finally, we adopt the conversion by \citet{RKennicutt1998} for solar metallicity and a \cite{PKroupa2001} IMF \citep[which is comparable to the][IMF]{GChabrier2003} to derive SFRs for all galaxies with H$\alpha$ coverage (see Table~\ref{tab:emissionlines}).
The majority of the galaxies do not have detected H$\alpha$ emission and thus we derive a 3\,$\sigma$ upper limit on the SFR.

Two galaxies stand out in Figure~\ref{fig:halpha}. First, galaxy 55878 has very strong emission lines, and we will further discuss this galaxy in Sect.~\ref{sec:sample_sfr}). Second, HM4-59375 stands out, as despite the low H$\alpha$ flux and resulting SFR, the galaxy has significantly detected emission lines. For this galaxy the spectrosopic redshift of $z_{\rm spec}=1.552$ is significantly lower than the photometric redshift used in the selection. Hence, H$\alpha$ does not fall in the K band, as do the other galaxies in its targeted redshift regime, but in the H band for which the observations are significantly deeper. For galaxy 60736, the spectroscopic redshift falls outside the selection window, and thus we have no coverage of the H$\alpha$ wavelength regions (see Fig.~\ref{fig:features}). Hence, this galaxy is missing from Figure~\ref{fig:halpha}.


\subsection{Structural Measurements and Dynamical Masses} \label{sec:dynamical_masses}

The Heavy Metal pointings overlap with the COSMOS/ACS-F814W \citep{NScoville2007} and the COSMOS-DASH/WFC3-IR-F160W imaging \citep{IMomcheva2017,LMowla2018,SCutler2022}, enabling structural measurements. We derive galaxy sizes from the F814W images by fitting  single-component S\'ersic models with {\sc Galfit} \citep{CPeng2002}, following the technique described in \citet{ABeverage2021}. For the COSMOS-DASH imaging, we adopt the structural measurements by \citet{SCutler2022}. 

For the $z\sim1.4$ galaxies, we use both the F814W and F160W structural parameters in our analyses, listed in Table~\ref{tab:struct}. We derive the structural parameters ($R_{\rm e, major}$, $n$, $q$) at rest-frame 5000\,\AA\, using interpolation. 
For HM1-213947, the F160W image results in a bad {\sc Galfit} fit, and thus for this galaxy we only use F814W. We use Equation 1 by \citet{AvanderWel2014} to correct the size to rest-frame 5000\,\AA.  For the $z\sim2.1$ galaxies we only use the F160W measurements, as these galaxies are not or barely detected in F814W. We also correct these size measurements, standardizing them to the rest-frame 5000\,\AA\ wavelength (see Table~\ref{tab:struct}), following \citet{AvanderWel2014}. These corrections are generally  subtle, fluctuating within the range of -0.01 to 0.009. For three galaxies, no F160W size measurements were available, as either the fit failed or there was no coverage. For three additional galaxies, the fit was qualified as ``bad''.  We nonetheless use these structural measurements in the subsequent analysis for the two galaxies (HM3-103236 and HM4-56163) for which no ACS measurements are available. Nonetheless, we flag these galaxies in the subsequent figures. For these galaxies no uncertainties are available in the catalogs by \citet{SCutler2022}, and thus we adopt uncertainties of 25\%, $\pm$1.0 and $\pm$0.1 for $R_{\rm e}$, $n$, and $q$, respectively.

We also use the {\sc Galfit} parameters to refine our stellar mass measurements and ensure their consistency with other structural measurements. For this refinement process, we derive a mass correction factor by comparing the integrated magnitude from {\sc Galfit} with the magnitude from the corresponding filter band in the photometric catalog. For the F814W and F160W filters, which are absent from the UltraVISTA catalog by \citep{AMuzzin2013a}, we compute their magnitudes by integrating the best-fit {\sc fast} model using the respective filter curves. For the $z\sim2.1$ galaxies, we combine the correction factors from F160W and F814W following their proximity to rest-frame 5000\,\AA. The {\sc Galfit} magnitudes are typically fainter than the catalog magnitudes by 0.094, resulting in an average mass correction of -0.038\,dex. However, for some galaxies, in particular blended systems such as HM1-213931, the mass corrections can be as large as -0.28\,dex. The corrected masses ($M_{*,\rm c}$) for the galaxies with structural measurements are listed in Table~\ref{tab:struct}.

The Heavy Metal spectra yield velocity dispersion measurements ($\sigma_v$) for all but two galaxies, as described in our accompanying paper \citep{ABeverage2024}. These measurements are derived using the absorption-line fitter ({\tt alf}) code \citep{CConroy2012d,JChoi2016,CConroy2018}. \citet{ABeverage2023} shows that the {\tt alf} velocity dispersions are in perfect agreement with those found by p{\sc pxf} \citep{MCappellari2004} for a large sample of $z\sim0.7$ quiescent sample of galaxies. We increase the measured velocity dispersion measurements ($\sigma_{v}$) by 4\% to obtain the velocity dispersion within 1 $r_e$ ($\sigma_{v,\rm e}$) \citep[see][]{JvandeSande2013}. 

The velocity dispersions and structural measurements together enable an estimate of the dynamical mass. We still have a poor understanding of the internal stellar dynamics within these galaxies. A few resolved investigations of three lensed distant quiescent galaxies have hinted at the presence of rotational support to varying degrees \citep{ANewman2015, ANewman2018b, SToft2017}. Nonetheless, due to our limited knowledge and to facilitate comparison with similar works, here we define dynamical mass as 
\begin{equation}
M_{\rm dyn} = \frac{\beta(n) \sigma_{v,e}^2 R_{\rm e}}{G}
\end{equation}
with $\beta(n) = 8.87 - 0.831n + 0.0241n^2$, the virial constant for a spherical isotropic model described by profile $R_e^{1/n}$ for different values of the S\'ersic index $n$ \citep{MCappellari2006}. For $R_{\rm e}$ we take the circularized radius ($R_{\rm e} = R_{\rm e, major} \sqrt{q}$) at rest-frame wavelength of 5000\,\AA. The resulting dynamical masses are listed in Table~\ref{tab:struct}.


\section{Results}\label{sec:results}

While quiescent galaxies have been studied extensively throughout cosmic time, the  majority of these investigations have relied on photometric data. The absence of spectroscopic information may lead to biases in our photometric redshifts, stellar masses, and stellar population properties. Consequently our studies of the buildup and growth of galaxies over cosmic time may be biased, as well. The Heavy Metal survey provides redshifts for a significant sample of distant quiescent galaxies, resulting in more accurate stellar population properties. Additionally, the presence of absorption lines facilitates kinematic and chemical composition studies, while emission lines offer an alternative avenue for examining their star formation characteristics. In Section~\ref{sec:sample} we examine our galaxy sample and compare it with the parent galaxy sample from which the spectroscopic sample was drawn. In Section~\ref{sec:sample_sfr} we present the star formation properties of the primary Heavy Metal galaxies and assess whether they indeed have quiescent stellar populations. Moving to Section~\ref{sec:structures}, we discuss their structural properties, and finally, in Section~\ref{sec:masscomp} we compare the stellar masses to the dynamical masses.

\begin{figure}
  \begin{center}  
\includegraphics[width=0.48\textwidth]{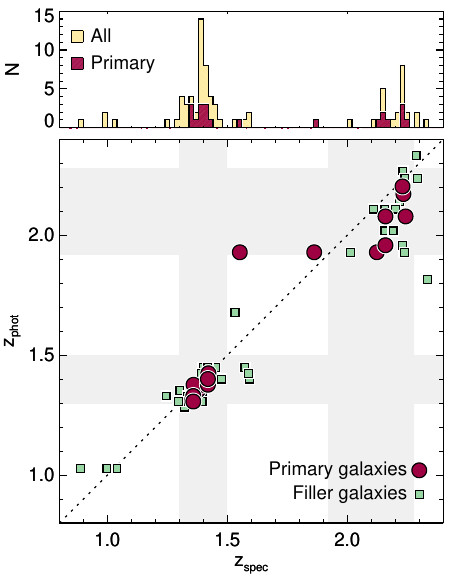}
  \caption{Photometric vs. spectroscopic redshift for all spectroscopically confirmed primary (circles) and filler (squares) targets in the Heavy Metal survey. We measured a spectroscopic redshift of all primary targets. For the filler galaxies the success rate was much lower with 65\%. The normalized medium absolute deviations \citep[$\sigma_{\rm nmad}$,][]{GBrammer2008} between the photometric and spectroscopic redshifts are 0.017 and 0.014 for the filler and primary galaxies, respectively. The shaded areas present the targeted redshift intervals used in the selection. Two of the primary targets scattered out the targeted redshift interval.
  \label{fig:redshifts}}
    \end{center}  
\end{figure}

\begin{figure*}
  \begin{center}  
\includegraphics[width=0.4\textwidth]{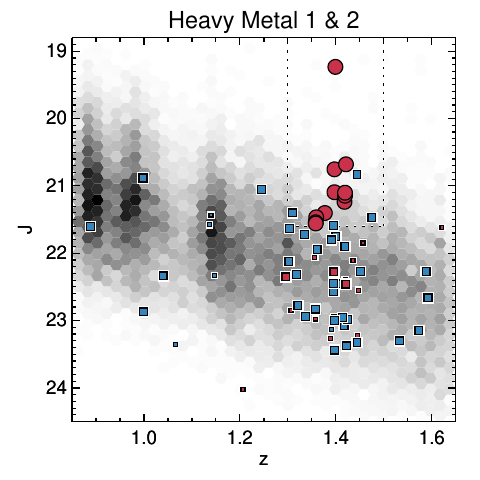}  
\includegraphics[width=0.4\textwidth]{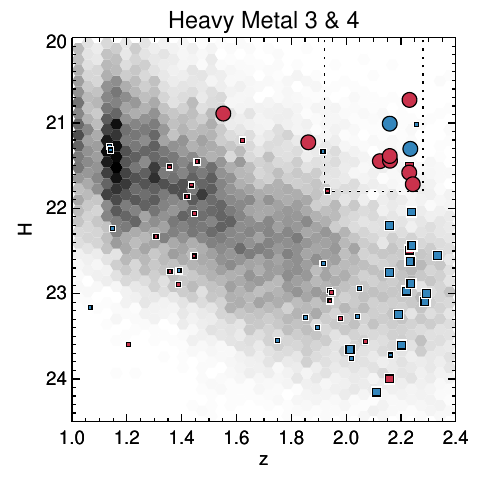} \\ \vspace{-0.2in}
\includegraphics[width=0.4\textwidth]{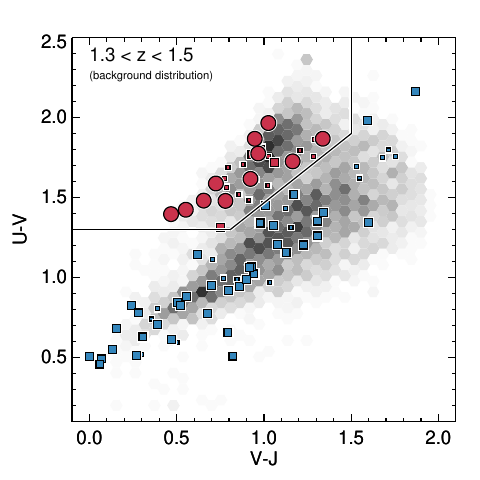}  
\includegraphics[width=0.4\textwidth]{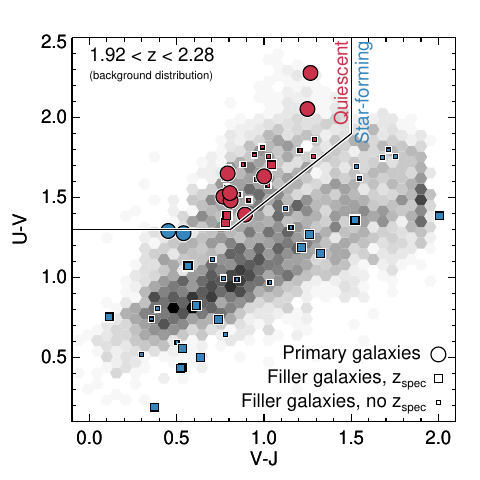}  
  \caption{Apparent magnitude vs. redshift (top panels) and rest-frame $U-V$ vs. $V-J$ colors (bottom panels) for all galaxies observed in the LRIS and MOSFIRE masks. The Heavy Metal 1 and 2 masks (left panels) primarily targeted bright ($J<21.6$) quiescent galaxies at $1.3<z<1.5$. The Heavy Metal 3 and 4 masks (right panels) primarily targeted bright ($H<21.8$) quiescent galaxies at $1.92<z<2.28$. Quiescent galaxies (red symbols) were selected by their red $U-V$ and blue $V-J$ colors, as in indicated by the selection box. However, several primary galaxies scattered out of the boxes when including the spectral information. The fillers are star-forming (blue symbols) and fainter quiescent galaxies at similar or higher/lower redshifts. The filler galaxies for which a spectroscopic redshift was measured are indicated by the larger symbols. We also show the parent UltraVISTA galaxy distribution from which the samples were drawn. The $UVJ$ panels only include the UltraVISTA galaxies in the targeted redshift intervals.
  \label{fig:selection}}
    \end{center}  
\end{figure*}

\subsection{Galaxy sample and success rate}\label{sec:sample}

Our primary galaxy sample is selected to have quiescent stellar populations, be relatively bright, and fall in two redshift intervals, \lowzr\ and \highzr.
In Figure~\ref{fig:redshifts} we show the photometric versus spectrospic redshifts of the primary (circles) and filler (squares) galaxies, as well as the distribution of the spectroscopic redshifts. Most primary galaxies fall in or very close to the selection windows, and their photometric and spectrscopic redshifts agree well with a normalized median absolute deviation in $\Delta z/(1+z_{\rm spec})$ of $\sigma_{\rm nmad}=0.014$. The only exception is HM4-59375, which has a significantly lower redshift than predicted by the photometry. This figure also shows the filler galaxies. These galaxies are drawn from a larger redshift distribution, though galaxies at similar redshifts were prioritized. The scatter for the filler galaxies is slightly larger with a $\sigma_{\rm nmad}=0.017$, which may be explained by their fainter magnitudes. 

The histogram in Figure~\ref{fig:redshifts} shows that the spectroscopic redshifts of the primary and filler targets are clustered, and several potential overdensities may exist, specifically at $z\sim1.40$ (HM1), $z\sim1.42$ (HM2), $z\sim2.16$ (HM4), and $z\sim2.23$ (HM3). This finding is not surprising, as we specifically selected pointings for which we can observe multiple quiescent galaxies in one field of view. A further investigation into these overdensities is beyond the scope of this paper. Nevertheless, when interpreting our results, it is important to keep in mind that the environments in which our galaxies reside may not be typical for distant quiescent galaxies.

In Figure~\ref{fig:selection} we show all primary and filler targets in magnitude versus redshift and rest-frame $U-V$ versus $V-J$ space. The left panels show the galaxies in Heavy Metal 1 and 2, while the right panels show galaxies in Heavy Metal 3 and 4. The boxes in the top panels enclosed by the dotted lines indicate the primary target selection in terms of magnitude and redshift. In contrast to Figure~\ref{fig:redshifts}, here we show both the confirmed filler galaxies (large squares) and the filler galaxies for which we did {\it not} measure a spectroscopic redshift (small squares). The top-left box in the bottom panels enclosed by the solid lines indicates our quiescent galaxy selection \citep[red symbols;][]{AMuzzin2013b}. Galaxies outside the box are generally identified as star-forming galaxies (blue symbols). 

While we measure spectroscopic redshifts of all primary targets, for the filler targets the success rate is lower with 71\% (42/59) and 53\% (17/32) for the $z\sim1.4$ masks and $z\sim2.1$ masks, respectively. There are several reasons for the lower success rate of the fillers. First, for Heavy Metal 1 and 2, most fillers are only targeted by either MOSFIRE {\it or} LRIS. Second, many fillers are faint quiescent targets, for which we do not detect clear absorption lines. The few faint quiescent fillers that are confirmed all have emission lines in their spectra. 
However, in the Heavy Metal 3 and 4 masks, there are several quiescent filler targets at $z\sim1.4$ that are as bright as the faintest primary targets. Unfortunately, we do not capture the 4000\,\AA\ region crucial for spectroscopic redshift measurements for these galaxies. Finally, for several star-forming fillers, the emission lines may be outside the atmospheric windows. For example, we find no confirmed star-forming galaxies below $z=2$ in the Heavy Metal 3 and 4 masks.

Based on the photometric redshifts, all primary targets were initially selected to be quiescent. However, when rederiving the rest-frame colors using the spectroscopic redshifts, two of the primary targets (HM3-107590 and HM4-55878) shift just outside the quiescent box. Given their location, though, we expect these galaxies to be post-starburst or young quiescent galaxies \citep[e.g.,][]{KWhitaker2012,SBelli2019,KSuess2021,MPark2023}, and thus still have quiescent populations. We will further assess their star-formation properties in the next section.

Finally, we compare our primary galaxies to the parent galaxy distributions at \lowzr\ and \highzr\ from which the targets are drawn.
At $z\sim1.4$ the primary targets  do sample nearly the full distribution along the quiescent sequence, though there is a bias toward bluer colors. The quiescent galaxies at $z\sim2.1$ span a larger range along the quiescent sequence, but on average are also biased toward the bluer and younger systems. This bias is expected, as our bright magnitude limit favors galaxies with lower mass-to-light ratios ($M/L$), which are generally bluer and younger. Obtaining a more representative sample would require significantly longer integration times and larger surveys, and thus necessitates  more efficient telescopes and spectrographs, such as NIRSpec on JWST.

\begin{figure*}
  \begin{center}  
\includegraphics[width=0.45\textwidth]{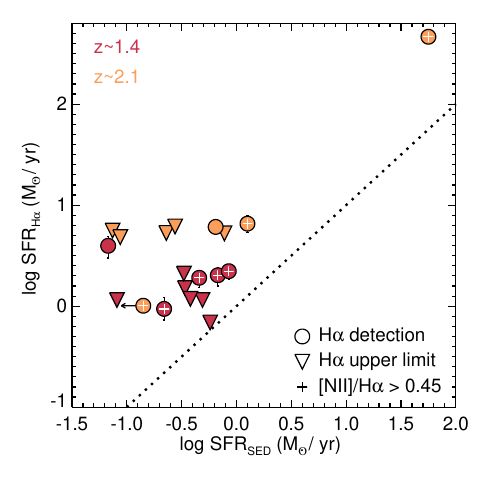}  \hspace{0.1in}
\includegraphics[width=0.45\textwidth]{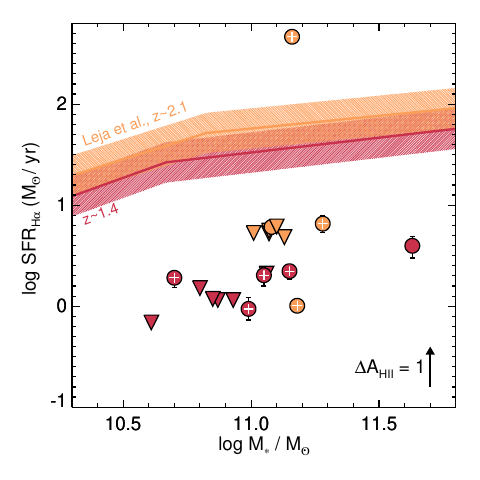}  
\vspace{-0.2in}
  \caption{SFR derived from the H$\alpha$ emission lines vs. the best-fit SED SFR and stellar mass for all primary galaxies with coverage of H$\alpha$. Orange and red symbols represent galaxies at $z\sim1.4$ (Heavy Metal 1 and 2) and $z\sim2.1$ (Heavy Metal 3 and 4), respectively. For galaxies without detected H$\alpha$ we show a 3$\sigma$ upper limit. For all but two galaxies with detected H$\alpha$, the [N\,{\sc ii}]/H$\alpha > 0.45$ (white plusses), implying that the H$\alpha$ flux is not dominated by star formation. Thus, for these galaxies, the H$\alpha$ SFR is overestimated and more comparable to an upper limit. Consequently, the H$\alpha$ SFRs (and limits) are larger than the SED SFRs, as illustrated in the left panel. In the right panel, we compare the H$\alpha$ SFRs with the star-forming main sequence from \cite{JLeja2022} at $z\sim1.4$ (red shaded area) and $z\sim2.1$ (orange shaded area). Except for HM4-55878, which has bright emission lines originating from a luminous AGN, all other galaxies are significantly below the star-forming main sequence. When using SED SFRs, they would shift to even lower values.}
  \label{fig:SFR}
    \end{center}  
\end{figure*}

\begin{figure*}
  \begin{center}  
  \includegraphics[width=0.8\textwidth]{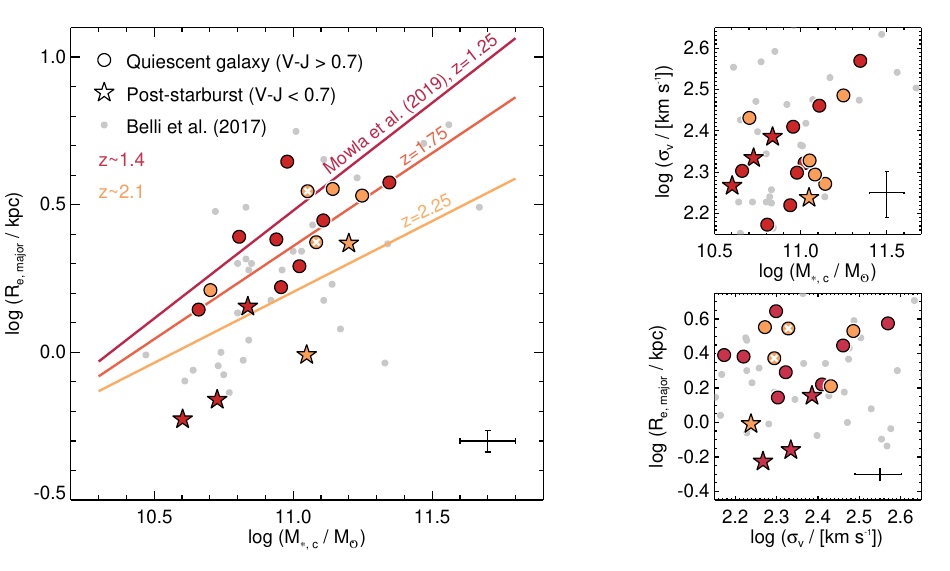} 
  \vspace{-0.1in}
  \caption{The left panel shows the effective radius (major axis) at rest-frame 5000\,\AA\ vs. stellar mass for the primary Heavy Metal galaxies. Red and orange symbols depict galaxies at $z\sim1.4$ and $z\sim2.1$, respectively. Stars and circles represent post-starburst ($V-J<0.7$) and older quiescent galaxies, respectively. The white crosses indicate the galaxies with bad Galfit fits. Heavy Metal galaxies at $z\sim1.4$ exhibit a bias toward smaller sizes, likely due to our selection criteria favoring younger galaxies. This bias indeed diminishes when excluding post-starburst galaxies. The right panels show how both $M_{\rm stellar}$ and $R_{\rm e, major}$ relate to the velocity dispersion ($\sigma_{v, \rm e}$). While the post-starburst galaxies are smaller, their velocity dispersions are comparable to those of older quiescent galaxies of the same stellar mass.}
  \label{fig:size_mass}
  \end{center}  
\end{figure*}

\subsection{Star formation constraints}\label{sec:sample_sfr}

All primary Heavy Metal galaxies are selected to have quiescent stellar populations based on their rest-frame $UVJ$ colors. In this section, we assess whether these galaxies indeed have low SFRs using both the stellar continuum emission and their emission-line properties. 

The SFRs derived from fitting the stellar spectra and photometry with SPS models are listed in Table~\ref{tab:sample}. Except for HM4-55878, all primary galaxies have best-fit SFRs of $<1\,\rm M_\odot\,yr^{-1}$. HM4-55878 has a significantly higher SFR than expected based on its $UVJ$ colors and the initial photometric analysis. This disparity is attributed to the influence of strong emission lines on the broadband SED. In our fitting procedure, we first adjusted the broadband photometry to account for the impact of these lines, as outlined in Section~\ref{sec:fit}. 

When examining the H$\alpha$ SFRs, we find a similar result. With the exception of HM4-55878, the primary galaxies exhibit either very faint or undetectable H$\alpha$ emission. Among the nine galaxies where H$\alpha$ is detected at $>3\sigma$, seven display [N\,{\sc ii}]/H$\alpha$ ratios exceeding 0.45, implying that star formation is likely not the primary ionization source \citep[e.g.,][]{JBaldwin1981,GKauffmann2003,LKewley2006}. Our study supports previous findings that high [N\,{\sc ii}]/H$\alpha$ ratios are common in distant quiescent galaxies \citep[e.g.,][]{MKriek2007,ANewman2018a,SBelli2017b}. Although such line ratios are commonly associated with photoionization by AGNs \citep[e.g.,][]{GKauffmann2003,LKewley2006}, in quiescent galaxies, they are thought to originate from the photoionization by hot evolved stars, including post-asymptotic giant branch stars \citep[e.g.,][]{RYan2012,FBelfiore2016}. For the majority of the galaxies, we do not have a meaningful measurement of [O\,{\sc iii}]/H$\beta$, and thus we cannot further assess the origin of the line emission in our sample. Only HM-55878 has a significant detection for all lines, with its line ratios suggesting an AGN (Y. Ma et al. in preparation). HM1-213931 and HM4-56163 have [N\,{\sc ii}]/H$\alpha<0.45$, and thus star formation is likely the dominant ionization source. These galaxies have low SFRs of 4-5 $M_\odot \rm yr^{-1}$, but the uncertainties are significant. For HM1-213931 we also do not see a clear emission line in the 2\,D spectrum (see Fig.~\ref{fig:halpha}).

In Figure~\ref{fig:SFR} (left panel) we compare the two SFR measurements. For consistency, both measurements assume solar metallicity and a similar IMF (Kroupa vs. Chabrier). For galaxies that have no detected H$\alpha$ emission, we show the 3$\sigma$ upper limit (triangles). For galaxies with detected H$\alpha$, we mark the ones for which star formation is not the primary ionization mechanism by a plus. For these galaxies the SFRs are overestimated, and the values should been regarded as upper limits. Additional attenuation toward H\,{\sc ii} regions, however, could potentially have resulted in an underestimation of the H$\alpha$ SFRs, as indicated by the arrow (right panel). 

Figure~\ref{fig:SFR} shows that all SFR upper limits from H$\alpha$ are not inconsistent with the SED SFRs. The galaxies for which H$\alpha$ does not originate from star formation are all located above the 1-to-1 line as well. Only for HM1-213931 and HM4-56163, for which H$\alpha$ is thought to originate from star formation, the two SFRs are inconsistent. HM1-213931 seems to be a merger of several galaxies (see Fig.~\ref{fig:speclz}), and thus it may not be surprising to find a low SFR of $4\pm 1\,M_\odot$ yr$^{-1}$. In particular, different physical regions for the stellar and nebular components may explain the discrepant values. The low [N\,{\sc II}]/H$\alpha$ ratio implies low metallicity, which makes it more likely that the star formation is either fueled by low-metallicity infalling gas or associated with a nearby smaller galaxy. \citet{SBelli2017b} find similarly low levels of (metal-poor) star formation activity in distant quiescent galaxies, which they attribute to rejuvenation events due minor mergers or inflowing gas. Higher spatial resolution spectra will be needed to examine the different components and further assess this galaxy. 

In the right panel of Figure~\ref{fig:SFR} we show the H$\alpha$ SFRs versus stellar mass, in comparison the star-forming main sequence (ridge) from \cite{JLeja2022} at similar redshifts (and for a similar IMF). Except for AGN HM4-55878, all primary quiescent targets are significantly below the star-forming main sequence at its redshift. Furthermore, the majority of these data points are upper limits, either because H$\alpha$ is undetected or because H$\alpha$ is not originating from star formation. Hence, except for HM4-55878, all galaxies have indeed strongly suppressed star formation.

\begin{figure*}
  \begin{center}  
  \includegraphics[width=1.0\textwidth]{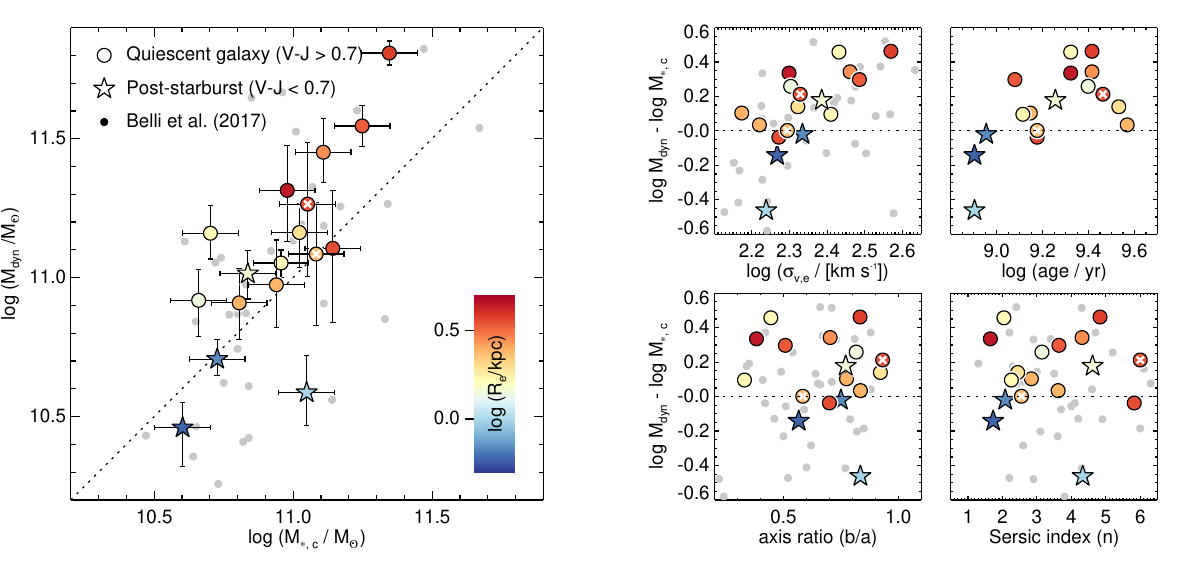}  \hspace{0.1in}
  \vspace{-0.2in}
  \caption{{\it Left:} dynamical vs. stellar mass for all primary Heavy Metal galaxies for which sizes and stellar velocity dispersion could be measured, assuming a \cite{GChabrier2003} IMF. The white crosses indicate the galaxies with bad Galfit fits. We also show the galaxies by \cite{SBelli2017} with comparable redshifts ($z>1.35$). {\it Right:} the difference between the dynamical and stellar mass vs. the velocity dispersion ($\sigma_{v,e}$), age of the galaxy, axis ratio $b/a$, and S\'ersic index $n$. The ages, adopted from \cite{ABeverage2024}, are derived using {\sc alf} and present the luminosity-weighted age. In all panels, the galaxies are color coded by their effective radii at rest-frame 5000\,\AA\ (major axis). For the majority of the galaxies, the dynamical mass exceeds the stellar mass, with a median dark matter fraction of 28\%. Three galaxies have stellar masses that exceed their dynamical masses. Interestingly, three galaxies with low $M_{\rm dyn}/M_{\rm stellar}$ are the smallest and youngest galaxies in the sample.}
  \label{fig:Mdyn}
  \end{center}  
\end{figure*}

\subsection{Galaxy structures}\label{sec:structures}

Quiescent galaxies follow a size-mass relationship, where galaxies with greater mass or luminosity exhibit larger effective radii \citep[e.g.,][]{JKormendy1977, SShen2003}. This relationship evolves over cosmic time, with galaxies at greater distances appearing more compact \citep[e.g.,][]{ITrujillo2006, PvanDokkum2008, AvanderWel2014, LMowla2018, KSuess2019a, KSuess2019b}. In Figure~\ref{fig:size_mass}, we compare the half-light radii at rest-frame 5000\,\AA\ of the Heavy Metal galaxies with the average size-mass relation at $z=1.25$, $z=1.75$, and $z=2.25$ as reported by \cite{LMowla2018} (using the same IMF) for a large representative sample of massive quiescent galaxies. To ensure consistency with prior research, we consider the major axis (noncircularized) as the half-light radius. In contrast to Figure~\ref{fig:SFR}, here we use the stellar masses that are corrected using the {\sc Galfit} magnitudes, to make them consistent with the size measurements (see Sect.~\ref{sec:dynamical_masses}). \cite{LMowla2018} also applied this correction.

When comparing the galaxies at $z\sim1.4$ to the relations at $z=1.25$ and $z=1.75$ (Fig.~\ref{fig:size_mass}), we find that, on average, the Heavy Metal galaxies are smaller. This trend can likely be attributed to our selection criteria favoring quiescent galaxies with lower $M/L$, indicative of younger ages. 
Several studies have indeed highlighted that younger quiescent galaxies have smaller half-light radii than their older counterparts of equivalent mass \citep[e.g.,][]{KWhitaker2012, SBelli2015, MYano2016, OAlmaini2017, DMatlby2018, PFWu2020,KSuess2020,KSuess2021,DSetton2022}. 
When excluding the youngest galaxies (stars), as identified by their blue $V-J$ ($<0.7$) colors  \citep[e.g.,][]{SBelli2019, ABeverage2021}, we find a good agreement between the relations by Mowla and the Heavy Metal galaxies at $z\sim1.4$.  
The sizes of the $z\sim2.1$ Heavy Metal galaxies are more challenging to compare, as there are only five galaxies with robust size measurements, of which two have blue $V-J$ colors. Our primary quiescent galaxies also have similar sizes to the spectroscopic galaxy sample by \cite{SBelli2017}, when including galaxies at similar redshifts ($1.35<z<2.45$).

In Figure~\ref{fig:size_mass} we also show the stellar masses and sizes in relation to their velocity dispersions. These panels show that the youngest galaxies, despite their small sizes, have similar velocity dispersions as older galaxies of the similar mass. The Heavy Metal galaxies follow a distribution roughly similar to that of the sample by \cite{SBelli2017} in both diagrams.

\begin{figure*}
  \begin{center}  
  \includegraphics[width=0.45\textwidth]{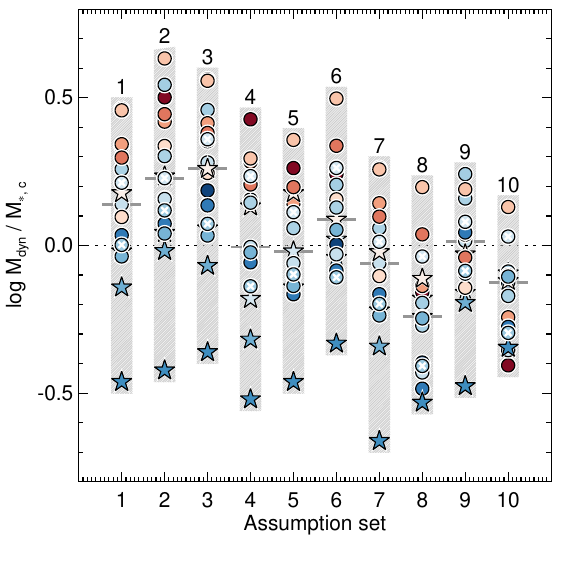}
  \hspace{0.1in}
  \includegraphics[width=0.45\textwidth]{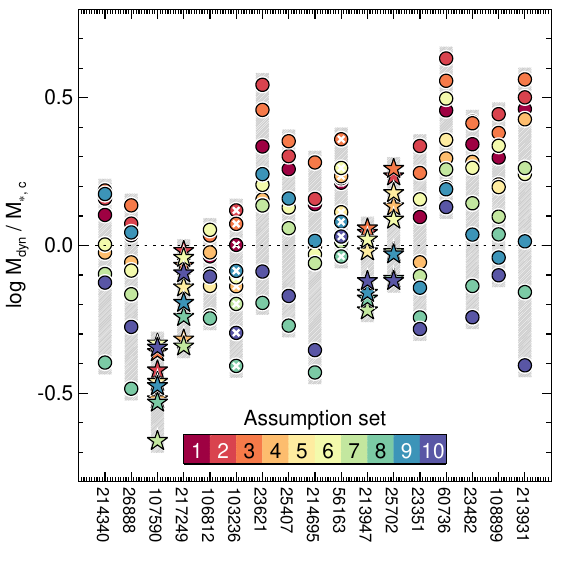}  
  \caption{{\it Left:} the distribution in \mr\ for different assumptions when calculating the dynamical and stellar mass (see Table~\ref{tab:assumptions}). Symbols are similar as in Figure~\ref{fig:Mdyn}. The galaxies are color coded according to their velocity dispersion following the color bar in Figure~\ref{fig:mdyn_ass_two}. The vertical light-gray stripes indicate the full range and the median is indicated by the dark-gray horizontal bar {\it Right:} the distribution in \mr\ for the individual galaxies for all ten assumption sets. The galaxies are ordered by increasing velocity dispersion from left to right. Each vertical stripe corresponds to a separate galaxy and each color to a different assumption set.}
  \label{fig:mdyn_ass}
  \end{center}  
\end{figure*}

\begin{figure}
    \centering      
  \includegraphics[width=0.45\textwidth]{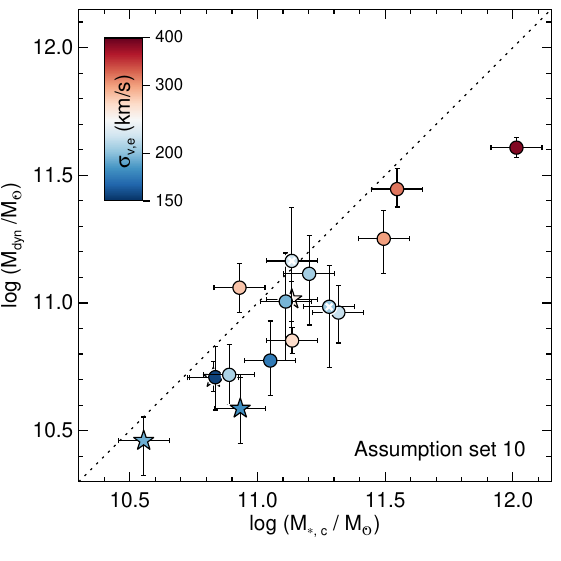}  \vspace{-0.1in}
  \caption{Dynamical vs. stellar mass when adopting the $\sigma_v$-dependent IMF by \cite{TTreu2010} for the cores of nearby early-type galaxies, subsolar metallicity, and half-mass radii. For this combination of assumptions, the scatter in \mr\ is smallest. However, the stellar mass exceeds the dynamical mass by 34\%. This tension may imply that distant quiescent galaxies do not simply grow inside-out into present-day massive early-type galaxies.  }
  \label{fig:mdyn_ass_two}
\end{figure}

\subsection{Comparison of dynamical and stellar masses}\label{sec:masscomp}

The combination of deep Keck spectra with high-resolution HST imaging enables dynamical mass measurements for the majority of the primary Heavy Metal galaxies, as listed in Table~\ref{tab:struct}. In addition to the stellar content, the dynamical mass also includes the dark matter and gas components. Thus, in theory, the dynamical masses should give us insights into these dark components. In practice, however, this is extremely challenging, as both the stellar and dynamical mass measurements rely on many assumptions (see Sect.~\ref{sec:dynamical_masses}). Nonetheless, the boundary condition that the stellar mass should not exceed the dynamical mass provides an independent check on our stellar mass measurements, and may give us insights into assumptions that went into our mass estimates.

In Figure~\ref{fig:Mdyn} we show the dynamical vs. stellar mass for the primary Heavy Metal galaxies. For the majority of the galaxies, the dynamical mass exceeds the stellar mass, with a median dark matter fraction of 28\%. For two galaxies the dynamical masses are below their stellar masses, with one galaxy (HM3-107590) being off by $>3 \sigma$. These galaxies are among the smallest (based on the light-weighted size) and youngest, as shown in the top-right panel of Figure~\ref{fig:Mdyn}. Interestingly, \cite{JRunco2022} found a similar result for a post-starburst galaxy at redshift $z=1.89$, with the stellar mass also being significantly larger than the dynamical mass. \cite{MCappellari2023} also found a trend with age for $z\sim0.7$ galaxies in the LEGA-C survey, with the younger galaxies having lower dynamical-to-stellar mass ratios.

\begin{table*}
\caption{Dynamical-to-stellar mass ratios for varying assumptions}
\label{tab:assumptions}
\centering
\hspace{-0.8in}\begin{tabular}{l l c r c r r}
\hline \hline
\multicolumn{1}{c}{No} & \multicolumn{2}{c}{\underline{$M_{\rm dyn}$ assumptions}} & \multicolumn{2}{c}{\underline{$M_*$ assumptions}} & \multicolumn{2}{c}{\underline{log\,\mr}} \\
& \multicolumn{1}{c}{$R_e$} & \multicolumn{1}{c}{Virial coefficent} & \multicolumn{1}{c}{Z} & \multicolumn{1}{c}{IMF$^{a}$} & \multicolumn{1}{c}{median} & \multicolumn{1}{c}{$\sigma$}\\
\hline
  1 & circ, light & $\beta(n)$ & 0.0190 & Chabrier &  0.14  &   0.23 \\
  2 & sma, light & $\beta(n)$ & 0.0190 & Chabrier &    0.23  &   0.25   \\
  3 & sma, light & $\beta(n)K(q)$ & 0.0190 & Chabrier &  0.26  &  0.24  \\
  4 & circ, light & 5 & 0.0190 & Chabrier &  -0.01  &   0.24  \\
  5 & circ, mass & $\beta(n)$ & 0.0190 & Chabrier &  -0.02  &   0.20  \\
  6 & circ, light & $\beta(n)$ & 0.0096 & Chabrier &   0.09  &   0.20 \\
  7 & circ, light & $\beta(n)$ & 0.0190 & Salpeter &  -0.06  &   0.23 \\
  8 & circ, mass & $\beta(n)$ & 0.0096 & Salpeter &  -0.24  &   0.19 \\
  9 & circ, light & $\beta(n)$ & 0.0190 & $\sigma_v$ &   0.01   &  0.17 \\
 10 & circ, mass & $\beta(n)$ & 0.0096 & $\sigma_v$ & -0.13 &  0.14\\
\hline
\multicolumn{7}{l}{$^a$ $\sigma_v$ corresponds to the $\sigma_v$-dependent IMF by \cite{TTreu2010}} \\
\end{tabular}
\end{table*}

In order to assess how the masses compare when adopting different assumptions, and to understand why some galaxies have $M_{*,c}>M_{\rm dyn}$, we discuss the different assumptions below. First, when deriving the dynamical mass, we circularize the effective radius and use a S\'ersic-dependent virial coefficient $\beta(n)$. If we had not circularized the radius, \mr\ would increase by 0.023\,dex and the scatter would increase from 0.232 to 0.255 (assumption set 2, Fig.\ref{fig:mdyn_ass} and Table~\ref{tab:assumptions}). Instead of circularizing, we also explore the axis ratio correction by \cite{AvanderWel2022}, which is implemented using an additional virial coefficient $K(q)$. This combination increases the median \mr\ by 0.12 dex as well as the scatter (assumption set 3). Assuming a virial constant of 5 would also increase the scatter in \mr, but the median \mr\ would decrease by 0.15\,dex (assumption set 4).

Second, in our dynamical mass measurement, we assume that the galaxies are pressure supported. However, if they are (partially) rotationally supported, our dynamical mass measurement would be off. For example, for HM3-107590 the low dynamical mass could be explained by a face-on view or by a strong misalignment of the slit and the major axis of the galaxy. For such cases, part of the velocity field would not be included in our dispersion measurement and we would underestimate the mass. We check for this possibility by examining \mr\ as a function of the S\'ersic index and axis ratio ($b/a$) in the right panels of Figure~\ref{fig:Mdyn}. HM3-107590 is nearly round and the velocity dispersion is indeed lower compared to galaxies of similar mass (Fig.~\ref{fig:size_mass}). The S\'ersic index appears at odds with the galaxy being a disk, though this measurement is quite uncertain as this galaxy is just barely resolved.

In this context, it is interesting to note that  \citet{SBelli2017} find higher \mr\ for galaxies with low S\'ersic indices ($n<2.5$) and low axis ratios, and interpret this finding as evidence for a significant contribution of rotational motion. 
We do not see any indications that galaxies with the highest \mr\ preferentially have low $n$ and low $b/a$. However, in contrast to \citet{SBelli2017}, we circularize our effective radii when deriving the dynamical mass, which lowers \mr\ for galaxies with low axis ratios, and thus we partially account for inclination effects. Improving upon our simplified approach requires a forward modeling method, preferentially combined with spatially resolved spectroscopy, allowing for dynamical models with different levels of rotational support and correcting for inclination and aperture effects \citep[e.g.,][]{SPrice2016,SPrice2020,JvanHoudt2021,AdeGraaff2023}.

Third, dynamical masses depend on size measurements, which may have been biased due to stellar population gradients. Distant quiescent galaxies have redder centers, with the gradient being stronger in galaxies that are more massive, older, and at lower redshifts 
\citep[e.g.,][]{MMosleh2017, KSuess2019a, KSuess2019b, KSuess2020, KSuess2021, TMiller2023}. By applying the average size corrections by \citet[][$\sim$0.2\,dex and $\sim$0.1\,dex for the older quiescent galaxies at $z\sim1.4$ and $z\sim2.1$, respectively, and no corrections for post-starburst galaxies]{KSuess2019a,KSuess2021}, we find that median \mr\ decreases by 0.16 dex (see Fig.~\ref{fig:mdyn_ass}, assumption set 5). The color gradient correction also reduces the scatter \mr. The $M_{\rm dyn}$ of the two youngest galaxies remain unaffected, as post-starburst galaxies tend to display uniform color gradients \citep[e.g.,][]{DSetton2020,KSuess2020,KSuess2021}. HM1-217249, the sole post-starburst galaxy with robust size measurements in both F814W and F160W, supports this trend. Thus, stellar population gradients do not explain the low \mr\ of the few post-starburst galaxies. Instead, they further lower the median inferred dark matter fraction of our full distant quiescent galaxy sample. Finally, size underestimation could occur due to the presence of an AGN, although the full SEDs and spectra provide limited room for a power-law continuum contribution.

The stellar masses could also be biased. First, as we assume a simple delayed exponential star-formation history, we likely miss older and low $M/L$ stellar populations in our stellar mass. This ``outshining'' problem has been discussed in many works \citep[e.g.,][]{CPapovich2001,SWuyts2007,JLeja2019,CGimenez2023}. However, for distant massive quiescent galaxies this effect is small, and thus the  {\sc fast} and {\sc Prospector} masses \citep{BJohnson2021}, assuming nonparameterized star formation histories, are very similar \citep{JLeja2019}. 
Second, we assume solar metallicity while the galaxies, on average, have subsolar iron abundances \citep{ABeverage2024}. Assuming a half-solar metallicity (Z=0.0096) would increase the median stellar masses by 13\% (see Fig.~\ref{fig:mdyn_ass}) and decreases the scatter in \mr\ by 0.035. Third, we assume a \citet[][]{GChabrier2003} IMF, which, similar to a Kroupa IMF, is relatively bottom light. Assuming a \citet{ESalpeter1955} IMF would increase the stellar masses by 0.2 dex (Fig.~\ref{fig:mdyn_ass}), such that they exceed $M_{\rm dyn}$ for the majority of the galaxies. Thus, the IMF assumption causes the largest (systematic) uncertainty in our stellar mass estimates \citep[see also][]{BWang2023}. Combining both stellar mass effects {\it and} the color gradient correction (assumption set 8 in Fig.~\ref{fig:mdyn_ass}), would lead to stellar masses vastly exceeding the dynamical masses for nearly all galaxies. Hence, given our dynamical masses, we infer that a Chabrier IMF is more likely than a Salpeter IMF for distant quiescent galaxies. We will further explore the IMF in the next section. 


\section{Discussion}\label{sec:discussion}

\subsection{Implications for Photometric Studies}

While the number of distant star-forming galaxies with spectroscopic redshifts has increased tremendously in the past decade \citep[e.g.,][]{CSteidel2014,EWisnioski2015,MKriek2015}, the number of distant quiescent galaxies with spectroscopic redshifts or other spectroscopic information is still very small. Detecting absorption lines requires significantly longer integration times than observing nebular emission lines. Thus, the majority of studies of quiescent galaxies over cosmic time, including the buildup of the stellar mass function, still rely on photometric data. 

Our study presents a reassuring picture. The initial photometric redshifts of our primary quiescent galaxies agree well with their spectroscopic redshifts (see Sect.~\ref{sec:sample}) and nearly all galaxies have quiescent stellar populations with their SFRs significantly below the star-forming main sequence (see Sect.~\ref{sec:sample_sfr}). We do find, however, that photometric redshifts become less accurate beyond $z=2$. Furthermore, we show that for one galaxy, the contribution from strong AGN emission lines mimics the shape of a quiescent galaxy. \citet{CSchreiber2018} shows that the success rate of the $UVJ$ selection criteria further declines to about 80\% when going to $3<z<4$. \citet{BForrest2020} presents a less optimistic picture with a spectroscopic confirmation rate of about 50\% for quiescent galaxy candidates beyond $z=3$.  Nonetheless, out to $z\sim2$, we do not expect that mass functions of quiescent galaxies will be strongly biased by incorrect photometric redshifts or quiescent galaxy classifications.

\subsection{Implications for the evolution of massive quiescent galaxies}

One popular explanation for the size evolution of quiescent galaxies over cosmic time is growth by minor mergers. This scenario is supported by the finding that the central mass densities of quiescent galaxies remain roughly constant, while their (blue) outskirts are building up over time \citep[e.g.,][]{RBezanson2009,PvanDokkum2010,GBarro2017,KSuess2021}. Furthermore, distant quiescent galaxies have many small companions \citep{ANewman2012,KSuess2023}. Thus, in this scenario, distant quiescent galaxies are the cores of massive galaxies today. These same cores are also found to have a bottom-heavy IMF  \citep[e.g.,][]{TTreu2010,CConroy2012d}, with the highest velocity dispersion galaxies having a larger excess of low-mass stars. Thus, for this inside-growth scenario, the IMF in the high-dispersion galaxies should already be bottom heavy at these early times.

In Figure~\ref{fig:Mdyn} we indeed find that \mr\ correlates with $\sigma_{v,e}$, which could imply that the IMF may be more bottom heavy in higher-dispersion galaxies. This trend was already visible in the $M_{\rm dyn} - M_{*}$ diagrams in several distant quiescent galaxy studies \citep[e.g.,][]{JvandeSande2013,SBelli2017,BForrest2022}, and discussed in detail in \cite{JMendel2020}. \cite{JMendel2020} argue that this trend is due to a varying IMF and that the IMF-$\sigma_v$ relation was already in place at these early times.

To further assess this theory, we show \mr\ assuming a $\sigma_v$-dependent IMF for our distant quiescent galaxies in Figure~\ref{fig:mdyn_ass} (assumption set 9). We use the relation by \citet[][Equation 4]{TTreu2010}, in which the IMF is more bottom heavy than the Salpeter IMF for galaxies with $\sigma_v > 250\rm \, km\,s^{-1}$.  This IMF results in a median \mr\ of 1.033. The scatter in \mr\ is strongly reduced, which is expected as we are (partially) removing the trend with $\sigma_{v,e}$. Interestingly, the scatter is smallest when also assuming subsolar metallicity (Z=0.0096) and correcting for color gradients (Fig.~\ref{fig:mdyn_ass}, assumption set 10). However, this assumption set results in stellar masses that exceed the dynamical masses for all but one galaxy, with a median \mr\ of 0.75 (Fig.~\ref{fig:mdyn_ass_two}). Thus, our dynamical masses may suggest that compact distant quiescent galaxies do not ``passively'' evolve into the cores of massive elliptical galaxies today and that the evolution is more complicated \citep[e.g.,][]{SWellons2015}. Major mergers (with galaxies with a different IMF) and/or late-time central star formation could have affected the average IMF in today's cores. 

We come to a similar conclusion based on our elemental abundance measurements presented in our accompanying paper \citep[][]{ABeverage2024}. The iron abundances in distant quiescent galaxies are much lower than found in the cores of nearby massive early-type galaxies \citep[][]{MGu2022}. Neither minor mergers nor progenitor bias can explain this evolution, and thus late-time star formation and/or major mergers are needed to explain the increase in iron abundance. We do note, though, that minor mergers are still needed to explain the structural and size evolution of massive quiescent galaxies over cosmic time.

Interestingly, \cite{PvanDokkum2023} came to the opposite conclusion, based on a perfect lensing system \citep[see also][]{WMercier2023}. They find that a bottom-heavy IMF must already be in place for a distant quiescent galaxy at $z\sim 1.9$, because the stellar mass, assuming the \cite{GChabrier2003} IMF, would lead to an unrealistically large dark matter fraction within the Einstein radius. Obtaining spectroscopic redshifts for both galaxies as well as a dynamical mass measurement for the quiescent galaxy lens would be needed to directly compare our results. 

To further unravel this puzzle, we need progress on several fronts. First, we need to measure stellar population gradients and half-mass radii for our spectroscopic samples. This should preferentially be done from spectroscopic data, as age, metallicity, and dust gradients result in different $M/L$ gradients \citep[e.g.,][]{JvandeSande2015}. We would also have to redetermine the stellar masses, taking into account these stellar population gradients. Second, we need to resolve the kinematics of distant quiescent galaxies, such that we can model their stellar dynamics. Third, we need a direct spectroscopic measurement of the IMF in distant quiescent galaxies \citep[using gravity-sensitive absorption features; e.g.,][]{PvanDokkum2010IMF}, to obtain more accurate stellar masses and understand whether the bottom-heavy IMF was already in place at these early times. Finally, we need larger samples of galaxy spectra. JWST will enable advances in all these areas and has already collected spectra of a handful of distant quiescent galaxies \citep{TNanayakkara2022,ACarnall2023,DMarchesini2023,FDEugenio2023,SBelli2023}.

\section{Summary}

In this paper, we present an overview of the Heavy Metal survey, an ultradeep rest-frame optical spectroscopic survey of 21 distant quiescent galaxy candidates at $1.4\lesssim z \lesssim 2.2$.
The Heavy Metal survey was executed with MOSFIRE and LRIS on the Keck I telescope and overlaps with the UltraVISTA and COSMOS-DASH surveys.
Our primary targets were selected across two redshift intervals, \lowzr\ and \highzr, allowing the observation of multiple Balmer and metal (Ca, Mg, Fe) absorption lines in atmospheric windows. The extensive sky coverage enabled galaxy pointings for which we observe 5-6 ``bright'' quiescent candidates in one pointing, with two pointings per redshift interval. The remaining slits were placed on fainter quiescent and star-forming galaxies at similar redshifts. The $z\sim1.4$ and $z\sim2.1$ targets were observed for a total of $\sim18$ and $\sim32$ hr, respectively. The Heavy Metal survey is unique for its wavelength coverage and presents the first statistical sample of $z\gtrsim1.4$ quiescent galaxies with ultradeep spectra covering rest-frame $\sim$3700--5400\,\AA. 
 
We measure spectroscopic redshifts for all primary targets, and nearly all show clear Balmer and metal absorption lines in their spectra. 20 out of the 21 quiescent candidates indeed have quiescent stellar populations; the SFRs determined from H$\alpha$ {\it and} spectrophotometric fitting are both significantly below the star-forming main sequence. For 11 out of the 20 quiescent galaxies, we detect no H$\alpha$ and derive upper limits on the SFR from H$\alpha$. For nine targets, we do detect faint H$\alpha$ emission, but seven of them have emission-line ratios that indicate that star formation is not the primary ionization source; instead, they may be powered by hot evolved stars or low-luminosity AGNs. Hence, for these galaxies the H$\alpha$ SFRs are more comparable to upper limits, as well. For the remaining two galaxies with detected H$\alpha$, the SFRs are very low, and for one of them \niiha\ suggests that the star formation is likely associated with a nearby smaller galaxy. Finally, one of the quiescent candidates appeared to be an AGN, with strong (asymmetric) emission lines mimicking the SED shape of a quiescent galaxy. This galaxy will be discussed in detail in Y. Ma et al. (2024, in preparation). 

The primary goal of the Heavy Metal survey is to measure chemical compositions and ages from the stellar absorption-line spectra. These measurements are discussed in our accompanying paper \citep{ABeverage2024}. The stellar population fitting, presented in that paper, also yields accurate stellar velocity dispersion measurements for 19 out of the 21 primary galaxies. These measurements, combined with the structural parameters derived from HST F814W and F160W imaging, enable us to derive dynamical masses for the majority of the primary Heavy Metal galaxies.

In this paper, we compare our dynamical masses with the stellar masses from spectrophotometric modeling, considering various assumptions for both masses. Interestingly, for a fixed IMF, \mdyn/$M_*$ shows a positive correlation with $\sigma_v$. This correlation may suggest that a varying IMF, which is more bottom heavy for high-$\sigma_v$ galaxies, was already in place at these early times \citep[see also][]{JMendel2020}. When implementing the $\sigma_v$-dependent IMF found in the cores of nearby massive early-type galaxies, and also correcting for biases in our stellar mass and size measurements, we find a low scatter in \mdyn/$M_*$ of only 0.14\,dex and a median \mdyn/$M_*$ of 0.75. Thus, for these assumptions, the stellar mass measurements exceed the dynamical masses for nearly all quiescent galaxies. This result may imply that distant quiescent galaxies do not simply grow inside-out into massive early-type galaxies in today's Universe and late-time evolution (major mergers and/or late-time star formation) may be needed. In \cite{ABeverage2024} we come to a similar conclusion based on the difference in iron abundance between our distant quiescent galaxies and the cores of nearby massive early-type galaxies.

In order to fully characterize the distant quiescent galaxy population and solve this possible tension with the studies of the cores in nearby massive galaxies, we need to make progress on several fronts. First, we need a statistical sample of distant quiescent galaxies with resolved stellar kinematics, ages, elemental abundances, and robust stellar mass profiles. Moreover, we need to directly measure the IMF in distant quiescent galaxies using gravity-sensitive absorption lines. JWST will be able to make progress on all these fronts and thus will be transformative for our understanding of the formation histories of  distant quiescent galaxies and their evolutionary link to the massive early-type galaxies in the present-day Universe.\\

We acknowledge support from NSF AAG grants AST-1908748 and 1909942. C.C. acknowledges support from NSF grant AST-131547. The authors wish to recognize and acknowledge the very significant cultural role and reverence that the summit of Maunakea has always had within the indigenous Hawaiian community. We are most fortunate to have the opportunity to conduct observations from this mountain.

\software{\textsc{fast} \citep{MKriek2009b}, \textsc{fsps}\citep{CConroy2009,CConroy2010},\textsc{Galfit}\citep{MCappellari2004}}

\bibliography{mybib}

\begin{thebibliography}{}
\expandafter\ifx\csname natexlab\endcsname\relax\def\natexlab#1{#1}\fi
\providecommand{\url}[1]{\href{#1}{#1}}
\providecommand{\dodoi}[1]{doi:~\href{http://doi.org/#1}{\nolinkurl{#1}}}
\providecommand{\doeprint}[1]{\href{http://ascl.net/#1}{\nolinkurl{http://ascl.net/#1}}}
\providecommand{\doarXiv}[1]{\href{https://arxiv.org/abs/#1}{\nolinkurl{https://arxiv.org/abs/#1}}}

\bibitem[{{Almaini} {et~al.}(2017){Almaini}, {Wild}, {Maltby}, {Hartley},
  {Simpson}, {Hatch}, {McLure}, {Dunlop}, \& {Rowlands}}]{OAlmaini2017}
{Almaini}, O., {Wild}, V., {Maltby}, D.~T., {et~al.} 2017, \mnras, 472, 1401,
  \dodoi{10.1093/mnras/stx1957}

\bibitem[{{Antwi-Danso} {et~al.}(2023){Antwi-Danso}, {Papovich}, {Esdaile},
  {Nanayakkara}, {Glazebrook}, {Hutchison}, {Whitaker}, {Marsan}, {Diaz},
  {Marchesini}, {Muzzin}, {Tran}, {Setton}, {Kaushal}, {Speagle}, \&
  {Cole}}]{JAntwi-Danso2023}
{Antwi-Danso}, J., {Papovich}, C., {Esdaile}, J., {et~al.} 2023, arXiv
  e-prints, arXiv:2307.09590, \dodoi{10.48550/arXiv.2307.09590}

\bibitem[{{Baldwin} {et~al.}(1981){Baldwin}, {Phillips}, \&
  {Terlevich}}]{JBaldwin1981}
{Baldwin}, J.~A., {Phillips}, M.~M., \& {Terlevich}, R. 1981, \pasp, 93, 5,
  \dodoi{10.1086/130766}

\bibitem[{{Barro} {et~al.}(2017){Barro}, {Faber}, {Koo}, {Dekel}, {Fang},
  {Trump}, {P{\'e}rez-Gonz{\'a}lez}, {Pacifici}, {Primack}, {Somerville},
  {Yan}, {Guo}, {Liu}, {Ceverino}, {Kocevski}, \& {McGrath}}]{GBarro2017}
{Barro}, G., {Faber}, S.~M., {Koo}, D.~C., {et~al.} 2017, \apj, 840, 47,
  \dodoi{10.3847/1538-4357/aa6b05}

\bibitem[{{Belfiore} {et~al.}(2016){Belfiore}, {Maiolino}, {Maraston},
  {Emsellem}, {Bershady}, {Masters}, {Yan}, {Bizyaev}, {Boquien}, {Brownstein},
  {Bundy}, {Drory}, {Heckman}, {Law}, {Roman-Lopes}, {Pan}, {Stanghellini},
  {Thomas}, {Weijmans}, \& {Westfall}}]{FBelfiore2016}
{Belfiore}, F., {Maiolino}, R., {Maraston}, C., {et~al.} 2016, \mnras, 461,
  3111, \dodoi{10.1093/mnras/stw1234}

\bibitem[{{Belli} {et~al.}(2015){Belli}, {Newman}, \& {Ellis}}]{SBelli2015}
{Belli}, S., {Newman}, A.~B., \& {Ellis}, R.~S. 2015, \apj, 799, 206,
  \dodoi{10.1088/0004-637X/799/2/206}

\bibitem[{{Belli} {et~al.}(2017{\natexlab{a}}){Belli}, {Newman}, \&
  {Ellis}}]{SBelli2017}
---. 2017{\natexlab{a}}, \apj, 834, 18, \dodoi{10.3847/1538-4357/834/1/18}

\bibitem[{{Belli} {et~al.}(2019){Belli}, {Newman}, \& {Ellis}}]{SBelli2019}
---. 2019, \apj, 874, 17, \dodoi{10.3847/1538-4357/ab07af}

\bibitem[{{Belli} {et~al.}(2014){Belli}, {Newman}, {Ellis}, \&
  {Konidaris}}]{SBelli2014}
{Belli}, S., {Newman}, A.~B., {Ellis}, R.~S., \& {Konidaris}, N.~P. 2014,
  \apjl, 788, L29, \dodoi{10.1088/2041-8205/788/2/L29}

\bibitem[{{Belli} {et~al.}(2017{\natexlab{b}}){Belli}, {Genzel}, {F{\"o}rster
  Schreiber}, {Wisnioski}, {Wilman}, {Wuyts}, {Mendel}, {Beifiori}, {Bender},
  {Brammer}, {Burkert}, {Chan}, {Davies}, {Davies}, {Fabricius}, {Fossati},
  {Galametz}, {Lang}, {Lutz}, {Momcheva}, {Nelson}, {Saglia}, {Tacconi},
  {Tadaki}, {{\"U}bler}, \& {van Dokkum}}]{SBelli2017b}
{Belli}, S., {Genzel}, R., {F{\"o}rster Schreiber}, N.~M., {et~al.}
  2017{\natexlab{b}}, \apjl, 841, L6, \dodoi{10.3847/2041-8213/aa70e5}

\bibitem[{{Belli} {et~al.}(2023){Belli}, {Park}, {Davies}, {Mendel}, {Johnson},
  {Conroy}, {Benton}, {Bugiani}, {Emami}, {Leja}, {Li}, {Maheson}, {Mathews},
  {Naidu}, {Nelson}, {Tacchella}, {Terrazas}, \& {Weinberger}}]{SBelli2023}
{Belli}, S., {Park}, M., {Davies}, R.~L., {et~al.} 2023, arXiv e-prints,
  arXiv:2308.05795, \dodoi{10.48550/arXiv.2308.05795}

\bibitem[{{Beverage} {et~al.}(2021){Beverage}, {Kriek}, {Conroy}, {Bezanson},
  {Franx}, \& {van der Wel}}]{ABeverage2021}
{Beverage}, A.~G., {Kriek}, M., {Conroy}, C., {et~al.} 2021, \apjl, 917, L1,
  \dodoi{10.3847/2041-8213/ac12cd}

\bibitem[{{Beverage} {et~al.}(2023{\natexlab{a}}){Beverage}, {Kriek}, {Conroy},
  {Sandford}, {Bezanson}, {Franx}, {van der Wel}, \& {Weisz}}]{ABeverage2023}
---. 2023{\natexlab{a}}, \apj, 948, 140, \dodoi{10.3847/1538-4357/acc176}

\bibitem[{{Beverage} {et~al.}(2023{\natexlab{b}}){Beverage}, {Kriek}, {Suess},
  {Conroy}, {Price}, {Barro}, {Bezanson}, {Franx}, {Lorenz}, {Ma}, {Mowla},
  {Pasha}, {van Dokkum}, \& {Weisz}}]{ABeverage2024}
{Beverage}, A.~G., {Kriek}, M., {Suess}, K.~A., {et~al.} 2023{\natexlab{b}},
  arXiv e-prints, arXiv:2312.05307, \dodoi{10.48550/arXiv.2312.05307}

\bibitem[{{Bezanson} {et~al.}(2013){Bezanson}, {van Dokkum}, {van de Sande},
  {Franx}, \& {Kriek}}]{RBezanson2013}
{Bezanson}, R., {van Dokkum}, P., {van de Sande}, J., {Franx}, M., \& {Kriek},
  M. 2013, \apjl, 764, L8, \dodoi{10.1088/2041-8205/764/1/L8}

\bibitem[{{Bezanson} {et~al.}(2009){Bezanson}, {van Dokkum}, {Tal},
  {Marchesini}, {Kriek}, {Franx}, \& {Coppi}}]{RBezanson2009}
{Bezanson}, R., {van Dokkum}, P.~G., {Tal}, T., {et~al.} 2009, \apj, 697, 1290,
  \dodoi{10.1088/0004-637X/697/2/1290}

\bibitem[{{Brammer} {et~al.}(2008){Brammer}, {van Dokkum}, \&
  {Coppi}}]{GBrammer2008}
{Brammer}, G.~B., {van Dokkum}, P.~G., \& {Coppi}, P. 2008, \apj, 686, 1503,
  \dodoi{10.1086/591786}

\bibitem[{{Bruzual} \& {Charlot}(2003)}]{GBruzual2003}
{Bruzual}, G., \& {Charlot}, S. 2003, \mnras, 344, 1000,
  \dodoi{10.1046/j.1365-8711.2003.06897.x}

\bibitem[{{Cappellari}(2023)}]{MCappellari2023}
{Cappellari}, M. 2023, \mnras, 526, 3273, \dodoi{10.1093/mnras/stad2597}

\bibitem[{{Cappellari} \& {Emsellem}(2004)}]{MCappellari2004}
{Cappellari}, M., \& {Emsellem}, E. 2004, \pasp, 116, 138,
  \dodoi{10.1086/381875}

\bibitem[{{Cappellari} {et~al.}(2006){Cappellari}, {Bacon}, {Bureau}, {Damen},
  {Davies}, {de Zeeuw}, {Emsellem}, {Falc{\'o}n-Barroso}, {Krajnovi{\'c}},
  {Kuntschner}, {McDermid}, {Peletier}, {Sarzi}, {van den Bosch}, \& {van de
  Ven}}]{MCappellari2006}
{Cappellari}, M., {Bacon}, R., {Bureau}, M., {et~al.} 2006, \mnras, 366, 1126,
  \dodoi{10.1111/j.1365-2966.2005.09981.x}

\bibitem[{{Carnall} {et~al.}(2022){Carnall}, {McLure}, {Dunlop}, {Hamadouche},
  {Cullen}, {McLeod}, {Begley}, {Amorin}, {Bolzonella}, {Castellano},
  {Cimatti}, {Fontanot}, {Gargiulo}, {Garilli}, {Mannucci}, {Pentericci},
  {Talia}, {Zamorani}, {Calabro}, {Cresci}, \& {Hathi}}]{ACarnall2022}
{Carnall}, A.~C., {McLure}, R.~J., {Dunlop}, J.~S., {et~al.} 2022, \apj, 929,
  131, \dodoi{10.3847/1538-4357/ac5b62}

\bibitem[{{Carnall} {et~al.}(2023){Carnall}, {McLure}, {Dunlop}, {McLeod},
  {Wild}, {Cullen}, {Magee}, {Begley}, {Cimatti}, {Donnan}, {Hamadouche},
  {Jewell}, \& {Walker}}]{ACarnall2023}
---. 2023, \nat, 619, 716, \dodoi{10.1038/s41586-023-06158-6}

\bibitem[{{Chabrier}(2003)}]{GChabrier2003}
{Chabrier}, G. 2003, \pasp, 115, 763, \dodoi{10.1086/376392}

\bibitem[{{Choi} {et~al.}(2016){Choi}, {Dotter}, {Conroy}, {Cantiello},
  {Paxton}, \& {Johnson}}]{JChoi2016}
{Choi}, J., {Dotter}, A., {Conroy}, C., {et~al.} 2016, \apj, 823, 102,
  \dodoi{10.3847/0004-637X/823/2/102}

\bibitem[{{Cimatti} {et~al.}(2004){Cimatti}, {Daddi}, {Renzini}, {Cassata},
  {Vanzella}, {Pozzetti}, {Cristiani}, {Fontana}, {Rodighiero}, {Mignoli}, \&
  {Zamorani}}]{ACimatti2004}
{Cimatti}, A., {Daddi}, E., {Renzini}, A., {et~al.} 2004, \nat, 430, 184,
  \dodoi{10.1038/nature02668}

\bibitem[{{Conroy} \& {Gunn}(2010)}]{CConroy2010}
{Conroy}, C., \& {Gunn}, J.~E. 2010, \apj, 712, 833,
  \dodoi{10.1088/0004-637X/712/2/833}

\bibitem[{{Conroy} {et~al.}(2009){Conroy}, {Gunn}, \& {White}}]{CConroy2009}
{Conroy}, C., {Gunn}, J.~E., \& {White}, M. 2009, \apj, 699, 486,
  \dodoi{10.1088/0004-637X/699/1/486}

\bibitem[{{Conroy} \& {van Dokkum}(2012)}]{CConroy2012d}
{Conroy}, C., \& {van Dokkum}, P.~G. 2012, \apj, 760, 71,
  \dodoi{10.1088/0004-637X/760/1/71}

\bibitem[{{Conroy} {et~al.}(2018){Conroy}, {Villaume}, {van Dokkum}, \&
  {Lind}}]{CConroy2018}
{Conroy}, C., {Villaume}, A., {van Dokkum}, P.~G., \& {Lind}, K. 2018, \apj,
  854, 139, \dodoi{10.3847/1538-4357/aaab49}

\bibitem[{{Cutler} {et~al.}(2022){Cutler}, {Whitaker}, {Mowla}, {Brammer}, {van
  der Wel}, {Marchesini}, {van Dokkum}, {Momcheva}, {Song}, {Akhshik},
  {Nelson}, {Bezanson}, {Franx}, {Kriek}, {Lange-Vagle}, {Leja}, {MacKenty},
  {Muzzin}, \& {Shipley}}]{SCutler2022}
{Cutler}, S.~E., {Whitaker}, K.~E., {Mowla}, L.~A., {et~al.} 2022, \apj, 925,
  34, \dodoi{10.3847/1538-4357/ac341c}

\bibitem[{{Daddi} {et~al.}(2005){Daddi}, {Renzini}, {Pirzkal}, {Cimatti},
  {Malhotra}, {Stiavelli}, {Xu}, {Pasquali}, {Rhoads}, {Brusa}, {di Serego
  Alighieri}, {Ferguson}, {Koekemoer}, {Moustakas}, {Panagia}, \&
  {Windhorst}}]{EDaddi2005}
{Daddi}, E., {Renzini}, A., {Pirzkal}, N., {et~al.} 2005, \apj, 626, 680,
  \dodoi{10.1086/430104}

\bibitem[{{de Graaff} {et~al.}(2023){de Graaff}, {Rix}, {Carniani}, {Suess},
  {Charlot}, {Curtis-Lake}, {Arribas}, {Baker}, {Boyett}, {Bunker}, {Cameron},
  {Chevallard}, {Curti}, {Eisenstein}, {Franx}, {Hainline}, {Hausen}, {Ji},
  {Johnson}, {Jones}, {Maiolino}, {Maseda}, {Nelson}, {Parlanti}, {Rawle},
  {Robertson}, {Tacchella}, {{\"U}bler}, {Williams}, {Willmer}, \&
  {Willott}}]{AdeGraaff2023}
{de Graaff}, A., {Rix}, H.-W., {Carniani}, S., {et~al.} 2023, arXiv e-prints,
  arXiv:2308.09742, \dodoi{10.48550/arXiv.2308.09742}

\bibitem[{{D'Eugenio} {et~al.}(2023){D'Eugenio}, {Perez-Gonzalez}, {Maiolino},
  {Scholtz}, {Perna}, {Circosta}, {Uebler}, {Arribas}, {Boeker}, {Bunker},
  {Carniani}, {Charlot}, {Chevallard}, {Cresci}, {Curtis-Lake}, {Jones},
  {Kumari}, {Lamperti}, {Looser}, {Parlanti}, {Rix}, {Robertson}, {Rodriguez
  Del Pino}, {Tacchella}, {Venturi}, \& {Willott}}]{FDEugenio2023}
{D'Eugenio}, F., {Perez-Gonzalez}, P., {Maiolino}, R., {et~al.} 2023, arXiv
  e-prints, arXiv:2308.06317, \dodoi{10.48550/arXiv.2308.06317}

\bibitem[{{Esdaile} {et~al.}(2021){Esdaile}, {Glazebrook}, {Labb{\'e}},
  {Taylor}, {Schreiber}, {Nanayakkara}, {Kacprzak}, {Oesch}, {Tran},
  {Papovich}, {Spitler}, \& {Straatman}}]{JEsdaile2021}
{Esdaile}, J., {Glazebrook}, K., {Labb{\'e}}, I., {et~al.} 2021, \apjl, 908,
  L35, \dodoi{10.3847/2041-8213/abe11e}

\bibitem[{{Forrest} {et~al.}(2020){Forrest}, {Annunziatella}, {Wilson},
  {Marchesini}, {Muzzin}, {Cooper}, {Marsan}, {McConachie}, {Chan}, {Gomez},
  {Kado-Fong}, {L Barbera}, {Labb{\'e}}, {Lange-Vagle}, {Nantais}, {Nonino},
  {Pe{\~n}a}, {Saracco}, {Stefanon}, \& {van der Burg}}]{BForrest2020}
{Forrest}, B., {Annunziatella}, M., {Wilson}, G., {et~al.} 2020, \apjl, 890,
  L1, \dodoi{10.3847/2041-8213/ab5b9f}

\bibitem[{{Forrest} {et~al.}(2022){Forrest}, {Wilson}, {Muzzin}, {Marchesini},
  {Cooper}, {Marsan}, {Annunziatella}, {McConachie}, {Zaidi}, {Gomez}, {Urbano
  Stawinski}, {Chang}, {de Lucia}, {La Barbera}, {Lubin}, {Nantais},
  {Pe{\~n}a}, {Saracco}, {Surace}, \& {Stefanon}}]{BForrest2022}
{Forrest}, B., {Wilson}, G., {Muzzin}, A., {et~al.} 2022, \apj, 938, 109,
  \dodoi{10.3847/1538-4357/ac8747}

\bibitem[{{Franx} {et~al.}(2003){Franx}, {Labb{\'e}}, {Rudnick}, {van Dokkum},
  {Daddi}, {F{\"o}rster Schreiber}, {Moorwood}, {Rix}, {R{\"o}ttgering}, {van
  der Wel}, {van der Werf}, \& {van Starkenburg}}]{MFranx2003}
{Franx}, M., {Labb{\'e}}, I., {Rudnick}, G., {et~al.} 2003, \apjl, 587, L79,
  \dodoi{10.1086/375155}

\bibitem[{{Gim{\'e}nez-Arteaga} {et~al.}(2023){Gim{\'e}nez-Arteaga}, {Oesch},
  {Brammer}, {Valentino}, {Mason}, {Weibel}, {Barrufet}, {Fujimoto}, {Heintz},
  {Nelson}, {Strait}, {Suess}, \& {Gibson}}]{CGimenez2023}
{Gim{\'e}nez-Arteaga}, C., {Oesch}, P.~A., {Brammer}, G.~B., {et~al.} 2023,
  \apj, 948, 126, \dodoi{10.3847/1538-4357/acc5ea}

\bibitem[{{Glazebrook} {et~al.}(2004){Glazebrook}, {Abraham}, {McCarthy},
  {Savaglio}, {Chen}, {Crampton}, {Murowinski}, {J{\o}rgensen}, {Roth}, {Hook},
  {Marzke}, \& {Carlberg}}]{KGlazebrook2004}
{Glazebrook}, K., {Abraham}, R.~G., {McCarthy}, P.~J., {et~al.} 2004, \nat,
  430, 181, \dodoi{10.1038/nature02667}

\bibitem[{{Glazebrook} {et~al.}(2017){Glazebrook}, {Schreiber}, {Labb{\'e}},
  {Nanayakkara}, {Kacprzak}, {Oesch}, {Papovich}, {Spitler}, {Straatman},
  {Tran}, \& {Yuan}}]{KGlazebrook2017}
{Glazebrook}, K., {Schreiber}, C., {Labb{\'e}}, I., {et~al.} 2017, \nat, 544,
  71, \dodoi{10.1038/nature21680}

\bibitem[{{Grogin} {et~al.}(2011){Grogin}, {Kocevski}, {Faber}, {Ferguson},
  {Koekemoer}, {Riess}, {Acquaviva}, {Alexander}, {Almaini}, {Ashby}, {Barden},
  {Bell}, {Bournaud}, {Brown}, {Caputi}, {Casertano}, {Cassata}, {Castellano},
  {Challis}, {Chary}, {Cheung}, {Cirasuolo}, {Conselice}, {Roshan Cooray},
  {Croton}, {Daddi}, {Dahlen}, {Dav{\'e}}, {de Mello}, {Dekel}, {Dickinson},
  {Dolch}, {Donley}, {Dunlop}, {Dutton}, {Elbaz}, {Fazio}, {Filippenko},
  {Finkelstein}, {Fontana}, {Gardner}, {Garnavich}, {Gawiser}, {Giavalisco},
  {Grazian}, {Guo}, {Hathi}, {H{\"a}ussler}, {Hopkins}, {Huang}, {Huang},
  {Jha}, {Kartaltepe}, {Kirshner}, {Koo}, {Lai}, {Lee}, {Li}, {Lotz}, {Lucas},
  {Madau}, {McCarthy}, {McGrath}, {McIntosh}, {McLure}, {Mobasher},
  {Moustakas}, {Mozena}, {Nandra}, {Newman}, {Niemi}, {Noeske}, {Papovich},
  {Pentericci}, {Pope}, {Primack}, {Rajan}, {Ravindranath}, {Reddy}, {Renzini},
  {Rix}, {Robaina}, {Rodney}, {Rosario}, {Rosati}, {Salimbeni}, {Scarlata},
  {Siana}, {Simard}, {Smidt}, {Somerville}, {Spinrad}, {Straughn}, {Strolger},
  {Telford}, {Teplitz}, {Trump}, {van der Wel}, {Villforth}, {Wechsler},
  {Weiner}, {Wiklind}, {Wild}, {Wilson}, {Wuyts}, {Yan}, \&
  {Yun}}]{NGrogin2011}
{Grogin}, N.~A., {Kocevski}, D.~D., {Faber}, S.~M., {et~al.} 2011, \apjs, 197,
  35, \dodoi{10.1088/0067-0049/197/2/35}

\bibitem[{{Gu} {et~al.}(2022){Gu}, {Greene}, {Newman}, {Kreisch},
  {Quenneville}, {Ma}, \& {Blakeslee}}]{MGu2022}
{Gu}, M., {Greene}, J.~E., {Newman}, A.~B., {et~al.} 2022, \apj, 932, 103,
  \dodoi{10.3847/1538-4357/ac69ea}

\bibitem[{{Horne}(1986)}]{KHorne1986}
{Horne}, K. 1986, \pasp, 98, 609, \dodoi{10.1086/131801}

\bibitem[{{Jafariyazani} {et~al.}(2020){Jafariyazani}, {Newman}, {Mobasher},
  {Belli}, {Ellis}, \& {Patel}}]{MJafariyazani2020}
{Jafariyazani}, M., {Newman}, A.~B., {Mobasher}, B., {et~al.} 2020, \apjl, 897,
  L42, \dodoi{10.3847/2041-8213/aba11c}

\bibitem[{{Johnson} {et~al.}(2021){Johnson}, {Leja}, {Conroy}, \&
  {Speagle}}]{BJohnson2021}
{Johnson}, B.~D., {Leja}, J., {Conroy}, C., \& {Speagle}, J.~S. 2021, \apjs,
  254, 22, \dodoi{10.3847/1538-4365/abef67}

\bibitem[{{Kauffmann} {et~al.}(2003){Kauffmann}, {Heckman}, {Tremonti},
  {Brinchmann}, {Charlot}, {White}, {Ridgway}, {Brinkmann}, {Fukugita}, {Hall},
  {Ivezi{\'c}}, {Richards}, \& {Schneider}}]{GKauffmann2003}
{Kauffmann}, G., {Heckman}, T.~M., {Tremonti}, C., {et~al.} 2003, \mnras, 346,
  1055, \dodoi{10.1111/j.1365-2966.2003.07154.x}

\bibitem[{{Kennicutt}(1998)}]{RKennicutt1998}
{Kennicutt}, Jr., R.~C. 1998, \araa, 36, 189,
  \dodoi{10.1146/annurev.astro.36.1.189}

\bibitem[{{Kewley} {et~al.}(2006){Kewley}, {Groves}, {Kauffmann}, \&
  {Heckman}}]{LKewley2006}
{Kewley}, L.~J., {Groves}, B., {Kauffmann}, G., \& {Heckman}, T. 2006, \mnras,
  372, 961, \dodoi{10.1111/j.1365-2966.2006.10859.x}

\bibitem[{{Koekemoer} {et~al.}(2011){Koekemoer}, {Faber}, {Ferguson}, {Grogin},
  {Kocevski}, {Koo}, {Lai}, {Lotz}, {Lucas}, {McGrath}, {Ogaz}, {Rajan},
  {Riess}, {Rodney}, {Strolger}, {Casertano}, {Castellano}, {Dahlen},
  {Dickinson}, {Dolch}, {Fontana}, {Giavalisco}, {Grazian}, {Guo}, {Hathi},
  {Huang}, {van der Wel}, {Yan}, {Acquaviva}, {Alexander}, {Almaini}, {Ashby},
  {Barden}, {Bell}, {Bournaud}, {Brown}, {Caputi}, {Cassata}, {Challis},
  {Chary}, {Cheung}, {Cirasuolo}, {Conselice}, {Roshan Cooray}, {Croton},
  {Daddi}, {Dav{\'e}}, {de Mello}, {de Ravel}, {Dekel}, {Donley}, {Dunlop},
  {Dutton}, {Elbaz}, {Fazio}, {Filippenko}, {Finkelstein}, {Frazer}, {Gardner},
  {Garnavich}, {Gawiser}, {Gruetzbauch}, {Hartley}, {H{\"a}ussler},
  {Herrington}, {Hopkins}, {Huang}, {Jha}, {Johnson}, {Kartaltepe},
  {Khostovan}, {Kirshner}, {Lani}, {Lee}, {Li}, {Madau}, {McCarthy},
  {McIntosh}, {McLure}, {McPartland}, {Mobasher}, {Moreira}, {Mortlock},
  {Moustakas}, {Mozena}, {Nandra}, {Newman}, {Nielsen}, {Niemi}, {Noeske},
  {Papovich}, {Pentericci}, {Pope}, {Primack}, {Ravindranath}, {Reddy},
  {Renzini}, {Rix}, {Robaina}, {Rosario}, {Rosati}, {Salimbeni}, {Scarlata},
  {Siana}, {Simard}, {Smidt}, {Snyder}, {Somerville}, {Spinrad}, {Straughn},
  {Telford}, {Teplitz}, {Trump}, {Vargas}, {Villforth}, {Wagner}, {Wandro},
  {Wechsler}, {Weiner}, {Wiklind}, {Wild}, {Wilson}, {Wuyts}, \&
  {Yun}}]{AKoekemoer2011}
{Koekemoer}, A.~M., {Faber}, S.~M., {Ferguson}, H.~C., {et~al.} 2011, \apjs,
  197, 36, \dodoi{10.1088/0067-0049/197/2/36}

\bibitem[{{Kormendy}(1977)}]{JKormendy1977}
{Kormendy}, J. 1977, \apj, 218, 333, \dodoi{10.1086/155687}

\bibitem[{{Kriek} \& {Conroy}(2013)}]{MKriek2013}
{Kriek}, M., \& {Conroy}, C. 2013, \apjl, 775, L16,
  \dodoi{10.1088/2041-8205/775/1/L16}

\bibitem[{{Kriek} {et~al.}(2009){Kriek}, {van Dokkum}, {Labb{\'e}}, {Franx},
  {Illingworth}, {Marchesini}, \& {Quadri}}]{MKriek2009b}
{Kriek}, M., {van Dokkum}, P.~G., {Labb{\'e}}, I., {et~al.} 2009, \apj, 700,
  221, \dodoi{10.1088/0004-637X/700/1/221}

\bibitem[{{Kriek} {et~al.}(2006){Kriek}, {van Dokkum}, {Franx}, {Quadri},
  {Gawiser}, {Herrera}, {Illingworth}, {Labb{\'e}}, {Lira}, {Marchesini},
  {Rix}, {Rudnick}, {Taylor}, {Toft}, {Urry}, \& {Wuyts}}]{MKriek2006}
{Kriek}, M., {van Dokkum}, P.~G., {Franx}, M., {et~al.} 2006, \apjl, 649, L71,
  \dodoi{10.1086/508371}

\bibitem[{{Kriek} {et~al.}(2007){Kriek}, {van Dokkum}, {Franx}, {Illingworth},
  {Coppi}, {F{\"o}rster Schreiber}, {Gawiser}, {Labb{\'e}}, {Lira},
  {Marchesini}, {Quadri}, {Rudnick}, {Taylor}, {Urry}, \& {van der
  Werf}}]{MKriek2007}
---. 2007, \apj, 669, 776, \dodoi{10.1086/520789}

\bibitem[{{Kriek} {et~al.}(2015){Kriek}, {Shapley}, {Reddy}, {Siana}, {Coil},
  {Mobasher}, {Freeman}, {de Groot}, {Price}, {Sanders}, {Shivaei}, {Brammer},
  {Momcheva}, {Skelton}, {van Dokkum}, {Whitaker}, {Aird}, {Azadi}, {Kassis},
  {Bullock}, {Conroy}, {Dav{\'e}}, {Kere{\v s}}, \& {Krumholz}}]{MKriek2015}
{Kriek}, M., {Shapley}, A.~E., {Reddy}, N.~A., {et~al.} 2015, \apjs, 218, 15,
  \dodoi{10.1088/0067-0049/218/2/15}

\bibitem[{{Kriek} {et~al.}(2016){Kriek}, {Conroy}, {van Dokkum}, {Shapley},
  {Choi}, {Reddy}, {Siana}, {van de Voort}, {Coil}, \& {Mobasher}}]{MKriek2016}
{Kriek}, M., {Conroy}, C., {van Dokkum}, P.~G., {et~al.} 2016, \nat, 540, 248,
  \dodoi{10.1038/nature20570}

\bibitem[{{Kriek} {et~al.}(2019){Kriek}, {Price}, {Conroy}, {Suess}, {Mowla},
  {Pasha}, {Bezanson}, {van Dokkum}, \& {Barro}}]{MKriek2019}
{Kriek}, M., {Price}, S.~H., {Conroy}, C., {et~al.} 2019, \apjl, 880, L31,
  \dodoi{10.3847/2041-8213/ab2e75}

\bibitem[{{Kroupa}(2001)}]{PKroupa2001}
{Kroupa}, P. 2001, \mnras, 322, 231, \dodoi{10.1046/j.1365-8711.2001.04022.x}

\bibitem[{{Leja} {et~al.}(2019){Leja}, {Johnson}, {Conroy}, {van Dokkum},
  {Speagle}, {Brammer}, {Momcheva}, {Skelton}, {Whitaker}, {Franx}, \&
  {Nelson}}]{JLeja2019}
{Leja}, J., {Johnson}, B.~D., {Conroy}, C., {et~al.} 2019, \apj, 877, 140,
  \dodoi{10.3847/1538-4357/ab1d5a}

\bibitem[{{Leja} {et~al.}(2022){Leja}, {Speagle}, {Ting}, {Johnson}, {Conroy},
  {Whitaker}, {Nelson}, {van Dokkum}, \& {Franx}}]{JLeja2022}
{Leja}, J., {Speagle}, J.~S., {Ting}, Y.-S., {et~al.} 2022, \apj, 936, 165,
  \dodoi{10.3847/1538-4357/ac887d}

\bibitem[{{Lonoce} {et~al.}(2015){Lonoce}, {Longhetti}, {Maraston}, {Thomas},
  {Mancini}, {Cimatti}, {Ciocca}, {Citro}, {Daddi}, {di Serego Alighieri},
  {Gargiulo}, {Maiolino}, {Mannucci}, {Moresco}, {Pozzetti}, {Quai}, \&
  {Saracco}}]{ILonoce2015}
{Lonoce}, I., {Longhetti}, M., {Maraston}, C., {et~al.} 2015, \mnras, 454,
  3912, \dodoi{10.1093/mnras/stv2150}

\bibitem[{{Maltby} {et~al.}(2018){Maltby}, {Almaini}, {Wild}, {Hatch},
  {Hartley}, {Simpson}, {Rowlands}, \& {Socolovsky}}]{DMatlby2018}
{Maltby}, D.~T., {Almaini}, O., {Wild}, V., {et~al.} 2018, \mnras, 480, 381,
  \dodoi{10.1093/mnras/sty1794}

\bibitem[{{Maraston}(2005)}]{CMaraston2005}
{Maraston}, C. 2005, \mnras, 362, 799, \dodoi{10.1111/j.1365-2966.2005.09270.x}

\bibitem[{{Marchesini} {et~al.}(2023){Marchesini}, {Brammer}, {Morishita},
  {Bergamini}, {Wang}, {Bradac}, {Roberts-Borsani}, {Strait}, {Treu},
  {Fontana}, {Jones}, {Santini}, {Vulcani}, {Acebron}, {Calabr{\`o}},
  {Castellano}, {Glazebrook}, {Grillo}, {Mercurio}, {Nanayakkara}, {Rosati},
  {Tubthong}, \& {Vanzella}}]{DMarchesini2023}
{Marchesini}, D., {Brammer}, G., {Morishita}, T., {et~al.} 2023, \apjl, 942,
  L25, \dodoi{10.3847/2041-8213/acaaac}

\bibitem[{{McCracken} {et~al.}(2012){McCracken}, {Milvang-Jensen}, {Dunlop},
  {Franx}, {Fynbo}, {Le F{\`e}vre}, {Holt}, {Caputi}, {Goranova}, {Buitrago},
  {Emerson}, {Freudling}, {Hudelot}, {L{\'o}pez-Sanjuan}, {Magnard}, {Mellier},
  {M{\o}ller}, {Nilsson}, {Sutherland}, {Tasca}, \& {Zabl}}]{HMcCracken2012}
{McCracken}, H.~J., {Milvang-Jensen}, B., {Dunlop}, J., {et~al.} 2012, \aap,
  544, A156, \dodoi{10.1051/0004-6361/201219507}

\bibitem[{{McDermid} {et~al.}(2015){McDermid}, {Alatalo}, {Blitz}, {Bournaud},
  {Bureau}, {Cappellari}, {Crocker}, {Davies}, {Davis}, {de Zeeuw}, {Duc},
  {Emsellem}, {Khochfar}, {Krajnovi{\'c}}, {Kuntschner}, {Morganti}, {Naab},
  {Oosterloo}, {Sarzi}, {Scott}, {Serra}, {Weijmans}, \&
  {Young}}]{RMcDermid2015}
{McDermid}, R.~M., {Alatalo}, K., {Blitz}, L., {et~al.} 2015, \mnras, 448,
  3484, \dodoi{10.1093/mnras/stv105}

\bibitem[{{McLean} {et~al.}(2012){McLean}, {Steidel}, {Epps}, {Konidaris},
  {Matthews}, {Adkins}, {Aliado}, {Brims}, {Canfield}, {Cromer}, {Fucik},
  {Kulas}, {Mace}, {Magnone}, {Rodriguez}, {Rudie}, {Trainor}, {Wang}, {Weber},
  \& {Weiss}}]{IMcLean2012}
{McLean}, I.~S., {Steidel}, C.~C., {Epps}, H.~W., {et~al.} 2012, in Society of
  Photo-Optical Instrumentation Engineers (SPIE) Conference Series, Vol. 8446,
  Society of Photo-Optical Instrumentation Engineers (SPIE) Conference Series,
  \dodoi{10.1117/12.924794}

\bibitem[{{McLeod} {et~al.}(2021){McLeod}, {McLure}, {Dunlop}, {Cullen},
  {Carnall}, \& {Duncan}}]{DMcLeod2021}
{McLeod}, D.~J., {McLure}, R.~J., {Dunlop}, J.~S., {et~al.} 2021, \mnras, 503,
  4413, \dodoi{10.1093/mnras/stab731}

\bibitem[{{Mendel} {et~al.}(2015){Mendel}, {Saglia}, {Bender}, {Beifiori},
  {Chan}, {Fossati}, {Wilman}, {Bandara}, {Brammer}, {F{\"o}rster Schreiber},
  {Galametz}, {Kulkarni}, {Momcheva}, {Nelson}, {van Dokkum}, {Whitaker}, \&
  {Wuyts}}]{JMendel2015}
{Mendel}, J.~T., {Saglia}, R.~P., {Bender}, R., {et~al.} 2015, \apjl, 804, L4,
  \dodoi{10.1088/2041-8205/804/1/L4}

\bibitem[{{Mendel} {et~al.}(2020){Mendel}, {Beifiori}, {Saglia}, {Bender},
  {Brammer}, {Chan}, {F{\"o}rster Schreiber}, {Fossati}, {Galametz},
  {Momcheva}, {Nelson}, {Wilman}, \& {Wuyts}}]{JMendel2020}
{Mendel}, J.~T., {Beifiori}, A., {Saglia}, R.~P., {et~al.} 2020, \apj, 899, 87,
  \dodoi{10.3847/1538-4357/ab9ffc}

\bibitem[{{Mercier} {et~al.}(2023){Mercier}, {Shuntov}, {Gavazzi},
  {Nightingale}, {Arango}, {Ilbert}, {Amvrosiadis}, {Ciesla}, {Casey}, {Jin},
  {Faisst}, {Andika}, {Drakos}, {Enia}, {Franco}, {Gillman}, {Gozaliasl},
  {Hayward}, {Huertas-Company}, {Kartaltepe}, {Koekemoer}, {Laigle}, {Le
  Borgne}, {Magdis}, {Mahler}, {Maraston}, {Martin}, {Massey}, {McCracken},
  {Moutard}, {Paquereau}, {Rhodes}, {Robertson}, {Sanders}, {Trebitsch},
  {Tresse}, \& {Vijayan}}]{WMercier2023}
{Mercier}, W., {Shuntov}, M., {Gavazzi}, R., {et~al.} 2023, arXiv e-prints,
  arXiv:2309.15986, \dodoi{10.48550/arXiv.2309.15986}

\bibitem[{{Miller} {et~al.}(2023){Miller}, {van Dokkum}, \&
  {Mowla}}]{TMiller2023}
{Miller}, T.~B., {van Dokkum}, P., \& {Mowla}, L. 2023, \apj, 945, 155,
  \dodoi{10.3847/1538-4357/acbc74}

\bibitem[{{Momcheva} {et~al.}(2017){Momcheva}, {van Dokkum}, {van der Wel},
  {Brammer}, {MacKenty}, {Nelson}, {Leja}, {Muzzin}, \&
  {Franx}}]{IMomcheva2017}
{Momcheva}, I.~G., {van Dokkum}, P.~G., {van der Wel}, A., {et~al.} 2017,
  \pasp, 129, 015004, \dodoi{10.1088/1538-3873/129/971/015004}

\bibitem[{{Mosleh} {et~al.}(2017){Mosleh}, {Tacchella}, {Renzini}, {Carollo},
  {Molaeinezhad}, {Onodera}, {Khosroshahi}, \& {Lilly}}]{MMosleh2017}
{Mosleh}, M., {Tacchella}, S., {Renzini}, A., {et~al.} 2017, \apj, 837, 2,
  \dodoi{10.3847/1538-4357/aa5f14}

\bibitem[{{Mowla} {et~al.}(2018){Mowla}, {van Dokkum}, {Brammer}, {Momcheva},
  {van der Wel}, {Whitaker}, {Nelson}, {Bezanson}, {Muzzin}, {Franx},
  {MacKenty}, {Leja}, {Kriek}, \& {Marchesini}}]{LMowla2018}
{Mowla}, L., {van Dokkum}, P., {Brammer}, G., {et~al.} 2018, ArXiv e-prints.
\newblock \doarXiv{1808.04379}

\bibitem[{{Muzzin} {et~al.}(2013{\natexlab{a}}){Muzzin}, {Marchesini},
  {Stefanon}, {Franx}, {McCracken}, {Milvang-Jensen}, {Dunlop}, {Fynbo},
  {Brammer}, {Labb{\'e}}, \& {van Dokkum}}]{AMuzzin2013b}
{Muzzin}, A., {Marchesini}, D., {Stefanon}, M., {et~al.} 2013{\natexlab{a}},
  \apj, 777, 18, \dodoi{10.1088/0004-637X/777/1/18}

\bibitem[{{Muzzin} {et~al.}(2013{\natexlab{b}}){Muzzin}, {Marchesini},
  {Stefanon}, {Franx}, {Milvang-Jensen}, {Dunlop}, {Fynbo}, {Brammer},
  {Labb{\'e}}, \& {van Dokkum}}]{AMuzzin2013a}
---. 2013{\natexlab{b}}, \apjs, 206, 8, \dodoi{10.1088/0067-0049/206/1/8}

\bibitem[{{Nanayakkara} {et~al.}(2022){Nanayakkara}, {Glazebrook}, {Jacobs},
  {Schreiber}, {Brammer}, {Esdaile}, {Kacprzak}, {Labbe}, {Lagos},
  {Marchesini}, {Marsan}, {Nateghi}, {Oesch}, {Papovich}, {Remus}, \&
  {Tran}}]{TNanayakkara2022}
{Nanayakkara}, T., {Glazebrook}, K., {Jacobs}, C., {et~al.} 2022, arXiv
  e-prints, arXiv:2212.11638, \dodoi{10.48550/arXiv.2212.11638}

\bibitem[{{Newman} {et~al.}(2015){Newman}, {Belli}, \& {Ellis}}]{ANewman2015}
{Newman}, A.~B., {Belli}, S., \& {Ellis}, R.~S. 2015, \apjl, 813, L7,
  \dodoi{10.1088/2041-8205/813/1/L7}

\bibitem[{{Newman} {et~al.}(2018{\natexlab{a}}){Newman}, {Belli}, {Ellis}, \&
  {Patel}}]{ANewman2018b}
{Newman}, A.~B., {Belli}, S., {Ellis}, R.~S., \& {Patel}, S.~G.
  2018{\natexlab{a}}, \apj, 862, 126, \dodoi{10.3847/1538-4357/aacd4f}

\bibitem[{{Newman} {et~al.}(2018{\natexlab{b}}){Newman}, {Belli}, {Ellis}, \&
  {Patel}}]{ANewman2018a}
---. 2018{\natexlab{b}}, \apj, 862, 125, \dodoi{10.3847/1538-4357/aacd4d}

\bibitem[{{Newman} {et~al.}(2012){Newman}, {Ellis}, {Bundy}, \&
  {Treu}}]{ANewman2012}
{Newman}, A.~B., {Ellis}, R.~S., {Bundy}, K., \& {Treu}, T. 2012, \apj, 746,
  162, \dodoi{10.1088/0004-637X/746/2/162}

\bibitem[{{Oke} \& {Gunn}(1983)}]{JOke1983}
{Oke}, J.~B., \& {Gunn}, J.~E. 1983, \apj, 266, 713, \dodoi{10.1086/160817}

\bibitem[{{Oke} {et~al.}(1995){Oke}, {Cohen}, {Carr}, {Cromer}, {Dingizian},
  {Harris}, {Labrecque}, {Lucinio}, {Schaal}, {Epps}, \& {Miller}}]{JOke1995}
{Oke}, J.~B., {Cohen}, J.~G., {Carr}, M., {et~al.} 1995, \pasp, 107, 375,
  \dodoi{10.1086/133562}

\bibitem[{{Onodera} {et~al.}(2015){Onodera}, {Carollo}, {Renzini},
  {Cappellari}, {Mancini}, {Arimoto}, {Daddi}, {Gobat}, {Strazzullo},
  {Tacchella}, \& {Yamada}}]{MOnodera2015}
{Onodera}, M., {Carollo}, C.~M., {Renzini}, A., {et~al.} 2015, \apj, 808, 161,
  \dodoi{10.1088/0004-637X/808/2/161}

\bibitem[{{Papovich} {et~al.}(2001){Papovich}, {Dickinson}, \&
  {Ferguson}}]{CPapovich2001}
{Papovich}, C., {Dickinson}, M., \& {Ferguson}, H.~C. 2001, \apj, 559, 620,
  \dodoi{10.1086/322412}

\bibitem[{{Park} {et~al.}(2023){Park}, {Belli}, {Conroy}, {Tacchella}, {Leja},
  {Cutler}, {Johnson}, {Nelson}, \& {Emami}}]{MPark2023}
{Park}, M., {Belli}, S., {Conroy}, C., {et~al.} 2023, \apj, 953, 119,
  \dodoi{10.3847/1538-4357/acd54a}

\bibitem[{{Peng} {et~al.}(2002){Peng}, {Ho}, {Impey}, \& {Rix}}]{CPeng2002}
{Peng}, C.~Y., {Ho}, L.~C., {Impey}, C.~D., \& {Rix}, H.-W. 2002, \aj, 124,
  266, \dodoi{10.1086/340952}

\bibitem[{{Price} {et~al.}(2016){Price}, {Kriek}, {Shapley}, {Reddy},
  {Freeman}, {Coil}, {de Groot}, {Shivaei}, {Siana}, {Azadi}, {Barro},
  {Mobasher}, {Sanders}, \& {Zick}}]{SPrice2016}
{Price}, S.~H., {Kriek}, M., {Shapley}, A.~E., {et~al.} 2016, \apj, 819, 80,
  \dodoi{10.3847/0004-637X/819/1/80}

\bibitem[{{Price} {et~al.}(2020){Price}, {Kriek}, {Barro}, {Shapley}, {Reddy},
  {Freeman}, {Coil}, {Shivaei}, {Azadi}, {de Groot}, {Siana}, {Mobasher},
  {Sanders}, {Leung}, {Fetherolf}, {Zick}, {{\"U}bler}, \& {F{\"o}rster
  Schreiber}}]{SPrice2020}
{Price}, S.~H., {Kriek}, M., {Barro}, G., {et~al.} 2020, \apj, 894, 91,
  \dodoi{10.3847/1538-4357/ab7990}

\bibitem[{{Runco} {et~al.}(2022){Runco}, {Shapley}, {Kriek}, {Cappellari},
  {Topping}, {Sanders}, {Kokorev}, {Price}, {Reddy}, {Coil}, {Mobasher},
  {Siana}, {Zick}, {Magdis}, {Brammer}, \& {Aird}}]{JRunco2022}
{Runco}, J.~N., {Shapley}, A.~E., {Kriek}, M., {et~al.} 2022, \mnras, 517,
  4405, \dodoi{10.1093/mnras/stac2863}

\bibitem[{{Salpeter}(1955)}]{ESalpeter1955}
{Salpeter}, E.~E. 1955, \apj, 121, 161, \dodoi{10.1086/145971}

\bibitem[{{Saracco} {et~al.}(2019){Saracco}, {La Barbera}, {Gargiulo},
  {Mannucci}, {Marchesini}, {Nonino}, \& {Ciliegi}}]{PSaracco2019}
{Saracco}, P., {La Barbera}, F., {Gargiulo}, A., {et~al.} 2019, \mnras, 484,
  2281, \dodoi{10.1093/mnras/sty3509}

\bibitem[{{Schreiber} {et~al.}(2018){Schreiber}, {Glazebrook}, {Nanayakkara},
  {Kacprzak}, {Labb{\'e}}, {Oesch}, {Yuan}, {Tran}, {Papovich}, {Spitler}, \&
  {Straatman}}]{CSchreiber2018}
{Schreiber}, C., {Glazebrook}, K., {Nanayakkara}, T., {et~al.} 2018, \aap, 618,
  A85, \dodoi{10.1051/0004-6361/201833070}

\bibitem[{{Scoville} {et~al.}(2007){Scoville}, {Aussel}, {Brusa}, {Capak},
  {Carollo}, {Elvis}, {Giavalisco}, {Guzzo}, {Hasinger}, {Impey}, {Kneib},
  {LeFevre}, {Lilly}, {Mobasher}, {Renzini}, {Rich}, {Sanders}, {Schinnerer},
  {Schminovich}, {Shopbell}, {Taniguchi}, \& {Tyson}}]{NScoville2007}
{Scoville}, N., {Aussel}, H., {Brusa}, M., {et~al.} 2007, \apjs, 172, 1,
  \dodoi{10.1086/516585}

\bibitem[{{Setton} {et~al.}(2020){Setton}, {Bezanson}, {Suess}, {Hunt},
  {Greene}, {Kriek}, {Spilker}, {Feldmann}, \& {Narayanan}}]{DSetton2020}
{Setton}, D.~J., {Bezanson}, R., {Suess}, K.~A., {et~al.} 2020, \apj, 905, 79,
  \dodoi{10.3847/1538-4357/abc265}

\bibitem[{{Setton} {et~al.}(2022){Setton}, {Verrico}, {Bezanson}, {Greene},
  {Suess}, {Goulding}, {Spilker}, {Kriek}, {Feldmann}, {Narayanan},
  {Hall-Hooper}, \& {Kado-Fong}}]{DSetton2022}
{Setton}, D.~J., {Verrico}, M., {Bezanson}, R., {et~al.} 2022, \apj, 931, 51,
  \dodoi{10.3847/1538-4357/ac6096}

\bibitem[{{Sharples} {et~al.}(2013){Sharples}, {Bender}, {Agudo Berbel},
  {Bezawada}, {Castillo}, {Cirasuolo}, {Davidson}, {Davies}, {Dubbeldam},
  {Fairley}, {Finger}, {F{\"o}rster Schreiber}, {Gonte}, {Hess}, {Jung},
  {Lewis}, {Lizon}, {Muschielok}, {Pasquini}, {Pirard}, {Popovic}, {Ramsay},
  {Rees}, {Richter}, {Riquelme}, {Rodrigues}, {Saviane}, {Schlichter},
  {Schmidtobreick}, {Segovia}, {Smette}, {Szeifert}, {van Kesteren}, {Wegner},
  \& {Wiezorrek}}]{RSharples2013}
{Sharples}, R., {Bender}, R., {Agudo Berbel}, A., {et~al.} 2013, The Messenger,
  151, 21

\bibitem[{{Shen} {et~al.}(2003){Shen}, {Mo}, {White}, {Blanton}, {Kauffmann},
  {Voges}, {Brinkmann}, \& {Csabai}}]{SShen2003}
{Shen}, S., {Mo}, H.~J., {White}, S. D.~M., {et~al.} 2003, \mnras, 343, 978,
  \dodoi{10.1046/j.1365-8711.2003.06740.x}

\bibitem[{{Steidel} {et~al.}(2014){Steidel}, {Rudie}, {Strom}, {Pettini},
  {Reddy}, {Shapley}, {Trainor}, {Erb}, {Turner}, {Konidaris}, {Kulas}, {Mace},
  {Matthews}, \& {McLean}}]{CSteidel2014}
{Steidel}, C.~C., {Rudie}, G.~C., {Strom}, A.~L., {et~al.} 2014, \apj, 795,
  165, \dodoi{10.1088/0004-637X/795/2/165}

\bibitem[{{Suess} {et~al.}(2019{\natexlab{a}}){Suess}, {Kriek}, {Price}, \&
  {Barro}}]{KSuess2019a}
{Suess}, K.~A., {Kriek}, M., {Price}, S.~H., \& {Barro}, G. 2019{\natexlab{a}},
  \apj, 877, 103, \dodoi{10.3847/1538-4357/ab1bda}

\bibitem[{{Suess} {et~al.}(2019{\natexlab{b}}){Suess}, {Kriek}, {Price}, \&
  {Barro}}]{KSuess2019b}
---. 2019{\natexlab{b}}, \apjl, 885, L22, \dodoi{10.3847/2041-8213/ab4db3}

\bibitem[{{Suess} {et~al.}(2020){Suess}, {Kriek}, {Price}, \&
  {Barro}}]{KSuess2020}
---. 2020, \apjl, 899, L26, \dodoi{10.3847/2041-8213/abacc9}

\bibitem[{{Suess} {et~al.}(2021){Suess}, {Kriek}, {Price}, \&
  {Barro}}]{KSuess2021}
---. 2021, \apj, 915, 87, \dodoi{10.3847/1538-4357/abf1e4}

\bibitem[{{Suess} {et~al.}(2023){Suess}, {Williams}, {Robertson}, {Ji},
  {Johnson}, {Nelson}, {Alberts}, {Hainline}, {D'Eugenio}, {{\"U}bler},
  {Rieke}, {Rieke}, {Bunker}, {Carniani}, {Charlot}, {Eisenstein}, {Maiolino},
  {Stark}, {Tacchella}, \& {Willott}}]{KSuess2023}
{Suess}, K.~A., {Williams}, C.~C., {Robertson}, B., {et~al.} 2023, \apjl, 956,
  L42, \dodoi{10.3847/2041-8213/acf5e6}

\bibitem[{{Tanaka} {et~al.}(2019){Tanaka}, {Valentino}, {Toft}, {Onodera},
  {Shimakawa}, {Ceverino}, {Faisst}, {Gallazzi}, {G{\'o}mez-Guijarro}, {Kubo},
  {Magdis}, {Steinhardt}, {Stockmann}, {Yabe}, \& {Zabl}}]{MTanaka2019}
{Tanaka}, M., {Valentino}, F., {Toft}, S., {et~al.} 2019, \apjl, 885, L34,
  \dodoi{10.3847/2041-8213/ab4ff3}

\bibitem[{{Thomas} {et~al.}(2005){Thomas}, {Maraston}, {Bender}, \& {Mendes de
  Oliveira}}]{DThomas2005}
{Thomas}, D., {Maraston}, C., {Bender}, R., \& {Mendes de Oliveira}, C. 2005,
  \apj, 621, 673, \dodoi{10.1086/426932}

\bibitem[{{Toft} {et~al.}(2017){Toft}, {Zabl}, {Richard}, {Gallazzi},
  {Zibetti}, {Prescott}, {Grillo}, {Man}, {Lee}, {G{\'o}mez-Guijarro},
  {Stockmann}, {Magdis}, \& {Steinhardt}}]{SToft2017}
{Toft}, S., {Zabl}, J., {Richard}, J., {et~al.} 2017, \nat, 546, 510,
  \dodoi{10.1038/nature22388}

\bibitem[{{Tomczak} {et~al.}(2014){Tomczak}, {Quadri}, {Tran}, {Labb{\'e}},
  {Straatman}, {Papovich}, {Glazebrook}, {Allen}, {Brammer}, {Kacprzak},
  {Kawinwanichakij}, {Kelson}, {McCarthy}, {Mehrtens}, {Monson}, {Persson},
  {Spitler}, {Tilvi}, \& {van Dokkum}}]{ATomczak2014}
{Tomczak}, A.~R., {Quadri}, R.~F., {Tran}, K.-V.~H., {et~al.} 2014, \apj, 783,
  85, \dodoi{10.1088/0004-637X/783/2/85}

\bibitem[{{Treu} {et~al.}(2010){Treu}, {Auger}, {Koopmans}, {Gavazzi},
  {Marshall}, \& {Bolton}}]{TTreu2010}
{Treu}, T., {Auger}, M.~W., {Koopmans}, L. V.~E., {et~al.} 2010, \apj, 709,
  1195, \dodoi{10.1088/0004-637X/709/2/1195}

\bibitem[{{Trujillo} {et~al.}(2006){Trujillo}, {F{\"o}rster Schreiber},
  {Rudnick}, {Barden}, {Franx}, {Rix}, {Caldwell}, {McIntosh}, {Toft},
  {H{\"a}ussler}, {Zirm}, {van Dokkum}, {Labb{\'e}}, {Moorwood},
  {R{\"o}ttgering}, {van der Wel}, {van der Werf}, \& {van
  Starkenburg}}]{ITrujillo2006}
{Trujillo}, I., {F{\"o}rster Schreiber}, N.~M., {Rudnick}, G., {et~al.} 2006,
  \apj, 650, 18, \dodoi{10.1086/506464}

\bibitem[{{van de Sande} {et~al.}(2015){van de Sande}, {Kriek}, {Franx},
  {Bezanson}, \& {van Dokkum}}]{JvandeSande2015}
{van de Sande}, J., {Kriek}, M., {Franx}, M., {Bezanson}, R., \& {van Dokkum},
  P.~G. 2015, \apj, 799, 125, \dodoi{10.1088/0004-637X/799/2/125}

\bibitem[{{van de Sande} {et~al.}(2011){van de Sande}, {Kriek}, {Franx}, {van
  Dokkum}, {Bezanson}, {Whitaker}, {Brammer}, {Labb{\'e}}, {Groot}, \&
  {Kaper}}]{JvandeSande2011}
{van de Sande}, J., {Kriek}, M., {Franx}, M., {et~al.} 2011, \apjl, 736, L9,
  \dodoi{10.1088/2041-8205/736/1/L9}

\bibitem[{{van de Sande} {et~al.}(2013){van de Sande}, {Kriek}, {Franx}, {van
  Dokkum}, {Bezanson}, {Bouwens}, {Quadri}, {Rix}, \&
  {Skelton}}]{JvandeSande2013}
---. 2013, \apj, 771, 85, \dodoi{10.1088/0004-637X/771/2/85}

\bibitem[{{van der Wel} {et~al.}(2014){van der Wel}, {Franx}, {van Dokkum},
  {Skelton}, {Momcheva}, {Whitaker}, {Brammer}, {Bell}, {Rix}, {Wuyts},
  {Ferguson}, {Holden}, {Barro}, {Koekemoer}, {Chang}, {McGrath},
  {H{\"a}ussler}, {Dekel}, {Behroozi}, {Fumagalli}, {Leja}, {Lundgren},
  {Maseda}, {Nelson}, {Wake}, {Patel}, {Labb{\'e}}, {Faber}, {Grogin}, \&
  {Kocevski}}]{AvanderWel2014}
{van der Wel}, A., {Franx}, M., {van Dokkum}, P.~G., {et~al.} 2014, \apj, 788,
  28, \dodoi{10.1088/0004-637X/788/1/28}

\bibitem[{{van der Wel} {et~al.}(2016){van der Wel}, {Noeske}, {Bezanson},
  {Pacifici}, {Gallazzi}, {Franx}, {Mu{\~n}oz-Mateos}, {Bell}, {Brammer},
  {Charlot}, {Chauk{\'e}}, {Labb{\'e}}, {Maseda}, {Muzzin}, {Rix}, {Sobral},
  {van de Sande}, {van Dokkum}, {Wild}, \& {Wolf}}]{AvanderWel2016}
{van der Wel}, A., {Noeske}, K., {Bezanson}, R., {et~al.} 2016, \apjs, 223, 29,
  \dodoi{10.3847/0067-0049/223/2/29}

\bibitem[{{van der Wel} {et~al.}(2021){van der Wel}, {Bezanson}, {D'Eugenio},
  {Straatman}, {Franx}, {van Houdt}, {Maseda}, {Gallazzi}, {Wu}, {Pacifici},
  {Barisic}, {Brammer}, {Munoz-Mateos}, {Vervalcke}, {Zibetti}, {Sobral}, {de
  Graaff}, {Calhau}, {Kaushal}, {Muzzin}, {Bell}, \& {van
  Dokkum}}]{AvanderWel2021}
{van der Wel}, A., {Bezanson}, R., {D'Eugenio}, F., {et~al.} 2021, \apjs, 256,
  44, \dodoi{10.3847/1538-4365/ac1356}

\bibitem[{{van der Wel} {et~al.}(2022){van der Wel}, {van Houdt}, {Bezanson},
  {Franx}, {D'Eugenio}, {Straatman}, {Bell}, {Muzzin}, {Sobral}, {Maseda}, {de
  Graaff}, \& {Holden}}]{AvanderWel2022}
{van der Wel}, A., {van Houdt}, J., {Bezanson}, R., {et~al.} 2022, \apj, 936,
  9, \dodoi{10.3847/1538-4357/ac83c5}

\bibitem[{{van Dokkum} {et~al.}(2023){van Dokkum}, {Brammer}, {Wang}, {Leja},
  \& {Conroy}}]{PvanDokkum2023}
{van Dokkum}, P., {Brammer}, G., {Wang}, B., {Leja}, J., \& {Conroy}, C. 2023,
  arXiv e-prints, arXiv:2309.07969, \dodoi{10.48550/arXiv.2309.07969}

\bibitem[{{van Dokkum}(2001)}]{PvanDokkum2001}
{van Dokkum}, P.~G. 2001, \pasp, 113, 1420, \dodoi{10.1086/323894}

\bibitem[{{van Dokkum} \& {Conroy}(2010)}]{PvanDokkum2010IMF}
{van Dokkum}, P.~G., \& {Conroy}, C. 2010, \nat, 468, 940,
  \dodoi{10.1038/nature09578}

\bibitem[{{van Dokkum} {et~al.}(2009){van Dokkum}, {Kriek}, \&
  {Franx}}]{PvanDokkum2009}
{van Dokkum}, P.~G., {Kriek}, M., \& {Franx}, M. 2009, \nat, 460, 717,
  \dodoi{10.1038/nature08220}

\bibitem[{{van Dokkum} {et~al.}(2008){van Dokkum}, {Franx}, {Kriek}, {Holden},
  {Illingworth}, {Magee}, {Bouwens}, {Marchesini}, {Quadri}, {Rudnick},
  {Taylor}, \& {Toft}}]{PvanDokkum2008}
{van Dokkum}, P.~G., {Franx}, M., {Kriek}, M., {et~al.} 2008, \apjl, 677, L5,
  \dodoi{10.1086/587874}

\bibitem[{{van Dokkum} {et~al.}(2010){van Dokkum}, {Whitaker}, {Brammer},
  {Franx}, {Kriek}, {Labb{\'e}}, {Marchesini}, {Quadri}, {Bezanson},
  {Illingworth}, {Muzzin}, {Rudnick}, {Tal}, \& {Wake}}]{PvanDokkum2010}
{van Dokkum}, P.~G., {Whitaker}, K.~E., {Brammer}, G., {et~al.} 2010, \apj,
  709, 1018, \dodoi{10.1088/0004-637X/709/2/1018}

\bibitem[{{van Houdt} {et~al.}(2021){van Houdt}, {van der Wel}, {Bezanson},
  {Franx}, {d'Eugenio}, {Barisic}, {Bell}, {Gallazzi}, {de Graaff}, {Maseda},
  {Pacifici}, {van de Sande}, {Sobral}, {Straatman}, \& {Wu}}]{JvanHoudt2021}
{van Houdt}, J., {van der Wel}, A., {Bezanson}, R., {et~al.} 2021, \apj, 923,
  11, \dodoi{10.3847/1538-4357/ac1f29}

\bibitem[{{Vernet} {et~al.}(2011){Vernet}, {Dekker}, {D'Odorico}, {Kaper},
  {Kjaergaard}, {Hammer}, {Randich}, {Zerbi}, {Groot}, {Hjorth}, {Guinouard},
  {Navarro}, {Adolfse}, {Albers}, {Amans}, {Andersen}, {Andersen}, {Binetruy},
  {Bristow}, {Castillo}, {Chemla}, {Christensen}, {Conconi}, {Conzelmann},
  {Dam}, {de Caprio}, {de Ugarte Postigo}, {Delabre}, {di Marcantonio},
  {Downing}, {Elswijk}, {Finger}, {Fischer}, {Flores}, {Fran{\c{c}}ois},
  {Goldoni}, {Guglielmi}, {Haigron}, {Hanenburg}, {Hendriks}, {Horrobin},
  {Horville}, {Jessen}, {Kerber}, {Kern}, {Kiekebusch}, {Kleszcz}, {Klougart},
  {Kragt}, {Larsen}, {Lizon}, {Lucuix}, {Mainieri}, {Manuputy}, {Martayan},
  {Mason}, {Mazzoleni}, {Michaelsen}, {Modigliani}, {Moehler}, {M{\o}ller},
  {Norup S{\o}rensen}, {N{\o}rregaard}, {P{\'e}roux}, {Patat}, {Pena}, {Pragt},
  {Reinero}, {Rigal}, {Riva}, {Roelfsema}, {Royer}, {Sacco}, {Santin},
  {Schoenmaker}, {Spano}, {Sweers}, {Ter Horst}, {Tintori}, {Tromp}, {van
  Dael}, {van der Vliet}, {Venema}, {Vidali}, {Vinther}, {Vola}, {Winters},
  {Wistisen}, {Wulterkens}, \& {Zacchei}}]{JVernet2011}
{Vernet}, J., {Dekker}, H., {D'Odorico}, S., {et~al.} 2011, \aap, 536, A105,
  \dodoi{10.1051/0004-6361/201117752}

\bibitem[{{Wang} {et~al.}(2023){Wang}, {Leja}, {Atek}, {Labbe}, {Li},
  {Bezanson}, {Brammer}, {Cutler}, {Dayal}, {Furtak}, {Greene}, {Kokorev},
  {Pan}, {Price}, {Suess}, {Weaver}, {Whitaker}, \& {Williams}}]{BWang2023}
{Wang}, B., {Leja}, J., {Atek}, H., {et~al.} 2023, arXiv e-prints,
  arXiv:2310.06781, \dodoi{10.48550/arXiv.2310.06781}

\bibitem[{{Wellons} {et~al.}(2015){Wellons}, {Torrey}, {Ma}, {Rodriguez-Gomez},
  {Vogelsberger}, {Kriek}, {van Dokkum}, {Nelson}, {Genel}, {Pillepich},
  {Springel}, {Sijacki}, {Snyder}, {Nelson}, {Sales}, \&
  {Hernquist}}]{SWellons2015}
{Wellons}, S., {Torrey}, P., {Ma}, C.-P., {et~al.} 2015, \mnras, 449, 361,
  \dodoi{10.1093/mnras/stv303}

\bibitem[{{Whitaker} {et~al.}(2012){Whitaker}, {Kriek}, {van Dokkum},
  {Bezanson}, {Brammer}, {Franx}, \& {Labb{\'e}}}]{KWhitaker2012}
{Whitaker}, K.~E., {Kriek}, M., {van Dokkum}, P.~G., {et~al.} 2012, \apj, 745,
  179, \dodoi{10.1088/0004-637X/745/2/179}

\bibitem[{{Williams} {et~al.}(2009){Williams}, {Quadri}, {Franx}, {van Dokkum},
  \& {Labb{\'e}}}]{RWilliams2009}
{Williams}, R.~J., {Quadri}, R.~F., {Franx}, M., {van Dokkum}, P., \&
  {Labb{\'e}}, I. 2009, \apj, 691, 1879, \dodoi{10.1088/0004-637X/691/2/1879}

\bibitem[{{Wisnioski} {et~al.}(2015){Wisnioski}, {F{\"o}rster Schreiber},
  {Wuyts}, {Wuyts}, {Bandara}, {Wilman}, {Genzel}, {Bender}, {Davies},
  {Fossati}, {Lang}, {Mendel}, {Beifiori}, {Brammer}, {Chan}, {Fabricius},
  {Fudamoto}, {Kulkarni}, {Kurk}, {Lutz}, {Nelson}, {Momcheva}, {Rosario},
  {Saglia}, {Seitz}, {Tacconi}, \& {van Dokkum}}]{EWisnioski2015}
{Wisnioski}, E., {F{\"o}rster Schreiber}, N.~M., {Wuyts}, S., {et~al.} 2015,
  \apj, 799, 209, \dodoi{10.1088/0004-637X/799/2/209}

\bibitem[{{Wu} {et~al.}(2020){Wu}, {van der Wel}, {Bezanson}, {Gallazzi},
  {Pacifici}, {Straatman}, {Bari{\v{s}}i{\'c}}, {Bell}, {Chauke}, {D'Eugenio},
  {Franx}, {Muzzin}, {Sobral}, \& {van Houdt}}]{PFWu2020}
{Wu}, P.-F., {van der Wel}, A., {Bezanson}, R., {et~al.} 2020, \apj, 888, 77,
  \dodoi{10.3847/1538-4357/ab5fd9}

\bibitem[{{Wuyts} {et~al.}(2007){Wuyts}, {Labb{\'e}}, {Franx}, {Rudnick}, {van
  Dokkum}, {Fazio}, {F{\"o}rster Schreiber}, {Huang}, {Moorwood}, {Rix},
  {R{\"o}ttgering}, \& {van der Werf}}]{SWuyts2007}
{Wuyts}, S., {Labb{\'e}}, I., {Franx}, M., {et~al.} 2007, \apj, 655, 51,
  \dodoi{10.1086/509708}

\bibitem[{{Yan} \& {Blanton}(2012)}]{RYan2012}
{Yan}, R., \& {Blanton}, M.~R. 2012, \apj, 747, 61,
  \dodoi{10.1088/0004-637X/747/1/61}

\bibitem[{{Yano} {et~al.}(2016){Yano}, {Kriek}, {van der Wel}, \&
  {Whitaker}}]{MYano2016}
{Yano}, M., {Kriek}, M., {van der Wel}, A., \& {Whitaker}, K.~E. 2016, \apjl,
  817, L21, \dodoi{10.3847/2041-8205/817/2/L21}

\bibitem[{{Zhuang} {et~al.}(2023){Zhuang}, {Leethochawalit}, {Kirby},
  {Nightingale}, {Steidel}, {Glazebrook}, {Barone}, {Skobe}, {Sweet},
  {Nanayakkara}, {Allen}, {Vasan}, {Jones}, {Kacprzak}, {Tran}, \&
  {Jacobs}}]{ZZhuang2023}
{Zhuang}, Z., {Leethochawalit}, N., {Kirby}, E.~N., {et~al.} 2023, \apj, 948,
  132, \dodoi{10.3847/1538-4357/acc79b}

\end{thebibliography}

\end{document}